\documentclass[12pt,preprint]{aastex63}
\usepackage{amssymb}
\usepackage{amsmath}
\usepackage{romannum}
\usepackage{epstopdf}
\usepackage{xcolor}
\usepackage{amsmath}
\usepackage{graphicx}
\usepackage{hyperref}
\usepackage{graphics,graphicx,natbib}
\usepackage{longtable}
\usepackage{bm}
\usepackage{booktabs} % To thicken table lines

\usepackage{subfigure}
\def\lapp{\ifmmode\stackrel{<}{_{\sim}}\else$\stackrel{<}{_{\sim}}$\fi}
\def\gapp{\ifmmode\stackrel{>}{_{\sim}}\else$\stackrel{>}{_{\sim}}$\fi}

\newcommand{\degrees}{^{\circ}}

\usepackage{enumerate}
\usepackage{tabularx}
\begin{document}
\pagenumbering{arabic}
\title{Host Galaxies for Four Nearby CHIME/FRB Sources and the Local Universe FRB Host Galaxy Population}
%NGC 3252

% affiliations
\newcommand{\CU}{Cornell Center for Astrophysics and Planetary Science, Cornell University, Ithaca, NY 14853, USA}
\newcommand{\CMU}{McWilliams Center for Cosmology, Department of Physics, Carnegie Mellon University, Pittsburgh, PA 15213, USA}
\newcommand{\mcgillphysics}{Department of Physics, McGill University, 3600 rue University, Montr\'eal, QC H3A 2T8, Canada}
\newcommand{\tsi}{Trottier Space Institute, McGill University, 3550 rue University, Montr\'eal, QC H3A 2A7, Canada}
\newcommand{\wvuphysics}{Department of Physics and Astronomy, West Virginia University, P.O. Box 6315, Morgantown, WV 26506, USA}
\newcommand{\wvugws}{Center for Gravitational Waves and Cosmology, West Virginia University, Chestnut Ridge Research Building, Morgantown, WV 26505, USA}
\newcommand{\uoftphysics}{Department of Physics, University of Toronto, 60 St. George Street, Toronto, ON M5S 1A7, Canada}
\newcommand{\cita}{Canadian Institute for Theoretical Astrophysics, 60 St. George Street, Toronto, ON M5S 3H8, Canada}
\newcommand{\dunlapinstitute}{Dunlap Institute for Astronomy \& Astrophysics, University of Toronto, 50 St. George Street, Toronto, ON M5S 3H4, Canada}
\newcommand{\dunlapdep}{David A. Dunlap Department of Astronomy \& Astrophysics, University of Toronto, 50 St. George Street, Toronto, ON M5S 3H4, Canada}

\newcommand{\mitkavli}{MIT Kavli Institute for Astrophysics and Space Research, Massachusetts Institute of Technology, 77 Massachusetts Ave, Cambridge, MA 02139, USA}
\newcommand{\mitphysics}{Department of Physics, Massachusetts Institute of Technology, 77 Massachusetts Ave, Cambridge, MA 02139, USA}
\newcommand{\ubc}{Dept. of Physics and Astronomy, 6224 Agricultural Road, Vancouver, BC V6T 1Z1 Canada}
\newcommand{\sidrat}{Sidrat Research, PO Box 73527 RPO Wychwood, Toronto, Ontario, M6C 4A7, Canada}
\newcommand{\perimeter}{Perimeter Institute for Theoretical Physics, 31 Caroline Street N, Waterloo ON N2L 2Y5 Canada}
\newcommand{\tata}{Department of Astronomy and Astrophysics, Tata Institute of Fundamental Research, Mumbai, 400005, India}
\newcommand{\ncra}{National Centre for Radio Astrophysics, Post Bag 3, Ganeshkhind, Pune, 411007, India}
\newcommand{\drao}{Dominion Radio Astrophysical Observatory, Herzberg Astronomy \& Astrophysics Research Centre, National Research Council Canada, PO Box 248, Penticton, BC V2A 6J9, Canada}

\newcommand{\waterloo}{Department of Physics and Astronomy, University of Waterloo, Waterloo, ON N2L 3G1, Canada}
\newcommand{\maxico}{Instituto de Astronomía, Universidad Nacional Autónoma de México, Apdo. Postal 877, Ensenada, Baja California 22800, México}
\newcommand{\ioffe}{Ioffe Institute, 26 Politekhnicheskaya st., St. Petersburg 194021, Russia}
\newcommand{\maxicoinstitute}{Instituto Nacional de Astrof{\'\i}sica, \'Optica y Electr\'onica, Luis Enrique Erro 1, Tonantzintla 72840, Puebla, Mexico}
\newcommand{\csiro}{CSIRO Space \& Astronomy, Parkes Observatory, P.O. Box 276, Parkes NSW 2870, Australia}

\newcommand{\FRQNT}{Fonds de Recherche du Quebec -- Nature et Technologies~(FRQNT) Postdoctoral Fellow}
\newcommand{\BANTING}{Banting Fellow}
\newcommand{\MSIF}{McGill Space Institute Fellow}
\newcommand{\nhfp}{NASA Hubble Fellowship Program~(NHFP) Einstein Fellow}
\newcommand{\UCSC}{Department of Astronomy and Astrophysics, University of California Santa Cruz, Santa Cruz, CA 95064, USA}
\newcommand{\IPMU}{Kavli Institute for the Physics and Mathematics of the Universe (Kavli IPMU), 5-1-5 Kashiwanoha, Kashiwa, 277-8583, Japan}
\newcommand{\NAOJ}{Division of Science, National Astronomical Observatory of Japan, 2-21-1 Osawa, Mitaka, Tokyo 181-8588, Japan}
\newcommand{\ucbastro}{Department of Astronomy, University of California Berkeley, Berkeley, CA 94720, USA}

%\shorttitle{}
%\shortauthors{}

\author[0000-0002-3615-3514]{Mohit~Bhardwaj}
\affiliation{\CMU}
\affiliation{\tsi}
\affiliation{\mcgillphysics}

\author[0000-0002-2551-7554]{Daniele Michilli}
\affiliation{\mitkavli}
\affiliation{\mitphysics}

\author[0000-0002-8139-8414]{Aida~Yu.~Kirichenko}
\affiliation{\maxico}

\author{Obinna~Modilim}
\affiliation{\mitphysics}

\author[0000-0002-6823-2073]{Kaitlyn~Shin}
\affiliation{\mitkavli}
\affiliation{\mitphysics}

\author[0000-0001-9345-0307]{Victoria~M.~Kaspi}
\affiliation{\tsi}
\affiliation{\mcgillphysics}

\author[0000-0001-5908-3152]{Bridget~C.~Andersen}
\affiliation{\tsi}
\affiliation{\mcgillphysics}

\author[0000-0003-2047-5276]{Tomas~Cassanelli}
\affiliation{Department of Electrical Engineering, Universidad de Chile, Av. Tupper 2007, Santiago 8370451, Chile}

\author[0000-0002-1800-8233]{Charanjot~Brar}
\affiliation{\tsi}
\affiliation{\mcgillphysics}

\author[0000-0002-2878-1502]{Shami~Chatterjee}
\affiliation{\CU}

\author[0000-0001-6422-8125]{Amanda~M.~Cook}
\affiliation{\dunlapdep}
\affiliation{\uoftphysics}

\author[0000-0003-4098-5222]{Fengqiu~Adam~Dong}
\affiliation{\ubc}

\author[0000-0001-8384-5049]{Emmanuel Fonseca}
\affiliation{\wvugws}
\affiliation{\wvuphysics}

\author[0000-0002-3382-9558]{B. M. Gaensler}
\affiliation{\dunlapdep}
\affiliation{\uoftphysics}
\affiliation{Present address: Division of Physical and Biological Sciences, University of California Santa Cruz, Santa Cruz, CA 95064, USA}

\author[0000-0003-2405-2967]{Adaeze~L.~Ibik}
\affiliation{\dunlapdep}
\affiliation{\uoftphysics}

\author[0000-0003-4810-7803]{J.~F.~Kaczmarek}
\affiliation{\csiro}

\author[0000-0003-2116-3573]{Adam~E.~Lanman}
\affiliation{\mitkavli}
\affiliation{\mitphysics}

\author[0000-0002-4209-7408]{Calvin~Leung}
\affiliation{\ucbastro}
\affiliation{\nhfp}

\author[0000-0002-4279-6946]{K.~W.~Masui}
\affiliation{\mitkavli}
\affiliation{\mitphysics}

\author[0000-0002-8897-1973]{Ayush~Pandhi}
\affiliation{\dunlapdep}
\affiliation{\uoftphysics}

\author[0000-0002-8912-0732]{Aaron~B.~Pearlman}
\affiliation{\tsi}
\affiliation{\mcgillphysics}
\affiliation{\BANTING}
\affiliation{\MSIF}
\affiliation{\FRQNT}

\author[0000-0002-4795-697X]{Ziggy~Pleunis}
\affiliation{\dunlapdep}

\author{J.~Xavier~Prochaska}
\affiliation{\UCSC}
\affiliation{\IPMU}
\affiliation{\NAOJ}

\author[0000-0001-7694-6650]{Masoud~Rafiei-Ravandi}
\affiliation{\tsi}
\affiliation{\mcgillphysics}

\author[0000-0003-3154-3676]{Ketan~R.~Sand}
\affiliation{\tsi}
\affiliation{\mcgillphysics}

\author[0000-0002-7374-7119]{Paul~Scholz}
\affiliation{\dunlapdep}
\affiliation{Department of Physics and Astronomy, York University, 4700 Keele Street, Toronto, Ontario, ON MJ3 1P3, Canada}

\author[0000-0002-2088-3125]{Kendrick M.~Smith}
\affiliation{\perimeter}

%\email{mohitb@andrew.cmu.edu}
%\author{CHIME/FRB collaboration}

\correspondingauthor{Mohit Bhardwaj}
\email{mohitb@andrew.cmu.edu}
\begin{abstract}
We present the host galaxies of four apparently non-repeating fast radio bursts (FRBs), FRBs 20181223C, 20190418A, 20191220A, and 20190425A, reported in the first Canadian Hydrogen Intensity Mapping Experiment (CHIME/FRB) catalog. Our selection of these FRBs is based on a planned hypothesis testing framework where we search all CHIME/FRB Catalog-1 events that have low extragalactic dispersion measure ($<$ 100 pc cm$^{-3}$), with high Galactic latitude ($\lvert$b$\rvert> 10\degrees$) and saved baseband data. We 
%then robustly 
associate the selected FRBs
%, FRB 20181220A, FRB 20181223C, FRB 20190418A and FRB 20190425A, 
to galaxies with moderate to high star-formation rates located at redshifts between 0.027 and 0.071. We also search for possible multi-messenger counterparts, including persistent compact radio and gravitational wave (GW) sources, and find none. 
%Among the FRBs in our sample, FRB 20190425A was previously suggested to be linked with GW 190425, which has come under scrutiny recently. Here we present additional evidence refuting this association.
%We next use these four FRBs along with 16 published local Universe FRBs (z$<$0.1) to conduct a comprehensive host demographic analysis. 
Utilizing the four FRB hosts from this study along with the hosts of 14 published local Universe FRBs (z $<0.1$) with robust host association, we conduct an FRB host demographics analysis.
We find all 18 local Universe FRB hosts in our sample to be spirals (or late-type galaxies), including the host of FRB 20220509G, which was previously reported to be elliptical. Using this observation, we scrutinize proposed FRB source formation channels and argue that core-collapse supernovae are likely the dominant channel to form FRB progenitors. Moreover, we infer no significant difference in the host properties of repeating and apparently non-repeating FRBs in our local Universe FRB host sample. 
%We then examine all proposed FRB formation channels in light of the observed preponderance of late-type local Universe hosts and argue that core-collapse supernovae (CCSNe) is the dominant FRB progenitor formation channel. 
%We then compare the host galaxies of Type II CCSNe which we find statistically similar to the observed local Universe FRB hosts. 
Finally, we find the burst rates of these four apparently non-repeating FRBs to be consistent with those of the sample of localized repeating FRBs observed by CHIME/FRB. Therefore, we encourage further monitoring of these FRBs with more sensitive radio telescopes. 

\end{abstract}

%\keywords{pulsars: general --- pulsars: timing --- surveys, RRATs}

\section{Introduction}
\label{sec:intro}
%reported a catalog of 536 fast radio bursts (FRBs) detected between 400 and 800 MHz from 2018 July 25 to 2019 July 1, including 62 bursts from 18 previously reported repeating sources.  based on the low. However, only a fraction of them have baseband localizations that is required for the robust host assocition of nearby FRBs.

Fast radio bursts (FRBs) are energetic transients of coherent radio emission that last for $\sim$ few milliseconds \citep{lbm+07,tsb+13} and are observed out to cosmological distances \citep[for more detail, see][]{2022A&ARv..30....2P}. Since the discovery of the first FRB in 2007, $\sim 1000$ FRBs have been reported to date.\footnote{For a complete list of known FRBs, see \url{https://www.herta-experiment.org/frbstats/} \citep{2021ascl.soft06028S} or the TNS: \url{https://www.wis-tns.org/} \citep{2020TNSAN..70....1Y}.} However, their origin continues to be a subject of intense debate. So far, extragalactic FRBs have exclusively manifested as radio phenomena. Consequently, given the lack of prompt or afterglow counterparts at other wavelengths, it becomes imperative to investigate their host galaxies and local surroundings to unravel their progenitors \citep{2020ApJ...903..152H,2022AJ....163...69B,2023arXiv230205465G}. For example, analyzing the local environment of FRBs can help in understanding the necessary conditions for the formation of their progenitors \citep{2017ApJ...843L...8B,2020Natur.577..190M,2021ApJ...917...75M,2021ApJ...908L..12T}.

%Moreover, host galaxies provide a wealth of information on stellar population properties such as stellar population age, stellar mass, metallicity, and star formation rate. These in turn can give insight into properties of FRB progenitors, including how delay times, or time between formation and merger, evolve with redshift and how compact object systems track the star formation and stellar mass in the universe, which are difficult to constrain through other methods.

Numerous models have been proposed to explain the origins of FRBs, encompassing both cataclysmic and non-cataclysmic formation channels \citep[for a review of FRB models, see][]{pww+18}. However, because of the high volumetric rate of FRBs \citep{2020MNRAS.494..665L, 2022MNRAS.511.1961H,2023ApJ...944..105S}, it is likely that most FRBs are repeating sources \citep{2019raviNat, 2021ApJ...919L..24B,2023arXiv230617403J}. Moreover, the majority of proposed progenitor models invoke young, highly magnetized neutron stars as FRB sources \citep{2022arXiv221203972Z}. The neutron star origin hypothesis gained strong support after the discovery of radio bursts from the Galactic magnetar, SGR 1935+2154 \citep{2020SGR,Bochenek2020},
%and 1E 1547.0–5408 \citep{2021ApJ...907....7I}
which is reminiscent of FRBs. It is interesting to note that SGR 1935+2154 is likely formed via core-collapse supernovae \citep{2018ApJ...852...54K}, one of the prompt formation channels proposed for FRB progenitors \citep{2022arXiv221203972Z}. Moreover, several other Galactic magnetars are found in supernova remnants, e.g., 1E 2259+586 or 1E 1841-045, which supports this argument.\footnote{See \url{https://www.physics.mcgill.ca/~pulsar/magnetar/main.html}
or \cite{2014ApJS..212....6O}.}

However, the recent discovery of a repeating FRB 20200120E \citep{bhardwaj2021} in an old globular cluster (GC) of the galaxy Messier 81 (M81) challenges this hypothesis and suggests the possibility of dynamically forming FRB sources in dense cluster cores through delayed formation channels \citep{Kristan2021arXiv}, such as accretion-induced collapse of white dwarfs and binary white dwarf mergers \citep{2021ApJ...917L..11K}.

%Moreover, the high volumetric rate of FRBs and/or diverse properties of the localized FRB hosts disfavour certain prompt-formation, such as superluminous supernovae and long GRBs, and  as delayed-formation channels, such as short GRBs, as the channels to form even a considerable fraction of FRB sources.

It is also conceivable that both delayed and prompt formation channels contribute to the population of FRB sources. This notion finds support in the observation that FRBs have been identified across a wide range of galactic environments, encompassing both actively star-forming galaxies and quiescent ones. The diversity within these host environments implies a broad range of potential formation timescales for the progenitors of FRBs  \citep{2023arXiv230205465G,2023arXiv230703344L}. However, it is important to note that existing FRB host demographic studies have not yet taken into consideration potential biases arising from radio and optical selection effects. These biases could potentially influence the core conclusions drawn from such studies \citep[][]{2021arXiv211207639S,2023MNRAS.523.5006J}. Regardless, a statistically large sample of localized FRBs, particularly those located in the local Universe (z $\lesssim$ 0.1), is essential to test different proposed FRB formation channels. Due to the limited sensitivity of current telescopes, in-depth study of FRB local environments is mainly possible for nearby sources. Consequently, realizing the complete potential of multi-wavelength and multi-messenger follow-ups depends on the proximity of FRBs. This facilitates endeavors such as detecting prompt X-ray emission from FRB sources, a prediction of almost all magnetar-based models \citep{scholz2020simultaneous,2023arXiv230810930P}. Therefore, local Universe FRBs are inarguably the best sources to uncover the origins of FRBs.

%The origin of FRBs is an area under much debate and a number of formation channels of their progenitors have been suggested. The most obvious explanation given the discovery of FRB-like bursts from a Galactic magnetar which is known to be formed via a type II core-collapse supernovae, is a system involving a neutron star (prefererable a magnetar). 

The Canadian Hydrogen Intensity Mapping Experiment Fast Radio Burst (CHIME/FRB) Project \citep{abb+2018ApJ} published its first catalog (hereafter denoted as Catalog-1) of 536 FRBs detected between 400 and 800 MHz from 2018 July 25 to 2019 July 1, including 62 bursts from 18 previously reported repeating sources \citep{firstchimefrbcatalog2021}. The detected FRBs show DMs ranging between 102 and 3037 pc cm$^{-3}$. 
However, most of the Catalog-1 FRBs are not localized to their hosts because the CHIME/FRB real-time pipeline processes can localize FRBs to a sky region of around a few tens of arcminutes, which is not sufficient to robustly identify FRB host galaxies \citep{Eftekhari2017pcc}. Moreover, only three Catalog-1 FRBs, all of which are repeating sources in the local Universe with low extragalactic DM, namely, FRB 20181030A, FRB 20180814A, and FRB 20190303A \citep{abb+19c,fonseca2020nine}, have been localized to their host galaxies \citep{2021ApJ...919L..24B, 2020Natur.577..190M, 2022arXiv221211941M} using the CHIME/FRB baseband localization pipeline \citep{2021ApJ...910..147M}. 

In this study, we present a systematic search for the host galaxies of FRBs reported in CHIME/FRB Catalog-1. Our methodology for selecting Catalog-1 FRBs, as detailed in \S\ref{section:sample}, identifies FRBs characterized primarily by their low DM-excess and saved baseband data. Following this methodology, we identify four apparently non-repeating FRBs and find only one plausible host galaxy candidate within each of their respective baseband localization regions as described in \S\ref{sec:host_search}. Furthermore, the chance association probability of the identified plausible host galaxies, as explained in \S\ref{sec:pcc}, remains below 10\%, even after accounting for the look-elsewhere effect. After identifying likely hosts of the four Catalog-1 FRBs, we detail our search for multi-messenger and multi-wavelength counterparts, including compact persistent radio sources, in \S\ref{sec:multi-wavelength}. In \S\ref{sec:hostdemographicsanalysis}, we discuss the implication of the localized nearby FRB host sample (z $< 0.1$), which includes the four hosts presented in this study, and determine the dominant formation channel of FRB sources. Additionally, we report on a burst rate analysis for the four FRBs in \S\ref{subsec:repition_rate}.
Finally, we summarize and conclude in \S\ref{sec:conclusion}.

\section{Observations \& Results}

\subsection{CHIME/FRB sample selection}
\label{section:sample}

In this section, we describe the formalism adopted for selecting Catalog-1 FRBs for our host association study. It is worth noting that not all baseband localized FRBs can be associated robustly with a galaxy \citep[see, for instance,][]{2021ApJ...910..147M,2023arXiv230402638I}. Therefore, looking for hosts of every Catalog-1 FRB that has a baseband localization is not an optimal strategy. Moreover, this would further result in decreasing the statistical power of finding promising FRB hosts when we correct for the look-elsewhere effect in order to avoid false positives. Therefore, we have opted for a planned hypothesis testing framework wherein we select Catalog-1 FRBs based on predefined criteria. These criteria are as follows:

%This   
%we use certain criterion in order to increase our chances to find a robust host and at the same time, avoid any possibility of making false-order to make robust association and avoid 

\begin{enumerate}
    \item Existence of baseband localization with 1$\sigma$ precision $\lesssim$ $1 \arcmin$.
    \item The DM-excess of the FRB should be  $\leq$ 100 pc cm$^{-3}$, i.e., DM $-$ max(DM(MW; NE2001, YMW16) $\leq$ 100 pc cm$^{-3}$, where NE2001 \citep{cordes2002ne2001} and YMW16 \citep{yao2017new} are two widely used Galactic disk electron density distribution models.
    \item The FRB should not be behind the Galactic Plane (Galactic latitude, $\mid$b$\mid$ $>$ 10$^{\circ}$).
\end{enumerate}

\begin{figure}[ht]
\centering
\includegraphics[width=.60\linewidth]{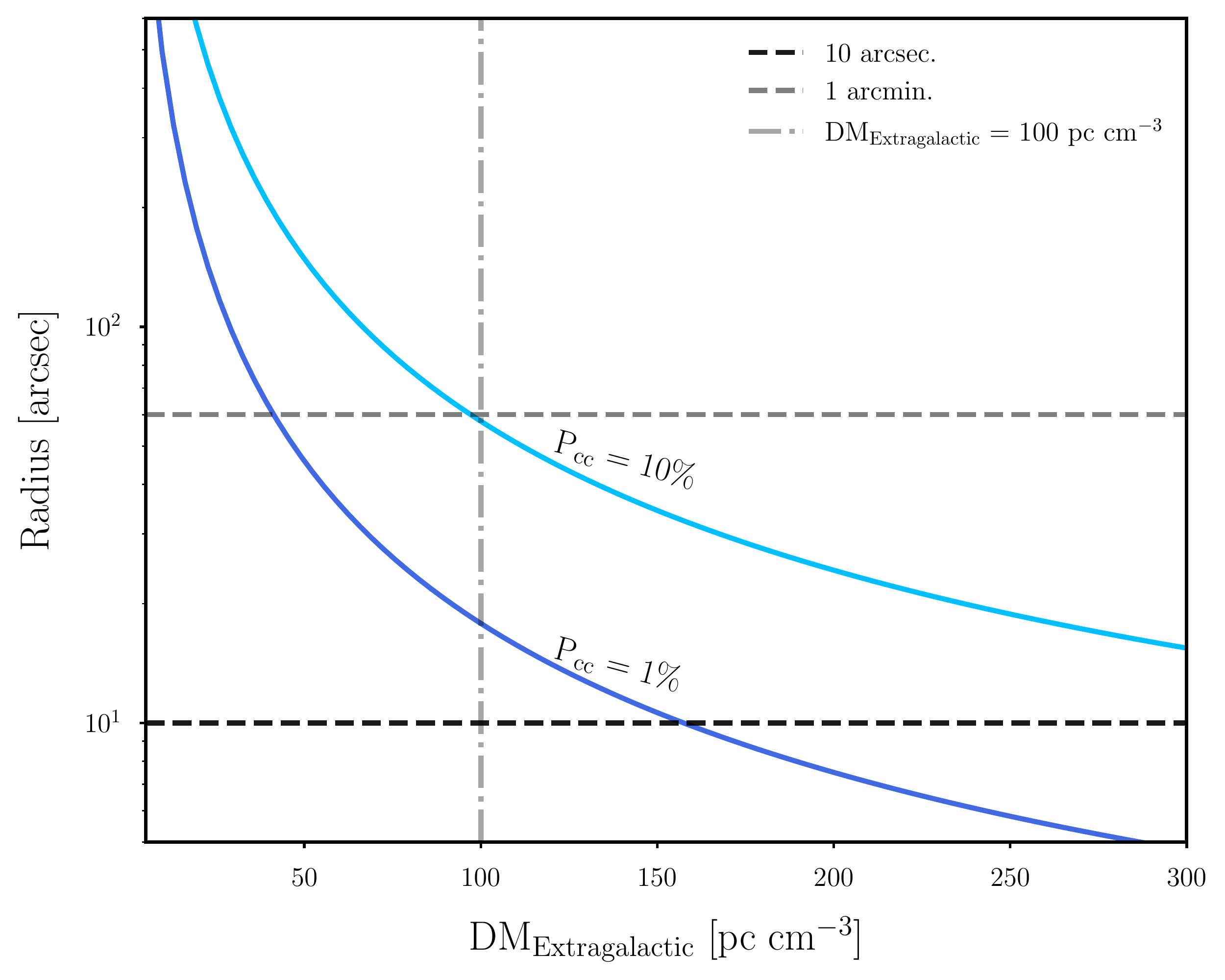}%\centering{}
\caption{Probability curves for P$_{\rm cc}$ = 0.01 and 0.1 as a function of extragalactic DM (or excess-DM)
and localization radius for the faintest FRB host discovered to date (M$_{\rm r}$ = $-$17 AB mag). Note that an FRB with an extragalactic DM $\approx$ 100 pc cm$^{-3}$ can be associated with a dwarf, star-forming galaxy, like the FRB 20121102A host, with P$_{\rm cc}$ of 0.01 and 0.1 using a localization precision of $\approx 10\arcsec$ and $1\arcmin$, respectively. The plot is produced using the formalism discussed in \S\ref{section:sample}.}
\label{fig:PCC-PLOT}
\end{figure}

The rationale for criteria 1 and 2 is as follows: As the maximum distance to the FRB hosts can be reasonably estimated from their
DM-excesses, FRBs with low DM-excess are expected to be nearby sources. For instance, using the Macquart relation \citep{mcquart-relation20}, an FRB with the DM-excess of 100 pc cm$^{-3}$ would have a maximum redshift of 0.1. If this low DM-excess FRB is localized using
the CHIME/FRB baseband pipeline \citep{2021ApJ...910..147M} to a localization region of radius $\sim 1\arcmin$, the number density of galaxies as faint as the faintest FRB host discovered to date \citep[FRB 20121102A with absolute r-band magnitude M$_{\rm r}$ = $-$17 AB mag;][]{2017ApJ...834L...7T} is expected to be small \citep{Eftekhari2017pcc}, hence making any plausible association with such a host a rare coincidence \citep[P$_{\rm cc}$ $\leq$ 10\%; the generally accepted threshold in the FRB community; see][]{2020ApJ...903..152H,2022AJ....163...69B}. This is shown in Figure \ref{fig:PCC-PLOT}. Therefore, CHIME/FRB baseband localizations are most promising to identify the host galaxies of low-DM FRB events (DM-excess $\leq$ 100 pc cm$^{-3}$). Moreover, this low DM-excess cutoff has an additional advantage. At z =  0.1, the faintest FRB host discovered to date, FRB 20121102A, would have an apparent r-band magnitude of $\lesssim$ 21 AB mag. There are several archival wide-sky optical surveys, such as the Sloan Digital Sky Survey \citep[SDSS;][]{2022ApJS..259...35A}, the Panoramic Survey Telescope and Rapid Response System (Pan-STARRS) survey \citep{2016arXiv161205560C}, and DESI (Dark Energy Spectroscopic Instrument) survey \citep{2016arXiv161100036D}, which are sufficiently deep to detect galaxies of r-band magnitude $\leq$ 21 AB mag. Therefore, the hosts of nearby FRBs can be identified in the aforementioned archival optical survey data.
We note that the choice of 100 pc cm$^{-3}$ is rather conservative as this excess includes the contributions from the FRB host and Milky Way circumgalactic medium. However, due to the lack of reliable constraints on their anticipated contributions, we do not consider these components in our DM-excess constraint. Nonetheless, from Figure \ref{fig:PCC-PLOT}, we note that it is possible to identify host galaxies with DM-excess $\lesssim$ 300 pc cm$^{-3}$ with the baseband localization precision of $\lesssim$ 20\arcsec. This will be explored in our future low-DM FRB localization papers.

Lastly, the inclusion of the final criterion, i.e., $\mid$b$\mid$ $>$ 10$^{\circ}$, is justified by the significant uncertainty in Milky Way DM models along Galactic plane sight lines \citep{2021PASA...38...38P,2023arXiv230101000R}.

Using these pre-planned criteria, we can evade the effect of the multiple testing problem or p-value hacking \citep{vidgen2016p}, which makes our host associations more robust.

\begin{figure}[ht]
\begin{center}
\includegraphics[width=.90\linewidth]{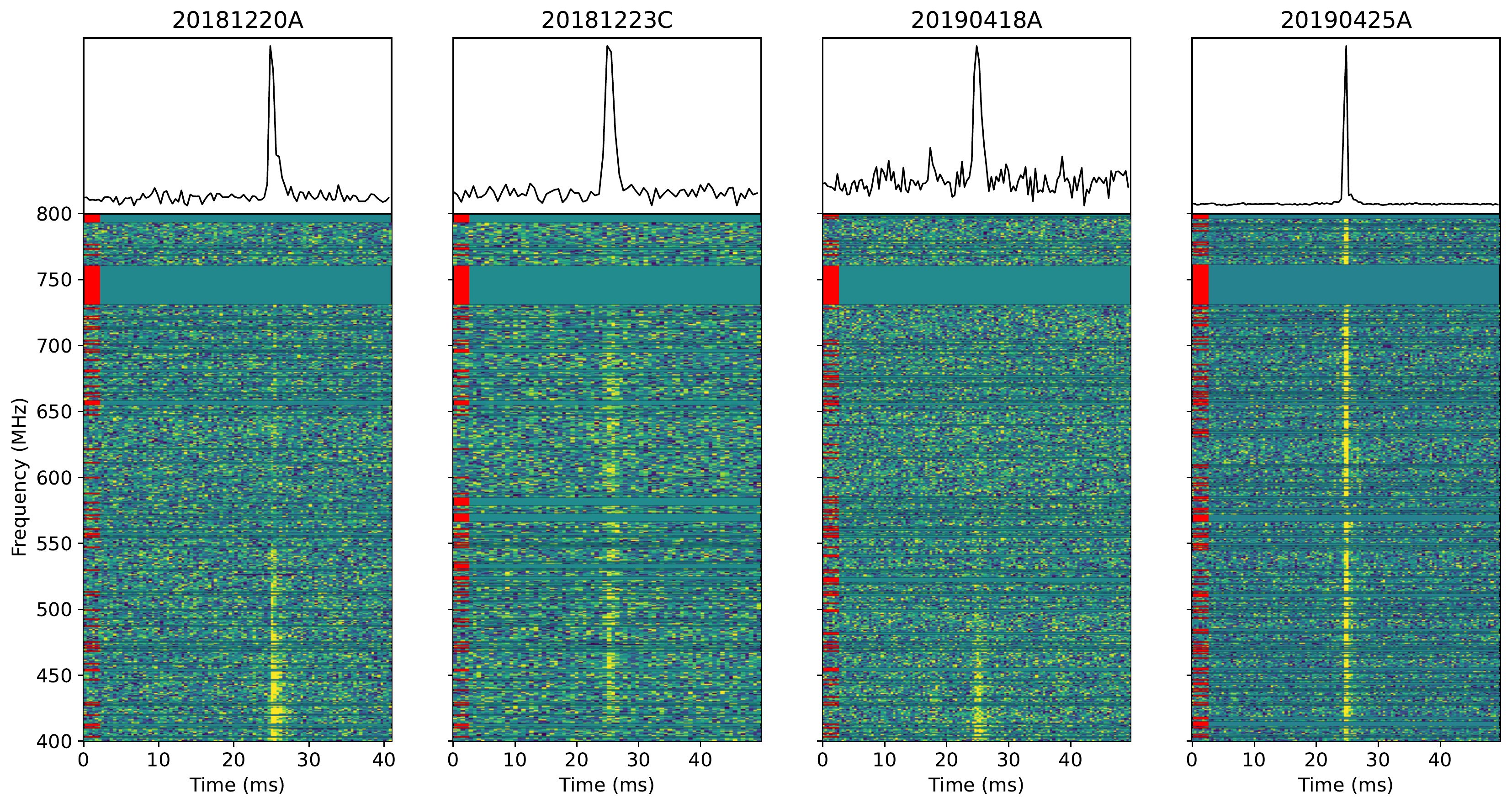}%\centering{}
\caption{Frequency versus time (``waterfall") plots of the dedispersed bursts detected from FRB 20181220A, FRB 20181223C, FRB 20190418A, and FRB 20190425A with saved baseband data. See Table \ref{tab:FRB-params} for their major burst properties. The waterfall plots are binned to have a temporal resolution of 0.39 ms and a spectral resolution of 0.391 MHz. Red lines represent bad frequency channels that were flagged in this analysis. %Note that FRB 20190425A shows sub-bursts which can be seen in Figure \ref{fig:190425} For more detail on this, refer \S\ref{subsection:GW}. Detailed analysis of these sub-bursts will be reported elsewhere
}
\label{fig:waterfall}
\end{center}
\end{figure}

Based on these criteria, we identify four Catalog-1 CHIME FRBs:  FRB 20181223C, FRB 20190418A, FRB 20191220A, and FRB 20190425A. We then run the CHIME/FRB baseband pipeline on the saved baseband data of these FRBs and
%FRB 20181223C, FRB 20190418A, FRB 20191220A, and FRB 20190425A, 
estimate their baseband localization regions. The procedure used to estimate the baseband localizations is detailed by \cite{2021ApJ...910..147M,2022arXiv221211941M}. 
%Briefly, the baseband system of CHIME/FRB stores $\sim 100$\,ms of channelized voltages around signals of interest \citep{abb+2018ApJ}. We have developed a pipeline to automatically process such baseband data and localize a burst on the sky with a precision of $\sim\frac{8}{\text{S/N}}$\,arcmin \citep{2021ApJ...910..147M}. This is achieved by mapping the signal strength with a grid of largely overlapping beams around an initial guess. The resulting S/N measured in each beam is fitted with a mathematical model of the formed beam of the telescope.
%Systematic effects have been corrected by using a sample of sources with known positions. 
The dedispersed baseband data waterfall plots and major characteristics of the four FRBs are shown in Figure \ref{fig:waterfall} and Table \ref{tab:FRB-params}, respectively. Other burst properties, such as fluence and flux density, along with a detailed description of the baseband data analysis of these FRBs will be presented elsewhere. Next, we describe the host identification procedure used in this work.
%As the reported baseband localization uncertainties are statistical in nature \citep{{2021ApJ...910..147M}, we combined the localization regions of the four FRB bursts using a weighted average with inverse variance weights and
%using a weighted mean of individual localization regions
%Using the baseband data  weocalized the FRB to a sky area of $\approx$ 5.3 arcmin$^{2}$ (90\% confidence region; see Table \ref{tab:params}). Next, we use the baseband localization region of FRB 20181030A to search for a potential host galaxy.  

\begin{table}[ht]
\begin{center}
\caption{Major Observables of the Selected CHIME FRBs.}
\begin{tabular}{@{} lcccc @{}}\toprule
\textbf{Parameter} & \textbf{FRB 20181220A} &  \textbf{FRB 20181223C} &  \textbf{FRB 20190418A} &  \textbf{FRB 20190425A}\\\midrule
R.A. (J2000)$^{a}$                & 23$^h$14$^m$52$^s$       & 12$^h$03$^m$43$^s$   & 04$^h$23$^m$16$^s$  & 17$^h$02$^m$42$^s$ \\
$\sigma$(R.A.) (\arcsec)$^{a}$     & 28     & 23    & 27 & 11 \\
Dec. (J2000)$^{a}$ & 48\degr20\arcmin25\arcsec & 27\degr33\arcmin09\arcsec & 16\degr04\arcmin02\arcsec & 21\degr34\arcmin35\arcsec \\
$\sigma$(Dec.) (\arcsec)$^{a}$     & 20   & 26  & 34 & 12 \\
l, b (deg, deg)                        & 106.82, $-$11.49 & 207.90, $+$79.40 & 179.30, $-$22.89 & 42.10, $+$33.08      \\
DM$^{b}$ (pc cm$^{-3}$)                 & 209.4                                & 112.5                             & 184.5            & 128.2             \\
%DM$^{c}$ (pc cm$^{-3}$)               & 209.5                                & 112.5                             & 184.5            & 128.1  %           \\
DM$_{\rm MW,NE2001}^{c}$ (pc cm$^{-3}$) & 126                                  & 20                                & 71               & 49                \\
DM$_{\rm MW,YMW16}^{c}$ (pc cm$^{-3}$)  & 123                                  & 20                                & 86               & 39                \\
%DM$_{\rm MW, halo}^{d}$ (pc cm$^{-3}$)  & 30                                   & 30                                & 30               & 30                \\
Max. redshift$^{d}$                         & 0.096                                & 0.089                             & 0.12            & 0.083             \\
Exposure$^{e}$ (hours)        & 71  $\pm$ 2                                & 64   $\pm$ 3                             & 68  $\pm$ 1             & 67   $\pm$ 2       \\     
 \hline
%\footnotetext{Thus far}
\end{tabular}
\label{tab:FRB-params}
\end{center}
$^a$Baseband localization region of the FRB along with 1$\sigma$ uncertainty.\\
%determined from baseband data saved for several \vk{``several'' is not sufficient information for a publication} FRB 20181030A bursts using the technique described by \cite{Daniele2020}. 
%The 95\% cl on the R.A and Dec. values are $35^{s}.6$ and 55$\arcsec$ , respectively.
$^b$From \cite{andersen2019chime}.

$^c$Maximum DM model prediction along this line-of-sight for the NE2001 \citep{cordes2002ne2001} and YMW16 \citep{yao2017new} Galactic disk electron density distribution models. %These values do not include the contribution from the Milky Way halo.

%$^d$Fiducial Milky Way halo prediction from the \cite{dolag2015constraints} hydrodynamic simulation and \cite{yamasaki2020galactic} Milky Way Halo model. \\
$^d$Estimated 90\% credible region upper limit estimated using the formalism discussed by \cite{2021ApJ...919L..24B}. \\
$^{e}$Total exposure for the upper transit of each source. The uncertainty in the exposure values is dominated by the corresponding source declination uncertainties since the widths of the synthesized beams vary significantly with declination.
\end{table}

\subsection{Host galaxy search}
\label{sec:host_search}

First, we argue below that % is unlikely that 
 all four FRBs are unlikely to be Galactic in origin. As shown in Table \ref{tab:FRB-params}, the DM-excess along respective FRB sight-lines is $>$ 50 pc cm$^{-3}$. This remains true even after accounting for the fiducial MW halo DM contribution of 30 pc cm$^{-3}$ using the MW halo DM model proposed by \cite{yamasaki2020galactic}; for more discussion on this, see \cite{2023ApJ...946...58C}.
 % (See Table \ref{Tab1}). 
 Moreover, there is no cataloged Galactic ionized region \citep{anderson2014wise}, molecular complex \citep{2001ApJ...547..792D}, satellite galaxy \citep{2019IAUS..344..381K, 2019IAUS..344..377K}, or Galactic globular cluster \citep{harris1996aj,2018yCat..36160012G,2019MNRAS.484.2832V} in the direction of the FRBs that can account for their observed FRB DM-excess. 
 %Finally, using the same argument as asserted by \cite{bhardwaj2021}, an FRB with a DM-excess of $<$ 50 pc cm$^{-3}$, if Galactic, would require a very distant ($\gtrsim$ 100 kpc) and unusually energetic neutron star as its source. As discussed below, we have found an extragalactic host with a low chance coincidence probability. Therefore,
Therefore, we argue for the extragalactic association of all four FRBs. Next, for the purpose of finding all promising FRB host galaxies, we estimate the 90\% credible upper limit on the maximum redshift of the four FRBs using the formalism discussed by \cite{2021ApJ...919L..24B}. Those values are reported in Table \ref{tab:FRB-params}. 
%On the other hand, 

We now report on our search of the host galaxies of the four selected FRBs, and for that, we employ archival datasets from Pan-STARRS, SDSS, and DESI, as outlined in \S\ref{section:sample}. When the FRB field-of-view (FOV) is covered by more than one of these surveys, priority is based on the survey's r-band completeness limit, which follows the following order of preference: DESI (most sensitive), SDSS, and Pan-STARRS. However, as discussed in \S\ref{section:sample}, all three surveys are complete to detect the faintest known FRB host at the maximum redshift of the four FRBs.

\subsubsection*{FRB 20181220A}
\label{sec:frb181220a}

Of the three optical surveys, only Pan-STARRS covers the FRB 20181220A localization region. The Pan-STARRS r-band image of the FRB FOV is shown in Figure \ref{fig:FOV}. Using the Pan-STARRS1 Source Types and Redshifts with Machine learning (PS1$-$STRM) catalog \citep[][]{2021MNRAS.500.1633B}, we find only one plausible extended source (completeness limit $\approx$ 21 AB mag), PSO J231447.57+482031.6, within the 2$\sigma$ baseband localization region of the FRB. The galaxy was first reported in the 2MASS-selected flat galaxy catalog as 2MFGC 17440 \citep{2004BSAO...57....5M} and classified as a Scd-type spiral using the observed morphology of the galaxy in 2MASS data \citep{2015AstBu..70...24M}.  The Pan-STARRS r-image of the FRB 2$\sigma$ localization region is shown in Figure \ref{fig:FOV}. 
%the source classification and photometric redshift (photo-z) catalog for PS1 3$\pi$ Data Release 1,

The spectroscopic redshift of 2MFGC 17440 is unavailable in any of the following major public astronomy databases: NASA/IPAC Extragalactic Database \citep[NED;][]{2017IAUS..325..379M}, Set of Identifications, Measurements, and Bibliography for Astronomical Data \citep[SIMBAD;][]{2000A&AS..143....9W}, and  VizieR.\footnote{\url{https://vizier.cds.unistra.fr/viz-bin/VizieR}} Therefore, we conducted spectroscopic observations on 2020 August 11 (Progam ID: GN-2020B-FT-201) using the Gemini Multi-Object Spectrograph (GMOS) on the Gemini-North telescope. Spectroscopy of the galaxy was reduced using the \texttt{PypeIt} reduction package \citep{2020JOSS....5.2308P}. This package employs optimal extraction techniques to generate a 1D spectrum from the flat-fielded and sky-subtracted 2D spectral images of the field. For flux calibration using \texttt{PypeIt}, we used the spectrophotometric standard star G191-B2B \citep{1995yCat.3116....0M} from the compilation of associated calibration observations accessible through the Gemini Observatory Archive Portal.\footnote{\url{https://archive.gemini.edu/}} We then fitted the calibrated 1D spectrum using the {\tt Specutils} package \citep{nicholas_earl_2023_8049033} and estimated the galaxy's spectroscopic redshift z$_{\rm spec}$ = 0.02746(3) based on the H$\alpha$, [NII], and [SII] line features. The fitted 1D spectra with a subset of these lines are shown in Figure \ref{fig:frb-redshift-estimates}. From the fitted spectrum, we estimate an H$\alpha$ flux density = (2.16 $\pm$ 0.18) $\times 10^{-13}$ erg s$^{-1}$, which we use to estimate the star-formation rate (SFR) of the galaxy as noted in Table \ref{tab: A-galaxy-properties}.

\subsubsection*{FRB 20181223C}
\label{sec:frb181223c}

The FRB 20181223C FOV is covered by all three surveys. Therefore, as per the preference order stated in \S\ref{sec:host_search}, we use DESI data to identify all plausible host galaxies within the 2$\sigma$ localization region of the FRB. Using the DESI data, \cite{Zhou2020ApJ} estimated photometric redshifts of all the identified galaxies with a 5$\sigma$ r-band completeness limit of 23.6 AB mag, which is sensitive to detect a dwarf host 5 times less luminous than any FRB host discovered to date (Mr = $-15$ AB mag). We identify nine plausible host galaxy candidate candidates shown in the Pan-STARRS r-band image of the FRB FOV in Figure \ref{fig:FOV}. We then determined their spectroscopic redshifts using multi-object spectroscopic observations from the 10.4-m Gran Telescopio Canarias (GTC), as detailed in \S\ref{sec:mos_frb20181223c}. We found that only one galaxy satisfies the maximum redshift constraint (i.e., z $<$ z$_{\rm max}$; See Table \ref{tab:FRB-params}), SDSS J120340.98+273251.4. 

%However, we only consider sources which are brighter than r-band magnitude $<$ 21.5 AB because first, these sources are too faint to obtain a reliable spectroscopic redshift with our observations, and second, the estimated photometric redshifts of all those sources are at least 5 times higher than the estimated maximum redshift of FRB 20181223C, i.e., 0.076. Note that at z = 0.076, the faintest FRB host detected to date would have an apparent r-band magnitude = 20.7 AB magnitude, which is $<$ 21.5 mag. After removing different imaging artifacts and multiple detections of the same extended sources, we identified nine host candidates within the 90\% confidence localization region of the FRB. As most of these galaxies (except two) in the field do not have a spectroscopic redshift reported in the literature. We used the GTC to obtain their redshifts as described in §2.4 and we report the values we obtained in Table 2

%With a redshift z = 0.03024(1), source 1 is the only galaxy in the field satisfying the condition z $<$ 0.078. 

%Therefore, we identify this galaxy as the
%most probable host for FRB 20180814A. 
We find the optical spectrum of the galaxy in the SDSS DR17 database \citep{2022ApJS..259...35A}.\footnote{\url{https://dr12.sdss.org/spectrumDetail?plateid=2226&mjd=53819&fiber=0509}} The galaxy is classified as a star-forming spiral at z$_{\rm spec}$ = 0.03024 $\pm$ 0.00001 \citep{2011ApJS..196...11S,2018MNRAS.476.3661D}. The notable physical properties of the galaxy from the SDSS database are presented in Table \ref{tab: A-galaxy-properties}.

\subsubsection*{FRB 20190418A}
\label{sec:frb190418a}

Among the three archival surveys we considered in this study, only Pan-STARRS and SDSS cover the baseband localization region of FRB 20190418A. Therefore, as per the preference order stated in \S\ref{sec:host_search}, we search for potential host candidates in the SDSS data and find only one galaxy within the 2$\sigma$ localization region of the FRB, SDSS J042314.96+160425.6. The Pan-STARRS r-image of the FRB 2$\sigma$ localization region is shown in Figure \ref{fig:FOV}. Moreover, while searching for deeper images, we find that the galaxy is also cataloged in the UKIDSS-DR9 GCS survey data release \citep{2007MNRAS.379.1599L} and has Petrosian K-band magnitude 14.42$\pm$0.04 mag (this is used in our {\tt Prospector} analysis; See Appendix \ref{app:prospector}). In the UKIDSS-DR9 GCS K-band data, the galaxy shows spiral arms-like features
%\footnote{For the image, use this link:\url{http://wsa.roe.ac.uk:8080/wsa/getImage_form.jsp}}
making it likely a spiral galaxy. 

As the spectroscopic redshift of the galaxy is not available in the public databases stated above, we conducted spectroscopic observations of SDSS J042314.96+160425.6 with the MOS instrument on the Gemini North Telescope on 2021 December 10 (Program ID = GN-2021B-Q-115). Spectroscopy was then reduced using the \texttt{PypeIt} reduction package \citep{2020JOSS....5.2308P}, which optimally extracts a 1D spectrum from the flat-fielded and sky-subtracted 2D
spectral image. For flux calibration within the framework of \texttt{PypeIt}, we used the spectrophotometric standard Feige 34 \citep{1990AJ.....99.1621O}, which was observed three days after the target. 
The calibrated 1D spectrum was then fitted using the {\tt Specutils} package as shown in Figure \ref{fig:frb-redshift-estimates}, which yields a spectroscopic redshift of z$_{\rm spec}$ = 0.07132(1) based on the H$\alpha$, [NII], and [SII] line features. We also estimate an H$\alpha$ flux density = (2.3 $\pm$ 0.3) $\times 10^{-15}$ erg s$^{-1}$, which we use to estimate the SFR of the galaxy as noted in Table \ref{tab: A-galaxy-properties}.

\subsubsection*{FRB 20190425A}
\label{sec:frb190425a}

\begin{figure*}
\centering     %%% not \center

\subfigure{\label{fig3:b}\includegraphics[width=0.49\textwidth, height=8cm]{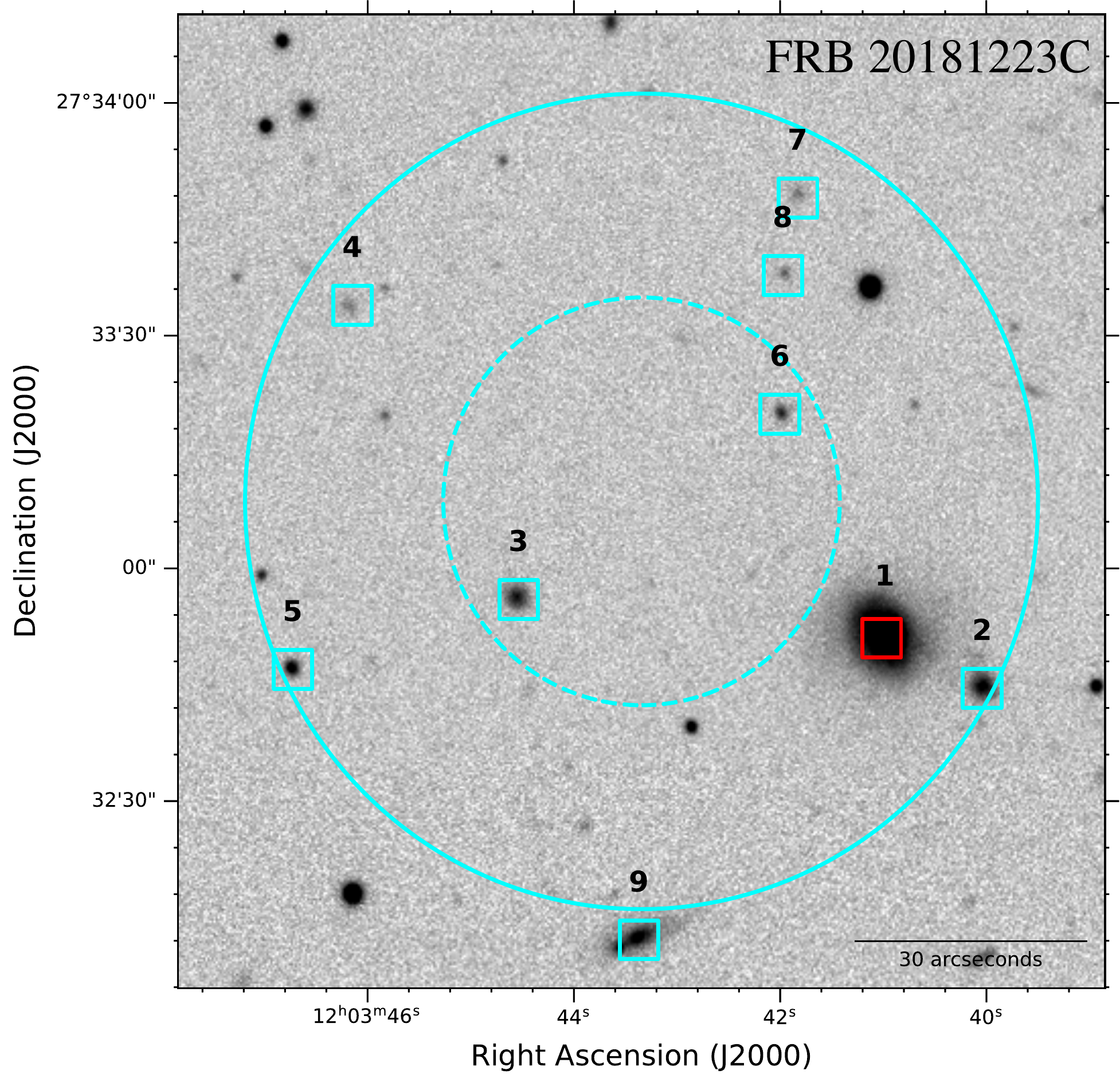}}
\subfigure{\label{fig3:c}\includegraphics[width=0.49\textwidth, height=8cm]{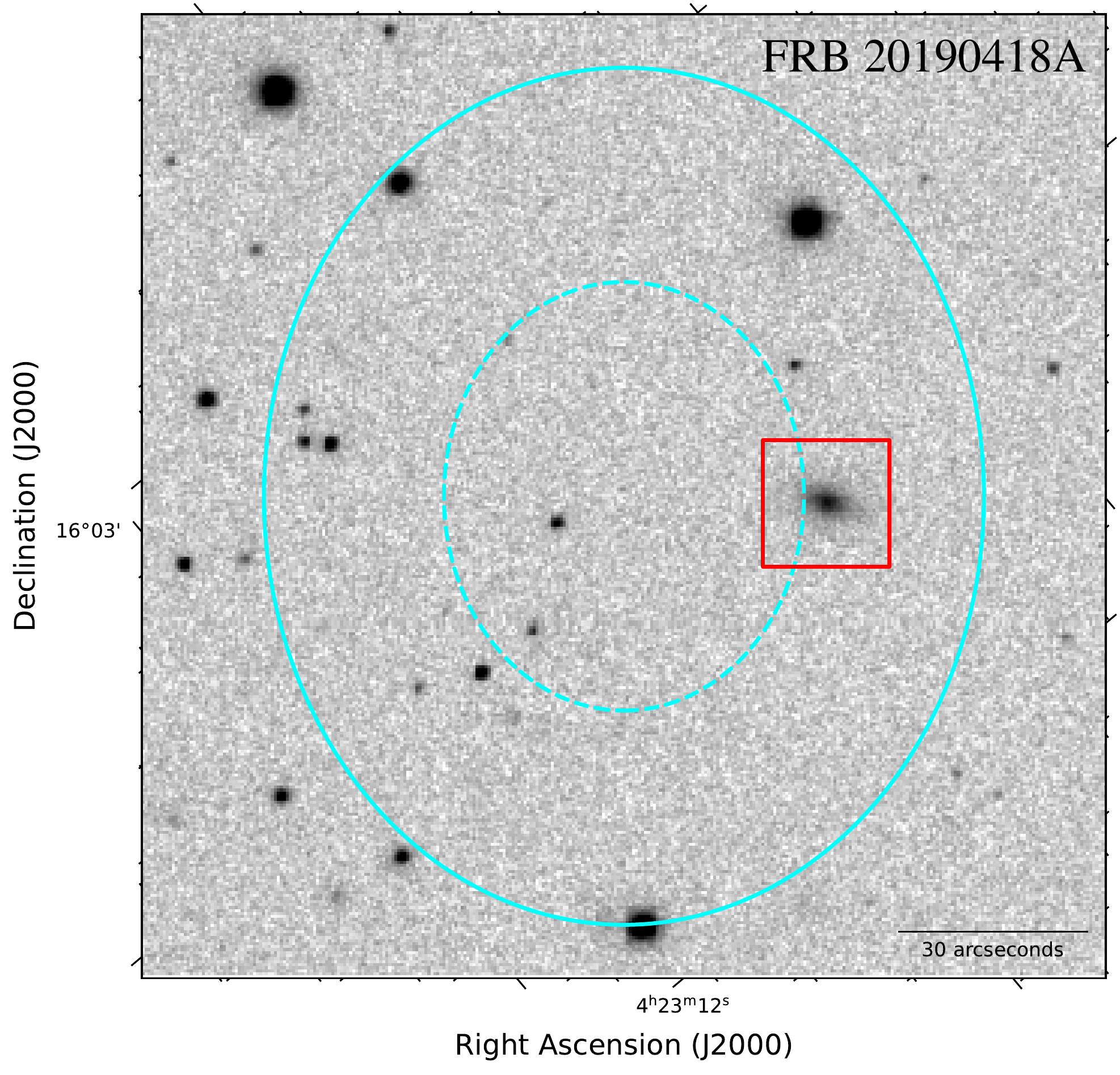}}\\

\subfigure{\label{fig3:a}\includegraphics[width=0.49\textwidth, height=5.55cm]{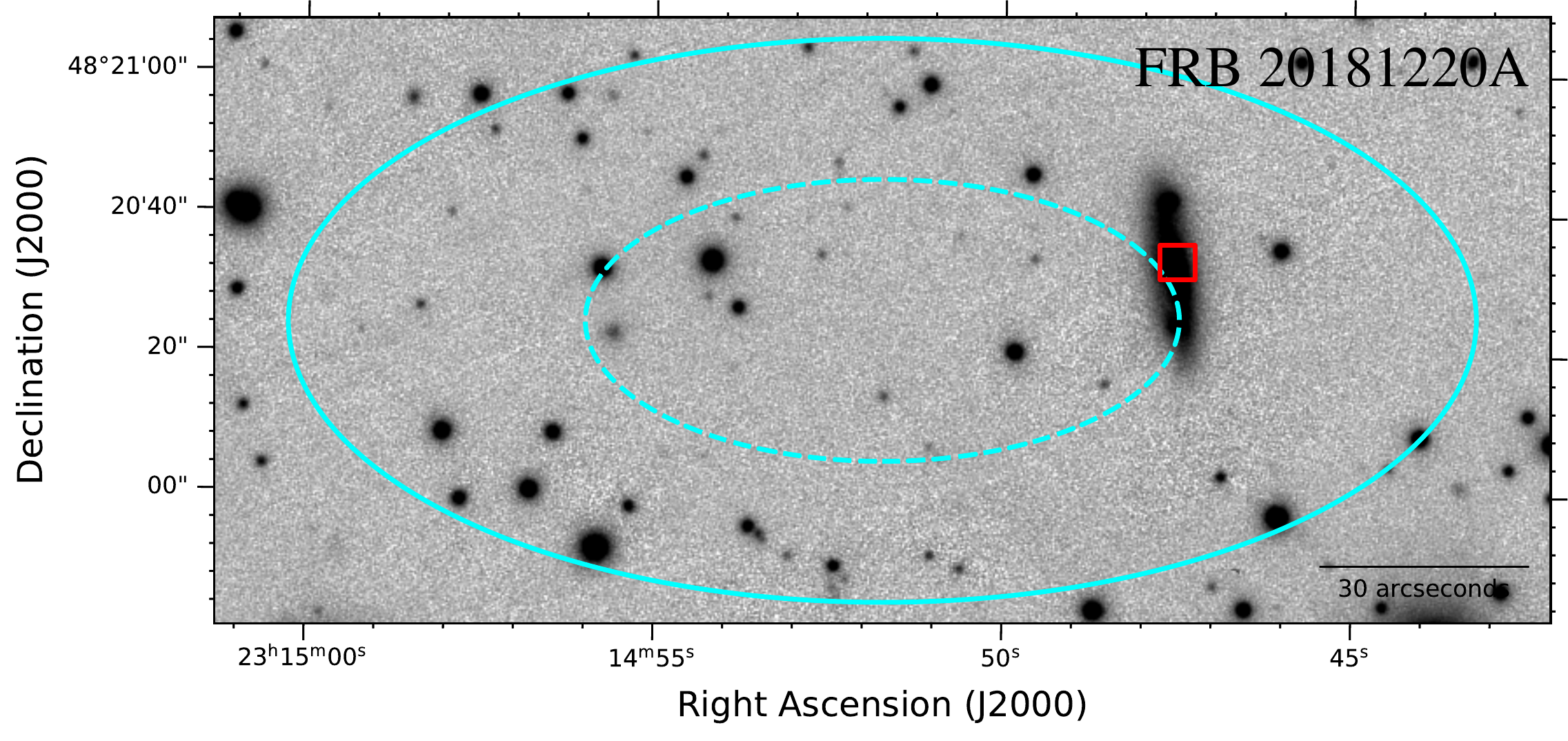}}
\subfigure{\label{fig3:d}\includegraphics[width=0.49\textwidth, height=5.5cm]{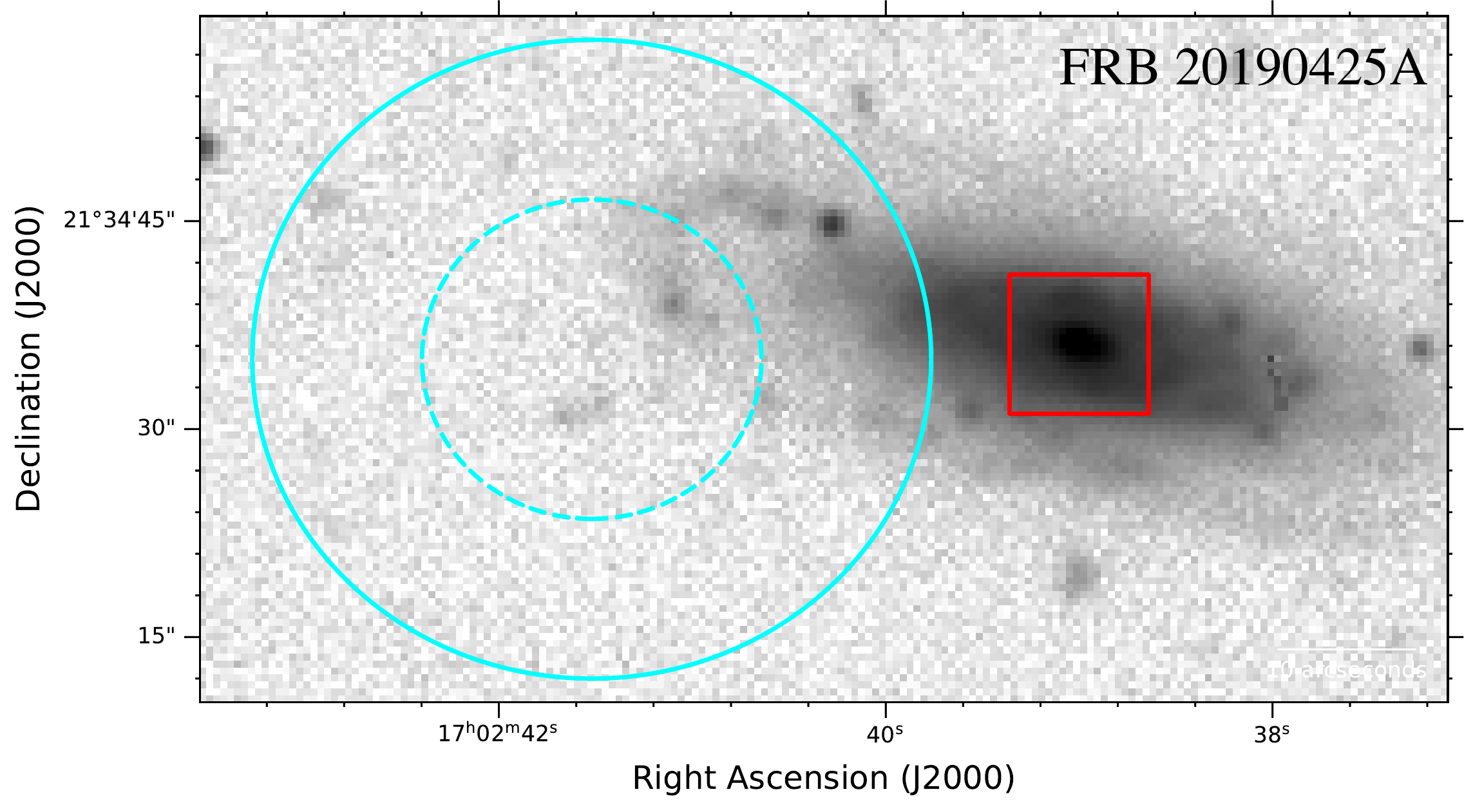}} 
\caption{Pan-STARRS r-band images of the baseband localization regions for FRB 20181223C (top left), FRB 20190418A (top right), FRB 20181220A (bottom left), and FRB 20190425A (bottom right). The 1$\sigma$ and 2$\sigma$ baseband localization regions are represented by dotted and solid cyan ellipses, respectively. While only one plausible host galaxy candidate is identified for FRB 20181220A, FRB 20190418A, and FRB 20190425A, we identify nine potential host galaxy candidates within the localization area of FRB 20181223C. However, following our multi-object spectroscopic analysis discussed in Appendix \ref{sec:mos_frb20181223c}, only one galaxy satisfies the estimated maximum redshift limit for FRB 20181223C. Finally, the most probable host galaxies for the four FRBs are delineated by red boxes.}
\label{fig:FOV}
\end{figure*}

\begin{figure*}
\centering     %%% not \center

\subfigure{\label{fig4:b}\includegraphics[width=0.49\textwidth, height=7cm]{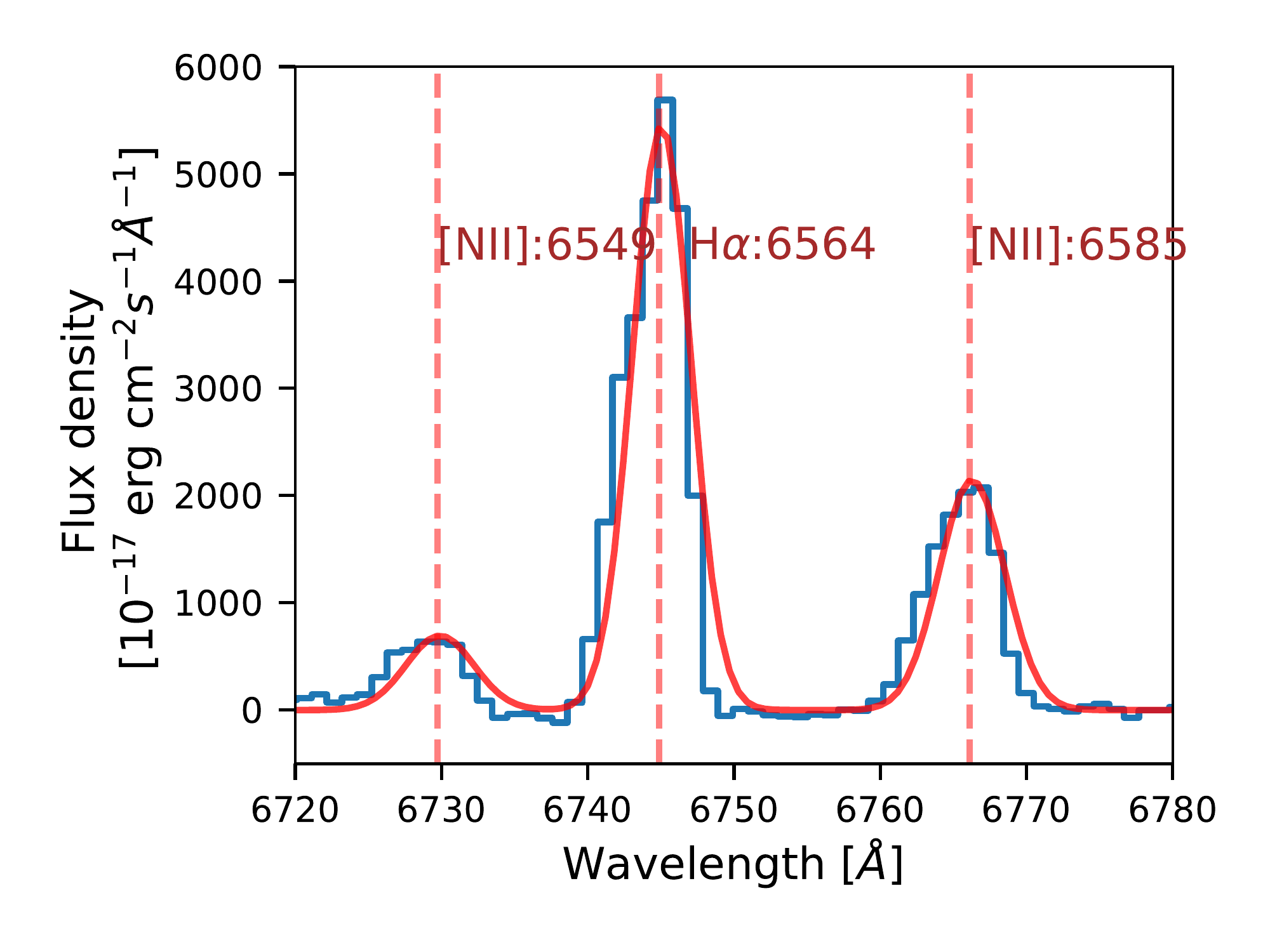}}
\subfigure{\label{fig4:c}\includegraphics[width=0.49\textwidth, height=7cm]{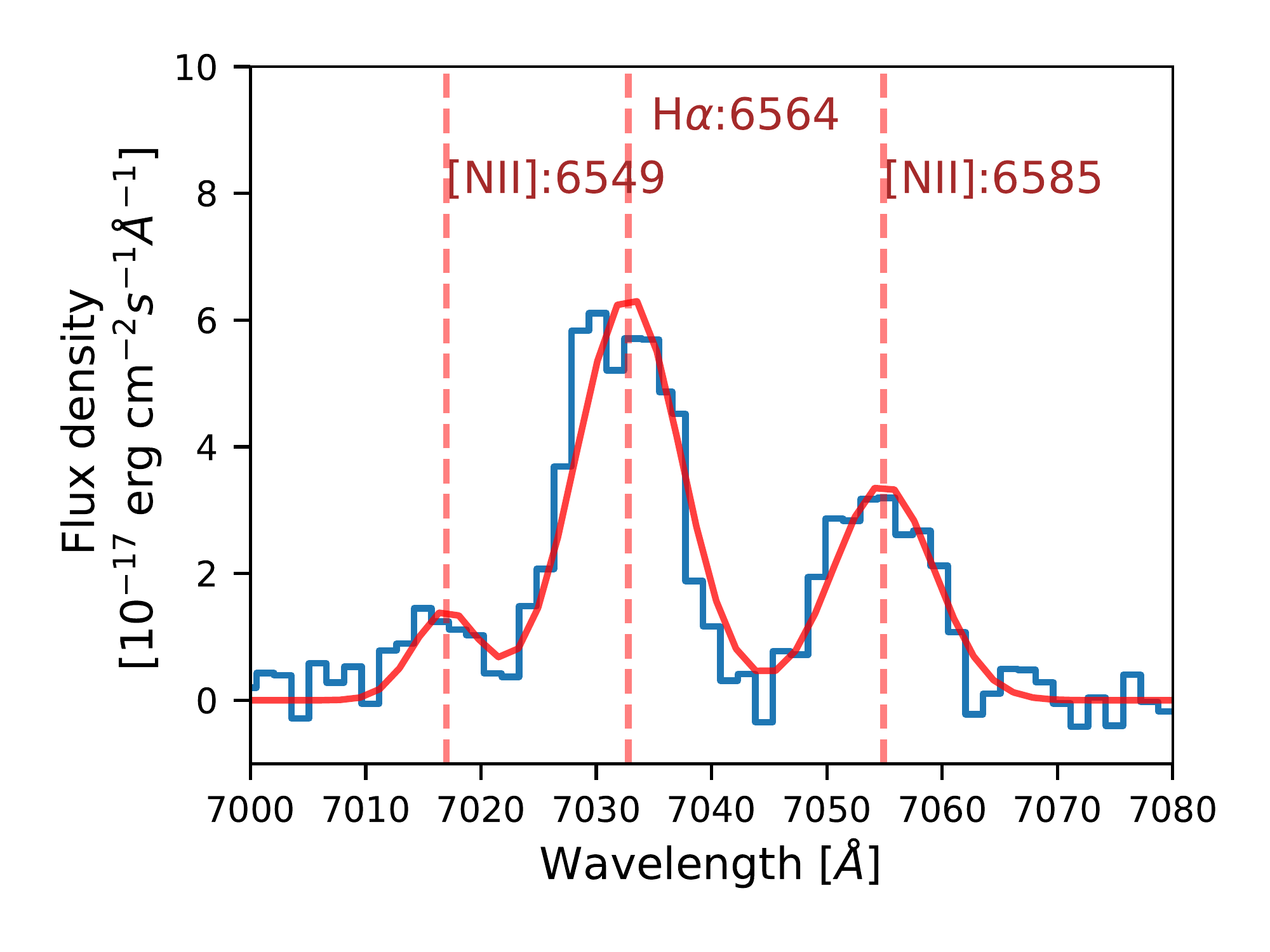}}\\
\caption{Galactic-extinction corrected spectra (in the observer's frame) of the host galaxies for FRB 20181220A (left) and FRB 20190418A (right). The solid red lines depict the best-fitted line profiles estimated using {\tt Specutils} for both galaxies, along with dotted vertical red lines denoting the prominent nebular emission lines, H$\alpha$ and [N II]. }
\label{fig:frb-redshift-estimates}
\end{figure*}

Similar to FRB 20181223C, the FRB 20190425A FOV is covered by all three imaging surveys. Therefore, we use DESI data to search for plausible host galaxies within the 2$\sigma$ localization region of the FRB presented in Figure \ref{fig:FOV}. We find only one cataloged galaxy in the DESI photo-z catalog \cite{Zou2019ApJS},
%within the FRB 2$\sigma$ baseband localization region
UGC 10667. UGC 10667 is a star-forming Sbc-type spiral galaxy \citep{2010AJ....139.2525H} located at z$_{\rm spec}$ = 0.03122 $\pm$ 0.00001 \citep{2022ApJS..259...35A}. The notable physical properties of the galaxy from the SDSS database are presented in Table \ref{tab: A-galaxy-properties}. We note that \cite{2023MNRAS.519.2235P} found this galaxy to be the most probable host of FRB 20190425A using the CHIME/FRB header localization region \citep[precision $\sim 10\arcmin$;][]{firstchimefrbcatalog2021} and Probabilistic Association of Transients to Hosts (PATH) package \citep[][]{2021ApJ...911...95A} (P(O\textbar x) = 0.79).   

\vspace{10pt}
\noindent Although we have identified only one potential host galaxy candidate within the 2$\sigma$ localization region of each of the four FRBs, claiming them as host galaxies requires validating that their presence is not a mere fortuitous alignment. To address this, we next estimate the probability of chance association for the identified host galaxy candidates.

%Based on the BPT \citep{2001ApJ...556..121K,2003MNRAS.346.1055K} and WISE AGN \citep{2011ApJ...735..112J} classification schemes

\subsection{Chance Association Probability}
\label{sec:pcc}

In accordance with established practices for estimating chance association probabilities (P$_{\rm cc}$) for various astrophysical transients \citep[for example,][]{2002AJ....123.1111B, 2015MNRAS.450.4221B, 2016ApJ...817..144B}, we calculate P$_{\rm cc}$ based on the angular separation between the FRB position and the center of a potential host galaxy, considering the uncertainty of the FRB's baseband localization region and the apparent magnitude of the galaxy. For a more comprehensive discussion, refer to \cite{Eftekhari2017pcc}. In essence, the derivation of P$_{\rm cc}$ relies on galaxy counts brighter than a certain magnitude threshold, accounting for the prevalence of faint galaxies on the sky. We adopt a Poisson distribution of galaxies across the sky and compute the likelihood of encountering one or more galaxies with apparent r-band magnitude (m${_\text{r}}$) equal to or less than that of the host galaxy by chance, within the 2$\sigma$ baseband localization region of the FRB. Using the areal number density of FRB host galaxies based on the formalism discussed by \cite{driver2016measurements}, we estimate P$_{\rm cc}$ using Equation 2 from \cite{Eftekhari2017pcc}. The estimated P$_{\rm cc}$ value for the proposed host galaxies of the four FRBs are presented in Table \ref{tab: A-galaxy-properties}.  Next, we apply a correction to the estimated P$_{\rm cc}$ to account for the look-elsewhere effect.

While adjusting the estimated P$_{\rm cc}$ values is not necessary for the look-elsewhere effect due to our use of a planned hypothesis testing framework in this study \citep[for more information, refer to][]{anderson2001empirical}, it is worth noting that this perspective is widely challenged in the literature \cite[for example, see ][]{Frane2015PlannedHT}. Hence, we proceed to adjust the estimated  P$_{\rm cc}$ value for each of our FRBs to account for the look-elsewhere effect.
To correct for the look-elsewhere effect, we use the Hochberg correction, also called Hochberg's step-up method \citep{10.1093/biomet/75.4.800}. It is an improvement over the Bonferroni correction employed by \cite{bhardwaj2021} as it provides increased statistical power, especially when the hypotheses being tested are either independent (which is true in our case) or positively correlated \citep{10.1093/biomet/asm067}. The Hochberg correction limits the risk of inflating the overall Type-I error rate (or false discovery rate) by adjusting the individual p-values for each hypothesis in a step-wise manner. 
To correct individual P$_{\rm cc}$ values, the Hochberg step-up procedure entails the following steps \cite[for more detailed discussion, refer][]{Dunnett1992ASM}:
\begin{enumerate}[Step 1:]
{
\item Arrange the estimated P$_{cc}$ value for the four host associations in ascending order from smallest to largest.
\item For each FRB host, calculate the adjusted P$_{\rm cc}$ value, or P$_{\rm cc, correct}$, using the following formula:
\begin{equation}
    \rm P_{cc, correct} = min((m - k + 1) \times P_{cc}, 1),  
\end{equation}
where m is the total number of hypotheses being tested, and k is the rank of the p-value in the ordered list. In our case, m = 4 and the P$_{\rm cc}$ value of FRB 20190425A being the smallest (3 $\times$ 10$^{-4}$) has rank 1, followed by those of FRB 20181220A (k=2), FRB 20181223C (k=3), and FRB 20190418A (k=4).
}\end{enumerate}

Following the above procedure, we estimate the P$_{\rm cc, correct}$ values for all 4 FRBs as listed in Table \ref{tab: A-galaxy-properties}. As the P$_{\rm cc, correct}$ values of all four low-DM CHIME FRBs in our sample are below the confidence threshold ($\alpha) = 0.1$ (as used in our planned hypothesis testing framework; see \S\ref{section:sample}), it suggests that the identified host galaxy candidates are indeed likely host galaxies. Thus, along with the fact that we found only one plausible host candidate despite the fact that our search is sensitive to identify hosts similar or fainter than the faintest FRB host to the estimated maximum redshift of each FRB, we concluded that for all four FRBs, the identified galaxies are indeed the most likely host galaxies. 

There also exists a Bayesian framework to identify the host galaxies of FRBs, called PATH \citep[][]{2021ApJ...911...95A}. A Bayesian framework can address the problem of the look-elsewhere effect by relying on associated prior distributions, assuming the priors accurately encapsulate the uncertainty surrounding the parameters being estimated \citep{Sjolander2019}. However, the extent to which the prior distributions used in the PATH framework conform to the aforementioned requirement remains unclear. Once this is accessed robustly, PATH offers a promising alternative to the existing P$_{\rm cc}$ formalism. However, in the absence of such an assessment, we decide not to use this framework in our study.
%There also exists a Bayesian framework to identify the host galaxies of FRBs, called PATH \citep[][]{2021ApJ...911...95A}. A Bayesian framework can address the problem of the look-elsewhere effect by relying on associated prior distributions, assuming the priors accurately encapsulate the uncertainty surrounding the parameters being estimated \citep{Sjolander2019}. However, the extent to which the prior distributions used in the PATH framework conform to the aforementioned requirement remains unclear. In the absence of a reliable method to assess this, we decide to not use this framework in our study.
%by leveraging associated prior distributions, under the assumption that associated priors accurately capture the uncertainty in our knowledge about the parameters being estimated, can resolve the problem of the look-elsewhere effect 
%how to correct the estimated posterior probability P(O | x) for the look-elsewhere effect, 
 Nonetheless, when employing PATH with its default priors, except for an undetected host prior P(U) = 0.1, the P(O\textbar x) for each FRB to the suggested host galaxy exceeds 0.90, supporting the results of the preceding analysis.

%-----------------------------------------------------------------
%--------------------EOS------------------------------------------
%\section{Star formation rate versus stellar mass}
%\label{sec:obs-sfr-m}

%https://arxiv.org/pdf/1502.01027.pdf

%https://ui.adsabs.harvard.edu/abs/2015ApJ...801L..29R/abstract

%https://tel.archives-ouvertes.fr/tel-03187486/file/98526_ZHOU_2020_diffusion.pdf

%http://candels-collaboration.blogspot.com/2013/02/star-formation-in-mountains.html

%https://academic.oup.com/mnras/article/470/1/1050/3813438

%https://www.ast.cam.ac.uk/sites/default/files/2018.07.Cambridge.PENG_.PDF

\subsection{Host Galaxy properties}
\label{sec:host-properties}

To estimate major physical properties of the host galaxies of the four low-DM FRBs,  we use a Bayesian inference spectral energy distribution (SED) fitting code, \texttt{Prospector} \citep{Leja2017,prospect2019}. Appendix \ref{app:prospector} describes the SED fitting analysis in detail. Major host properties of the four FRBs are presented in Table \ref{tab: A-galaxy-properties}.

\begin{table}[t]
\setlength{\tabcolsep}{0pt} % let LaTeX figure out amount of intercolumn whitespace

\caption{Major Observables of the CHIME FRB hosts.}
%\resizebox{\textwidth}{!}{%
%\begin{center}
%\begin{adjustbox}{width=1.0\textwidth, center}
\resizebox{\textwidth}{!}
{
\begin{tabular}{l|cccc}\toprule
\textbf{Parameter} & \textbf{FRB 20181220A} &  \textbf{FRB 20181223C} &  \textbf{FRB 20190418A} &  \textbf{FRB 20190425A}\\\midrule
R.A. (J2000)$^{a}$ (deg)                            & 348.6982     & 180.9207                & 65.8123 & 255.6625 \\
Dec. (J2000)$^{a}$ (deg)                           & 48.3421                     & 27.5476                  &  16.0738 & 21.5767  \\
Galaxy Name                                             & 2MFGC17440  & SDSSJ120340.98+273251.4\hspace{2 mm}  & SDSSJ042314.96+160425.6 & UGC10667      \\
Apparent r-band mag (m$_{r}$; AB)                 & 15.71                                & 16.78                             & 16.68            & 14.77             \\
Galactic r-band extinction (A$_{r}$; AB)                 &0.45                                & 0.05                             & 1.28            & 0.17             \\
P$_{\rm cc}$                 & 0.01                                & 0.02                             & 0.08            & 3$\times 10^{-4}$             \\
P$_{\rm cc, corrected}^{b}$                 &     0.03                            & 0.04                             & 0.08            & 1.2$\times 10^{-3}$              \\
Spectroscopic redshift (z$_{\rm spec}$) & 0.02746 $\pm$ 0.00003                                & 0.03024 $\pm$ 0.00001                            & 0.07132 $\pm$ 0.00001           & 0.03122 $\pm$ $\pm$ 0.00001           \\
Absolute r-band mag (M$_{r}$; AB)                & -19.71                                  & $-$18.85                                & $-$20.95               &  $-$20.93               \\
Effective radius, R$_{\rm eff}$ (kpc)$^{c}$             & 5.2                                  & 2.1                                &   3.7            & 4.6                \\
SFR (H$_{\alpha}$; M$_{\odot}$/yr)$^d$                  & 3.0 $\pm$ 0.3$^{e}$                                  & 0.054 $\pm$ 0.002$^{f}$                             & 0.15 $\pm$ 0.04$^{e}$               & 1.5 $\pm$ 0.5$^{g}$             \\
SFR (0$-$100 Myr; M$_{\odot}$/yr)$^h$                                        &2.9$_{-0.9}^{+1.6}$                                & 0.15$^{+0.12}_{-0.08}$                              & 0.8$^{+1.1}_{-0.6}$            & 1.6$^{+1.5}_{-0.9}$             \\
Stellar mass log(M/M$_{\odot}$)$^h$                              & 9.86$^{+0.14}_{-0.12}$                                  & 9.29$^{+0.16}_{-0.20}$                                & 10.27$^{+0.13}_{-0.17}$               & 10.26$^{0.09}_{-0.1}$                \\
Stellar metallicity log(Z/Z$_{\odot}$)$^h$                                            & $-0.03^{+0.24}_{-0.18}$                                    & -0.7$^{+0.4}_{-0.5}$                                 & -1.51$^{+0.47}_{-0.35}$                & -0.6$^{+0.4}_{-0.5}$                 \\
Mass weighted age (Gyr)$^h$                                 & 2.7$_{-1.6}^{+2.4}$                                   &  6.6$^{+0.2}_{-2.1}$                                  & 5.5$^{+2.1}_{-2.5}$                & 6.8$^{+2.0}_{-2.3}$                 \\
Log(specific SFR$_{\rm 0-100~Myr}$; yr$^{-1}$)$^h$                                            &  $-9.4^{+0.2}_{-0.2}$                                    & $-9.75^{+0.26}_{-0.33}$                                 & $-10.35^{+0.42}_{-0.71}$                & $-10.20^{+0.36}_{-0.43}$                 \\
A$_{\rm V}$$^h$                                    &  $0.79^{+0.20}_{-0.18}$                                  & $0.16^{+0.07}_{-0.05}$                                  & 0.21$^{+013}_{-0.09}$                & $0.66^{+0.16}_{-0.18}$                 \\
Host AGN$^h$                                                     & N                                    & N                                 & N                & N             \\
PRS luminosity (3$\sigma$ u.l.; 10$^{37}$ erg/s)   & 3.3                                    & 4.0                                 & 23.8                & 4.3             \\
Inclination angle (deg.) & 75$\pm$1$^{i1}$ & 36 $\pm$5$^{i2}$ & 61 $\pm$ 3$^{i1}$ & 69$\pm$1$^{i2}$ \\
 \hline
%\footnotetext{Thus far}
\end{tabular}%
}
%\end{adjustbox}
%}
\label{tab: A-galaxy-properties}
%\end{center}

\textbf{Notes.} \\
$^a$Centroid coordinate of the host galaxy.\\
$^{b}$P$_{\rm cc}$ corrected for the look-elsewhere effect using the formalism discussed in \S\ref{sec:pcc}.
$^c$Half-light radius values  of all except FRB 20181220A host are estimated using SDSS catalog r-band Petrosian magnitudes \citep{2022ApJS..259...35A} and Equation 7 of \cite{2005AJ....130.1535G}. The half-light radius of FRB 20181220A host is estimated using the {\tt Petrofit} package \citep{Geda_2022}.\\
$^d$Values obtained using the optical spectrum of the host galaxies. \\
$^{e}$To correct for host extinction, we use E(B$-$V) = A$_{v}$/R$_{v}$ = 0.19 and 0.06 for FRB 20181220A and FRB 20190418A, respectively. These values are estimated using \texttt{prospector}. For fair comparison with the {\tt Prospector} SFR estimates, we use the SFR$-$H$\alpha$ relation from \citep{1998ARA&A..36..189K}, but adopting the initial mass function (IMF) from \cite{2003PASP..115..763C}: SFR(M$_{\odot}$ y$^{-1}$) = 4.98 $\times$ 10$^{-42}$ L$_{\rm H\alpha}$ (erg s$^{-1}$).\\
$^{f}$From \cite{2023MNRAS.519.2235P}.\\
$^{g}$From \cite{2022A&A...666A.186D}.\\
$^h$Estimated using \texttt{Prospector}.\\
$^{i1}$The inclination angle is estimated using the \texttt{IncliNET} package \citep{2020ApJ...902..145K}.\\
$^{i2}$From \cite{2011ApJS..196...11S}.\\

\end{table}

\subsection{Multi-wavelength counterpart searches}
\label{sec:multi-wavelength}

We search for any promising association of the four low-DM FRBs with transients, such as supernovae, gamma-ray bursts, active galactic nucleus (AGN) flares, and gravitational wave (GW) events, which are proposed to be plausible progenitors for at least some apparently non-repeating FRBs.
%\citep[For a full list of models, see ][]{2019PhR...821....1P}. 
We first search the Transient Name Server\footnote{\url{http://www.wis-tns.org}} (TNS) database \citep{2020TNSAN..70....1Y} for transients (both classified and unclassified) up to 2023 August 1 for transients within the 2$\sigma$ baseband localization region of the four FRBs, and find none.  
We note that FRB 20190425A is proposed to be possibly associated with the GW 190425 event, which we discuss in \S\ref{subsection:GW}.
We next search the Zwicky Transient Facility \citep[ZTF;][]{2019PASP..131g8001G} public alert stream for transient candidates within the baseband localization of FRBs to check for any possible missed transients that are not reported to the TNS or are not classified there. Again, we find none. From this, we conclude that the four low-DM apparently non-repeating FRBs are unlikely to be temporally and spatially associated with any supernovae,  GRBs, and most luminous AGN flares ($>10^{45}$ erg) reported in the two databases.
%, as they are complete to detect these events within the local Universe (z $<$ 0.1). 
We now discuss our search for the presence of any persistent compact radio source, as are found to be spatially co-located with at least two FRB sources \citep{2017Natur.541...58C, 2022Natur.606..873N}.

\subsubsection{Persistent radio source search}
\label{subsection:radio}

We search archival radio data from the following catalogs to check for the presence of a persistent radio source within the four FRB %field-of-view: 
baseband localization regions: the NRAO VLA Sky Survey  \citep[NVSS;][]{condon1998nrao}, Faint Images of the Radio Sky at Twenty-cm survey \citep[FIRST;][]{1995ApJ...450..559B}, the VLA Sky Survey  \citep[VLASS;][]
%[, both 1.1 and 1.2 data releases]
{lacy2016vla}, the Rapid ASKAP Continuum Survey Data Release 1 \citep[RACS DR1;][]{2021PASA...38...58H}, the Westerbork Northern Sky Survey \citep[WENSS;][]{rengelink1997westerbork}, the Tata Institute of Fundamental Research Giant Metrewave Radio Telescope Sky Survey Alternative Data Release \citep[TGSS;][]{intema2017gmrt}, and the Low-Frequency ARray (LOFAR) Two-metre Sky Survey Data Release 2 \citep[LoTSS DR2;][]{2022A&A...659A...1S}. The results of our archival search are presented in Table \ref{tab:radio-limit}. As can be inferred from the Table, only in the case of FRB 20181220A do we find an extended radio source, NVSS J231448+482039, with an integrated flux density of $2.4 \pm 0.4$ mJy. The radio source is only detected in NVSS. Moreover, it is spatially coincident with the center of the host galaxy of FRB 20181220A, 2MFGC 17440. 
%From the non-detection in the TGSS data and assuming a power-law dependence of the NVSS radio source flux density i.e., S$_{\nu} \propto \mathrm{S}^{\alpha}$, we estimated a lower limit on $\alpha > -0.43$. This agrees well with the observed radio continuum spectral index of local star-forming galaxies \citep[between $-$0.1 and $-$0.7;][]{Marvil2015AJ}. 
%Moreover, as we have disfavored the presence of an AGN at the center of NGC 3252 in \S\ref{subsection:AGN}, 
The NVSS source is likely due to ongoing star formation in 2MFGC 17440. To test this, we estimate the SFR of the FRB host using the following radio–SFR relation from \cite{2011ApJ...737...67M}: log(SFR$_{\rm H\alpha}/\rm{M_{\odot} yr^{-1}}) =$ log(L$_{1.4}$(W/Hz)) $-  21.20  = 0.45$, i.e., 2.8 $\rm{M_{\odot} yr^{-1}}$, which matches with the SFR$_{\rm H\alpha}$ value reported in Table \ref{tab: A-galaxy-properties}. Hence, the observed 1.4 GHz radio emission in the FRB 20181220A host is likely due to ongoing star formation in the host. 

Based on this analysis, we find no credible persistent compact radio source within the 2$\sigma$ localization regions of the four low-DM FRBs. To estimate an upper limit on the isotropic luminosity of a putative compact radio source, we use VLASS 2.1 data due to its superior angular resolution (2.5\arcsec) among the radio surveys considered in this study. This enables us to derive a 3$\sigma$ upper limit on the luminosity of any potential compact radio source at 3 GHz, as stated in Table \ref{tab: A-galaxy-properties}.  For all four FRBs, we rule out a persistent radio source like the one found to be spatially coincident with the FRB 20121102A and FRB 20190520B sources \citep[$\approx 7 \times 10^{38}$ erg/s at 3 GHz;][]{chatterjee2017direct,resmi2020,2022Natur.606..873N}.

\begin{table}[h]
%\begin{center}
\hspace{+2.in}
\caption{Summary of the radio limits for the four FRBs derived using archival surveys.}
\label{tab:radio-limit}
%\begin{adjustbox}{max width=1.0\textwidth}{
\resizebox{\textwidth}{!}
{
\begin{tabular}{@{} *8c @{}}%{c|cccccc}
\toprule
\textbf{FRB Name}$^{a}$ & \textbf{LoTSS DR2} & \textbf{TGSS} & \textbf{WENSS} & \textbf{RACS DR1} & \textbf{NVSS} & \textbf{FIRST} & \textbf{VLASS}\\\hline
20181220A & $-$ & $<$ 10.5 mJy & $<$10.8 mJy & $-$ & 2.4 $\pm$ 0.4 mJy & $-$ & $<$ 0.48 mJy\\
20181223C & $<$ 0.250 mJy & $<$ 10.5 mJy & $-$ & $<$ 3 mJy & $<$ 1.3 mJy  & $<$ 0.45 mJy & $<$ 0.48 mJy\\
20190418A & $-$ & $<$ 10.5 mJy & $-$ & $<$ 3 mJy & $<$ 1.3 mJy & $-$ & $<$ 0.48 mJy\\
20190425A & $-$ & $<$ 10.5 mJy & $-$ & $<$ 3 mJy & $<$ 1.3 mJy  & $<$ 0.45 mJy & $<$ 0.48 mJy\\\hline
Frequency$^{b}$ (GHz) & 0.144 & 0.150 & 0.326 & 0.888 & 1.4 & 1.5 & 3 \\ 
\bottomrule
\end{tabular}
}
%\end{adjustbox}
\vspace{12pt}

$^a$For each survey, if the FRB localization region is covered but no source has been detected, a flux density upper limit of 3 $\times$ local rms noise is used. In addition, '$-$' is used to indicate that the survey does not cover the FRB localization region.\\
$^b$Central frequency of a given survey.\\
\end{table}
%C can anything between 27.37 and 27.95, depending on the age of the most dominant ionizing star population, and the NVSS integrated flux density of the radio source from Table \ref{tab:radio-data}, we estimate log(SFR) between $-0.1$ and $-0.7$ which 
%agrees with the SFR estimate in Table \ref{tab:host-properties}. %Hence, the observed 1.4 GHz radio emission is likely synchrotron radiation  generated  by  relativistic electrons accelerated mainly by supernovae in NGC 3252 \citep{cram1998ApJ}. 
%The NVSS . Assuming that the flux density is because of the ongoing SF in the galaxy, we estimate the SFR due 

\subsubsection{Proposed association between FRB 20190425A and GW 190425}
\label{subsection:GW}

\cite{2022arXiv220312038T} searched for gravitational-wave transients associated with FRBs detected by the CHIME/FRB project, during the first part of the third observing run of Advanced LIGO and Advanced Virgo (1 April 2019 15:00 UTC $-$ 1 Oct 2019 15:00 UTC). This duration includes two low-DM FRBs whose host galaxies are reported in this paper. Although the authors found no significant evidence for a gravitational-wave association (for time delay $\sim \pm$ few seconds) in their search, \cite{2023NatAs...7..579M} reported a possible association, at the 2.8$\sigma$ level, between GW 190425 \citep{2020ApJ...892L...3A}, the second binary neutron star (BNS) merger to be detected in gravitational waves, and FRB 20190425A, with UGC 10667 as the most probable host galaxy \citep{2023MNRAS.519.2235P}.
%argue that  of both FRB 20190425A and ge 190425A with an estimated true association probability of around 0.79, which we independently found to be the true host using the baseband localization region which is more than 200 times more than the FRB header localization region reported in \cite{firstchimefrbcatalog2021}) which they used in the paper. 
%However, \cite{2023arXiv230600948B} reassessed the association, taking into account the effect of highly dense and turbulent BNS ejecta and the expected off-axis nature of the merger if the two events are truly related, and found the association highly unlikely.
Nonetheless, \cite{2023arXiv230600948B} reevaluated the association, taking into account the effect of highly dense and turbulent BNS ejecta, as well as the expected off-axis nature of the merger if these events are related. Their findings suggest that the two events are very likely unrelated.

\section{Discussion}

\subsection{Host demographic analysis} 
\label{sec:hostdemographicsanalysis}

We now utilize the four FRB host galaxies reported in this work to explore the demographics of the FRB host population. Our aim is to derive meaningful constraints on the potential formation channels of FRBs. To broaden the scope of our investigation, we have incorporated all published and robustly associated FRB host galaxies to z = 0.1, which we will henceforth refer to as the `local Universe FRB host sample'. This specific redshift cutoff is based on our DM-excess $\leq$ 100 pc cm$^{-3}$ constraint, which effectively narrows down our FRB host search to within z $\approx$ 0.1, as discussed in \S\ref{section:sample}. As of July 2023, we found 18 published FRBs in the literature with robustly associated host galaxies (P$_{\rm cc} \leq$ 10\% or a P(O\textbar x) $>$ 90\%) located within z = 0.1. These are presented in Table \ref{tab:local_universe_frbs}. Additionally, there is no observed statistically significant distinction between the hosts of repeating and apparently non-repeating FRBs based on any of their major modelled physical properties \citep{2023arXiv230205465G}. This suggests that they arise from the same progenitors, or from multiple progenitors that form and evolve in similar environments. Therefore, we do not make any distinction between them in the subsequent analysis, except in \S\ref{subsec:comparison_r_nr_frbs}.

We acknowledge that our local Universe FRB sample comprises hosts localized by different telescopes, each with its own distinct selection effects. Consequently, the sample may exhibit unmodelled biases, potentially impacting the statistical analyses presented in the following sections. To address these concerns, we have undertaken the approach of performing only those analyses where either the selection bias is not important or where its inclusion would only strengthen our conclusions. By doing so, we aim to mitigate the impact of potential biases and ensure the robustness of our findings.

For example, about 33\% of the FRBs in our local Universe FRB host sample were detected using the DM-excess $\leq$ 100 pc cm$^{-3}$ constraint employed in this work. This particular constraint tends to favor FRBs in which the host's DM contribution is relatively low. 
Additionally, it should be noted that early-type galaxies, on average, would likely contribute less DM to the FRB host than late-type galaxies, such as spirals \citep{2015RAA....15.1629X,2022ApJ...927...35C}. Therefore, the low DM-excess constraint would introduce a bias against detecting FRBs in spiral galaxies. Hence, if we detect more spiral hosts, the significance of this observation would only be strengthened by including the selection bias stemming from the low DM-excess cutoff. Furthermore, we note that the low-DM excess cutoff can also introduce a bias against detecting FRBs in galaxies residing in massive halos ( $>$ 10$^{12}$ M$_{\odot}$) if their halos can contribute significantly to the observed host DMs. However, it is unclear how much significant this bias would be.
%this observation has recently encountered challenges from cosmological hydrodynamical simulations. 
For instance, in a study conducted by \cite{2023arXiv231001496K} using the SIMBA suite of simulations, it was demonstrated that various feedback processes, including AGN-driven jets, can result in the loss of more than 50\% of the circumgalactic medium mass from such massive halos into the intergalactic medium.
Even if such a bias against massive halos due to the low DM-excess cutoff were to exist, it seems improbable that it would result in an over-representation of late-type hosts in comparison to early-type hosts. To illustrate, untargeted surveys conducted with radio telescopes, such as the Australian Square Kilometre Array Pathfinder \citep[ASKAP;][]{2018Natur.562..386S} and the Deep Synoptic Array-110 \citep[DSA-110;][]{2023ApJ...949L...3R}, which do not impose a low DM-excess cutoff, have collectively identified over 30 FRB hosts to date \citep{2023arXiv230205465G,2023arXiv230703344L}. However, these surveys have not yet detected any FRBs associated with massive elliptical hosts, while multiple massive late-type hosts have been identified.

\subsubsection{Prevalence of local Universe FRBs in spiral/late-type galaxies}
\label{subsec:spiralhost}

To broadly determine and utilize the FRB host population using the local Universe FRB sample, we first identify the morphological class of FRB host galaxies in our sample. Using the widely accepted formalism in the literature \citep{2011arXiv1102.0550B}, we define two morphological classes: 1) early-type galaxies i.e., ellipticals (E), and transition E/S0 galaxies, which include those centrally concentrated galaxies that look spheroidal or ellipsoidal with a regular shape and no or weak hints of a disk morphology \citep[in short, single-component bulge-dominated galaxies;][]{2014MNRAS.444.1647K}, and 2) late-type galaxies, i.e., spirals (S), irregulars (Irr), and merging galaxies, i.e., galaxies with a clear disk and/or irregular structure or S or Irr with a hint of an ongoing merger process.

In Section \ref{sec:host_search}, we noted that all four low-DM FRB hosts reported in this work are spiral or disk-dominated galaxies based on their morphologies. For the remaining 14 local Universe FRB hosts, we look for existing morphological classification in the NED and SIMBAD databases, and find that the following eight local Universe FRB hosts are classified as `spiral' based on their morphology: FRB 20171020A, FRB 20180916B, FRB 2081030A, FRB 20190303A, FRB 20200120E, FRB 20200223B, FRB 20220207C, and FRB 20220319D.      

Next, we examine the host morphology of the remaining six local Universe FRB host galaxies for the clear presence of the well-formed disk, spiral arms, or bar features in available public images of the FRB hosts. 
%In case of only hints of the presence of a disk or spiral arms, we estimate the light profile of the concentration index (<3) and the best fitted sersic index of the galaxy ($\approx$ 1), both of which are closely correlated with the morphological type (Shimasaku et al.(2001)). 
Except for FRB 20220509G, we find that remaining five FRB host galaxies in our sample, FRB 20201124A, FRB 20210405I, FRB 20211127I, FRB 20211212A, and FRB 20220912A, show either well-formed disk, spiral arms, or bar-like morphology, as noted by the authors in their respective discovery papers as referenced in Table \ref{tab:local_universe_frbs}. For a detailed discussion on this, see Appendix \ref{app:hostsample}.

We note that FRB 20220509G is classified as elliptical by the authors in their discovery papers \citep{2023ApJ...950..175S, 2023ApJ...949L..26C}. However, we find a clear presence of bar-like structure and disk morphology in the Canada-France Imaging Survey (CFIS) r-band image \citep{2017ApJ...848..128I,2019ApJ...887..148F} which is at least 3 magnitude deeper than the Pan-STARRS image used by \cite{2023ApJ...950..175S}. For more discussion on this, see Appendix \ref{sec:FRB20220509G}. Therefore, we conclude that the FRB 20220509 host is a spiral galaxy. 
\begin{figure}[h]
\centering
\includegraphics[width=.55\linewidth]{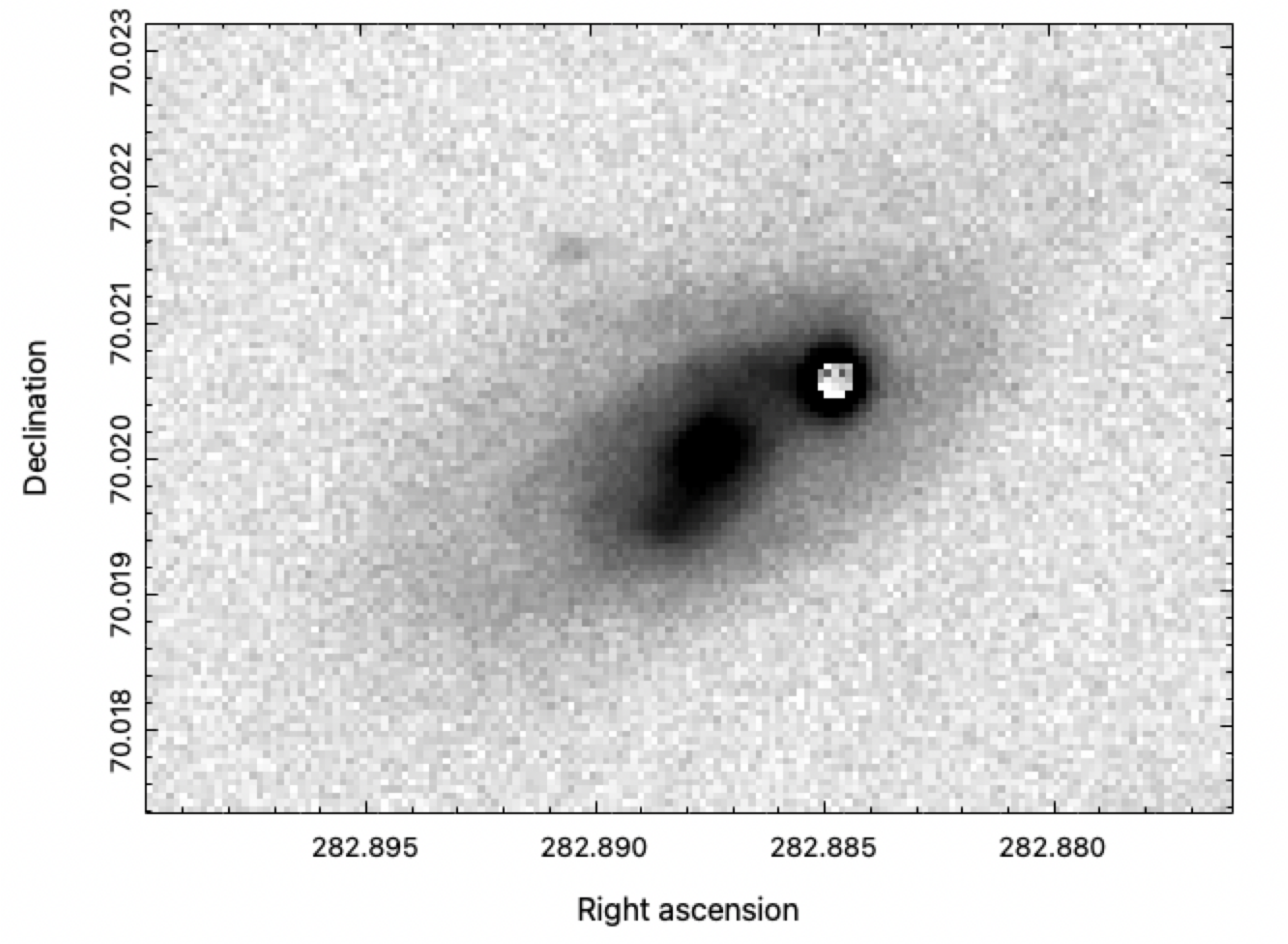}%\centering{}
\caption{The CFIS DR3 r-band image of the FRB 20220509G host (in ICRS reference frame). The star in close proximity to the FRB host is flagged for better visualization of the extended spiral-arm features. }
\label{fig:frb20220509g_cfis}
\end{figure}

\begin{figure}[h]
\centering
\includegraphics[width=.55\linewidth]{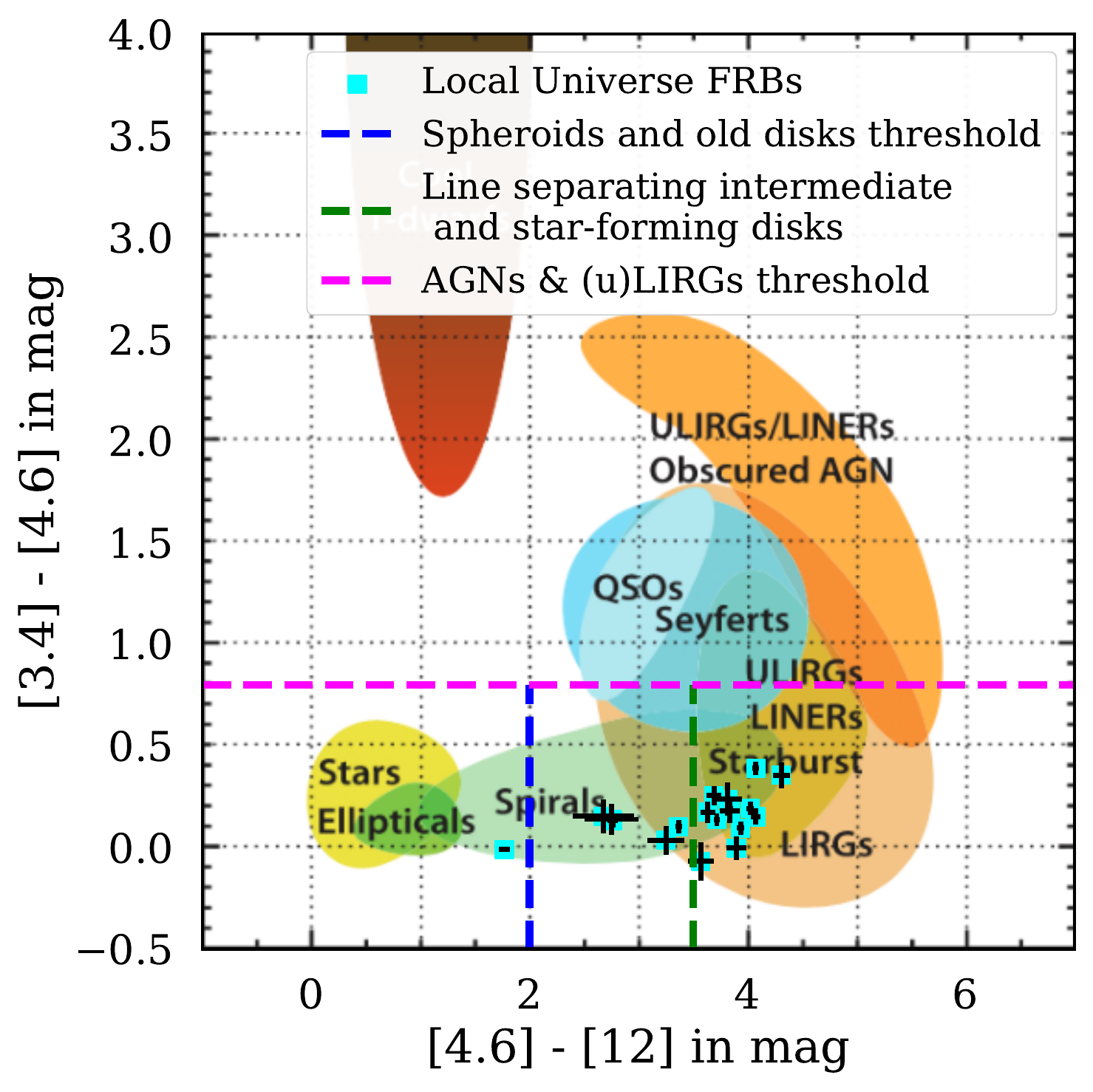}%\centering{}
\caption{WISE color-color plot  adapted from \cite{wright2010} for our sample of local Universe FRB hosts. It illustrates how galaxies separate by type, showing the simple divisions for spheroidal and early-type disks, intermediate disks, and late-type disk galaxies \citep{2017ApJ...836..182J}. The infrared colors of all 18 local Universe FRB
hosts support them being spiral or late-type galaxies.}
\label{fig:wise_classification}
\end{figure}

%which is not found in early-type galaxies, as noted by the authors in their respective discovery papers. 
Additionally, we look at the Wide-field Infrared Survey Explorer (WISE) color-color classification for the morphological classification of the FRB hosts. We find that all 18 FRB host galaxies are detected in WISE and are classified as spirals or star-forming disk galaxies based on the WISE color-color plot shown in Figure \ref{fig:wise_classification} \citep{wright2010}.  
%we show infrared WISE colors of 18 local Universe FRB hosts in three Wide-field Infrared Survey Explorer (WISE) bands, W1 ( 3.4 $\mu$um), W2 (4.6 $\mu$m), and W3 (12 $\mu$m) \citep{wright2010}. 
For this analysis, we use the AllWISE full-sky data release \citep{2014yCat.2328....0C} and extended source \citep{2020yCat..22450025J} catalogs to estimate WISE W1 ( 3.4 $\mu$um) $-$ W2 (4.6 $\mu$m) and W2 (4.6 $\mu$m) $-$ W3 (12 $\mu$m) colors of the 18 local Universe FRB hosts listed in Table \ref{tab:local_universe_frbs}. Overlaid on the same plot are classification thresholds, as estimated by \cite{2017ApJ...836..182J}, to separate early-type (spheroidal), intermediate-type (disk), and late-type (star-forming disk) galaxies. 
The figure clearly reveals that all 18 FRB host galaxies occupy the WISE color-color phase space associated with spiral galaxies or are located within the star-forming disk region. This observation, in conjunction with the morphological evidence discussed earlier, solidifies the classification of all 18 FRB hosts as spiral or late-type galaxies.
%As can be noted from the Figure, All 18 FRB host galaxies occupy the WISE color-color phase space associated with spiral galaxies or lies in the star-forming disk region. Based on this and in conjunction with their morphological evidence discussed above, 
%this solidifies the classification of all 18 FRB hosts as spiral galaxies.
%as discussed in Appendix \S\ref{sec:morphology}. The classification and other information of the FRBs in our list is presented in Table \ref{tab:local_universe_frbs}. Therefore,
%we classify the 18 local Universe FRB hosts as spiral galaxies. 

\begin{table}[ht]
\begin{center}
\caption{Sample of 18 local Universe FRBs (z$_{\rm host} \leq$ 0.1) considered in this study}
\label{tab:local_universe_frbs}
\begin{tabular}{@{} *5c @{}}\toprule
\textbf{FRB} & \textbf{Galaxy morphology} & \textbf{Redshift}  & \textbf{Repeater (Y/N)} & \textbf{Reference}\\\midrule 

20200120E & spiral  & $-$0.0001$^{a}$ & Y & \cite{bhardwaj2021}\\ 

20181030A & spiral  & 0.0039 & Y &  \cite{2021ApJ...919L..24B} \\ 

20171020A & spiral  & 0.0086 & N & \cite{2018ApJ...867L..10M}\\
 & & & & \cite{2023arXiv230517960L} \\		 

20220319D & spiral  & 0.011 & N & \cite{2023arXiv230101000R} \\	 

20181220A & spiral  & 0.0275 &	N & This work \\ 

20181223C & spiral  & 0.0302 & N & This work \\	 

20190425A & spiral  & 0.0312 & N & This work \\	 

20180916B & spiral  & 0.0337 & Y & \cite{2020Natur.577..190M}\\ 

20220207C & disk-dominated  & 0.043 & N &  \cite{2023arXiv230703344L} \\

20211127I & spiral  & 0.0469 & N & \cite{2023arXiv230205465G} \\ 
& & & & \cite{2023ApJ...949...25G} \\

%20201123A & spiral  & 0.0507 & \cite{2022MNRAS.514.1961R} \\

20200223B & spiral  & 0.0602 & Y & \cite{2023arXiv230402638I} \\ 

20190303A & spiral  & 0.064 & Y & \cite{2022arXiv221211941M} \\ 

20210405I & spiral  & 0.066 & N & \cite{2023arXiv230209787D}\\ 

%20180814A & spiral  & 0.0684  & \cite{2022arXiv221211941M}\\ 

20190418A &	spiral  & 0.0715 & N & This work \\	 

20211212A &	spiral  & 0.0715 & N & \cite{2023arXiv230205465G} \\	 

20220912A &	disk-dominated  & 0.077  & Y & \cite{2023ApJ...949L...3R}\\	 

20220509G & spiral & 0.0894  & N & This work (see Appendix D) \\	 

20201124A &	spiral & 0.098 & Y & \cite{2022Natur.609..685X}\\	 

\bottomrule 
 \hline
\end{tabular}
\end{center}
$^{a}$The line-of-sight velocity of M81 is dominated by its peculiar velocity. The comoving distance of M81 from us is $\approx$ 3.6 Mpc \citep{2019MNRAS.488..590B}.
\end{table}

The presence of 18 FRB sources in spiral hosts is an intriguing observation, as our local Universe host sample is likely biased against spirals (see \S\ref{sec:hostdemographicsanalysis}). We next examine if the distribution of FRB hosts aligns with the expected number of spirals in the local Universe \citep[$\sim$ 70$-$80\%;][]{2010A&A...509A..78D,2013RAA....13..651D,2016MNRAS.461.3663H,2017AJ....153..111G}. Using binomial statistics, we test this hypothesis and find a 1.2\% likelihood of this outcome by chance assuming the local Universe spiral fraction = 80\%. Therefore, the notable excess of spirals among FRB hosts is noteworthy, suggesting a strong preference for FRB progenitors within spiral or late-type galaxies. 

Note that the observed prevalence of late-type FRB host galaxies within the local Universe holds the potential to constrain the age of FRB progenitors. This is demonstrated by the distinction between early-type and late-type galaxies, where the former generally exhibit earlier star formation than the latter. Consequently, the stellar populations in early-type galaxies tend to be older than those in their late-type counterparts \citep{2013ApJ...777...18M,2017MNRAS.472.2833S}. This distinction in the average ages of the stellar populations, coupled with the inherent time delay between the peak star formation in the host galaxies (assuming that the FRB progenitors are most likely to form during that period) and FRB occurrences ultimately gives rise to discernible event rates between these two host categories. 
Such studies have been conducted for short GRBs (sGRBs), in which the probability distribution of the time delay for sGRB progenitors is assumed to follow a power law with an index $n$, such that P$(\tau) \propto \tau^{n}$ \citep{2007ApJ...665.1220Z}. If we employ this formalism for FRB progenitors using the early-to-late type host ratio $\sim$ 0.06, we would constrain the power law index for the FRB progenitors to be $n < -1.5$ \citep[from Figures 2 and 4 of][]{2007ApJ...665.1220Z}, a value steeper than what is observed for Type-Ia supernovae ($n$ $\approx-1$). Note that a steeper power-law index indicates that the majority of FRB progenitors exhibit shorter time delays (or lower ages for the progenitors of FRB sources) in comparison to the progenitors of sGRBs and Type-Ia supernovae.
We further elaborate on this in the following section.

%In conclusion, The observed excess of late-type galaxies can help us uncover the lifetime of FRB progenitors. 

%In this next section, we discuss in greater detail the implication of the observed 

\subsubsection{What is likely the most preferred FRB formation channel?}
\label{subsec:preferred_channel}

The observed preponderance of local Universe FRBs in spirals or late-type galaxies bears an important clue about their dominant progenitor formation channels. This preponderance might also extend to higher redshifts as no elliptical FRB host has been reported to date even at z $> 0.1$. In this section, based on the observed excess of spiral host galaxies, we aim to identify the dominant FRB formation channel. However, we note that this does not exclude the possibility that a small fraction of FRBs might be produced by channels other than the dominant one. 
%However, the relative significance of those channels is likely small enough to not significantly bias the main attributes of the FRB host population.

\textbf{Core collapse supernovae (CCSNe) formation channel}: This channel, encompassing Type Ib (SNIb), Type Ic (SNIc), and Type II (SNII) supernovae, is mostly observed in star-forming late-type galaxies, as noted in various studies \citep[e.g.,][]{2010MNRAS.405...57S,2012ApJ...759..107K,2020ApJ...904...35P,2021ApJS..255...29S}. Only a small fraction, less than 1\%, of all CCSNe occur within elliptical galaxies \citep{2022ApJ...927...10I}. This rarity can be attributed to their likely origin from the collapse of massive stars ($\geq$ 8 M$_{\odot}$). Considering that the lifetimes of these stars are expected to be shorter than 150 million years \citep{2017A&A...601A..29Z}, CCSNe should predominantly emerge in galaxies that are either young or actively undergoing star formation. These categories primarily encompass spiral galaxies, but also include irregular galaxies, providing a natural explanation for the observed prevalence of spiral hosts for local FRBs. 
Furthermore, observables such as host offsets, star formation rate (SFR), and stellar mass, derived from the existing albeit small sample of FRB host galaxies, align with the characteristics expected for neutron stars formed via prompt channels \citep{2020ApJ...899L...6L,2021ApJ...907L..31B, 2021MNRAS.508.1929C}. %Finally, \citep{heintz2020host} and  found that the projected offset of FRBs

\noindent However, there exist observations that potentially challenge this assertion, for instance, the detection of FRB 20200120E in an M81 globular cluster \citep{bhardwaj2021,Kristan2021arXiv} and the linkage of FRB 20180814A \citep[a repeating FRB;][]{2022arXiv221211941M} and FRB 20220509G \citep[an apparently non-repeating FRB;][]{2023ApJ...950..175S} to quiescent spiral galaxies. We delve into these specific cases below.
%\S\ref{subsec:sm_sfr}.
In the subsequent paragraphs, we explore whether alternative formation channels could also account for the excess of late-type FRB hosts within the local Universe.

\textbf{Long GRBs (LGRBs) and superluminous supernovae (SLSNe) formation channel}: These channels have been proposed to form potential progenitors for repeating FRBs, exemplified by FRB 20121102A \citep{2017ApJ...841...14M}. The formation channels are invoked to account for distinct features in the case of FRB 20121102A, including the presence of a low-metallicity dwarf host, a persistent radio source, and heightened activity in the repeating FRB source potentially linked to its young age \citep{2017ApJ...843...84N}. However, subsequent analyses of FRB host properties, such as the distribution of offsets between hosts and bursts, host stellar mass, and metallicity, seem inconsistent with LGRBs and SLSNe as the prevailing channels for FRB source formation \citep{bhandari2020host,2022AJ....163...69B}. Furthermore, the high volumetric rate of local Universe FRBs cannot be explained by the proposed formation of millisecond magnetars through these channels \citep{2021ApJ...919L..24B}.
Therefore, despite their occurrence in late-type star-forming galaxies (mostly dwarfs), we argue against LGRBs and SLSNe as the dominant FRB source formation channels.

\textbf{Globular cluster (GC) origin of FRB sources}: The discovery of FRB 20200120E within an old M81 GC suggests the potential for dynamically formed FRB sources within dense cluster cores, potentially via the accretion-induced collapse of white dwarfs and binary white dwarf mergers \citep{Kristan2021arXiv,2021ApJ...917L..11K}. It is now widely acknowledged that the populations of globular clusters in galaxies correlate with the stellar and dark halo mass of their host galaxies \citep{2014ApJ...787L...5H}. This suggests that if GCs are the principal sources of FRBs, the overall FRB rate should primarily follow stellar mass. 
In this scenario, the relative detection rates of FRBs in early- and late-type FRB host galaxies should align with the distribution of stellar mass in these galaxies. 
\cite{2014MNRAS.444.1647K} analyzed a morphologically classified sample of 2711 local Universe galaxies (z $< 0.1$) selected from the Galaxy and Mass Assembly (GAMA) survey with global stellar mass
((log(M$_{\ast}$/M$_{\odot} > 9.0$), and find that $\approx 79$\% of the stellar mass in the local Universe is currently found within early-type galaxies. 
However, our local Universe FRB host sample does not have any early-type hosts. Using binomial statistics, we find that the probability of finding no early-type host in our sample strongly disfavors (p-value $\approx$ 0) the null hypothesis that FRBs trace stellar mass alone. This remains true even when we consider only those local Universe FRB hosts in our sample that were not identified using the low DM-excess cutoff (i.e., 12 hosts). Therefore, we conclude that globular cluster sources are not the dominant formation channel of FRBs. 

\textbf{Delayed formation channels linked with Type Ia supernovae}: Several delayed formation pathways for FRB sources have been proposed, including binary white dwarf mergers \citep{2013ApJ...762L..17P,2018Ap&SS.363..242L,2021ApJ...907..111Z} and accretion-induced collapse of white dwarfs \citep{2020ApJ...893....9Z,2023MNRAS.525L..22K}. These scenarios are theorized to either lead to type Ia supernovae (SNIa) or share comparable rates and local environments to SNIa \citep{1999ApJ...516..892F, 2000ARA&A..38..191H, 2010AIPC.1314..233R, 2013ApJ...776L..39K}. Hence, in this section, we consider SNIa as a representative of these models. It is important to note that in this context, the cataclysmic channels mentioned earlier do not take place within globular clusters and consequently, they are not anticipated to solely follow the distribution of stellar mass \citep{2021ApJ...917...28K}. However, unlike CCSNe, SNIa originate in a mix of early-type and late-type galaxies \citep{2014MNRAS.438.1391P}. 
To assess if SNIa explain the observed late-type galaxy fraction, we analyze late-type SNIa host galaxies (z $\leq$ 0.1) using the Pantheon cosmological supernova catalog \citep{2018ApJ...859..101S}, yielding a fraction of 0.76. For a detailed discussion, see Appendix \ref{sec:typeIa_frb}. Employing binomial statistics, we test the null hypothesis of an inherent fraction of the late-type hosts of 0.76. The observed late-type FRB host fraction of 1 results in a p-value of 0.007, hence disfavoring the null hypothesis. Thus, FRB formation channels that trace SNIa appear unlikely to be the dominant FRB source formation channel.

\textbf{Delayed formation channels linked to binary neutron star (BNS) mergers}: BNS mergers have been proposed as possible FRB source formation channels \citep{2013PASJ...65L..12T,2014A&A...562A.137F,2018ApJ...864..117M,2022arXiv220312038T}. Moreover, alongside neutron star-black hole mergers, they are also believed to produce sGRBs \citep{2022PhRvD.105h3004S}. 
While the similarity in the offset distribution of FRB and sGRB from their host nuclei is noteworthy \citep{2020ApJ...903..152H, 2022AJ....163...69B}, sGRBs, like SNIa, are observed in both early and late galaxies, differing from what we found in our local Universe FRB host sample \citep{2022ApJ...940...56F}. Moreover, BNS mergers struggle to explain the high all-sky FRB rate \citep[$\{525 \pm 30\}^{+142}_{-131}$ FRBs/sky/day;][]{2021ApJS..257...59C} even if most of their remnants become repeating FRBs \citep{2020ApJ...893...44Z,2023arXiv230600948B}. Therefore, BNS mergers are unlikely to be the dominant formation channel of FRB sources.

%The GZ data also revealed a significant population of red spirals (Bamford et al. 2009; Skibba et al. 2009). With a simple selection of all spiral/diskgalaxies, these works suggested that around18 per cent of these galaxies are to be found in the red sequence (with an increasing fraction in intermediate density regions). The fraction of red spirals was seen to increase with environment, even at fixed stellar mass, suggesting that some environmental process is suppressing star formation in spiral galaxies, beyond a simple change in the mass function with environment (Bamford et al. 2009).

After examining all major FRB progenitor formation channels studied in the literature, we argue that CCSNe are the only channel that can explain the observed excess of late-type galaxies in the local Universe FRB host sample along with other FRB observables. Therefore, we argue that core-collapse supernovae are likely the dominant FRB source formation channel. However, we note that the rate of CCSNe is known to trace recent star formation \citep{2009ARA&A..47...63S,2022ApJ...927...10I}, which may not be the case for FRBs \citep{2022ApJ...924L..14Z, 2022MNRAS.511.1961H}. In the next section, we describe the implication of this observation on the nature of FRB sources formed via CCSNe. 

%, consistent with an evolved progenitor and an event rate that traces both stellar mass and star formation (Sullivan et al. 2006).
%

\subsubsection{Local Universe FRBs' star-formation rates and stellar masses: implications for the progenitors}
\label{subsec:sm_sfr}

In \S\ref{subsec:preferred_channel}, we argued, based on the preponderance of local Universe spiral hosts, that core-collapse supernovae are likely the dominant channel to form FRB sources. In this section, we continue to examine the similarities between the local Universe FRB host sample and CCSN hosts based on their host galaxies' physical properties, such as star-formation rate and stellar mass. Furthermore, among various core-collapse supernova channels, we focus here on Type II CCSNe (SNII) because of their relatively large volumetric rate\citep[$\sim$ 10$^{5}$ Gpc$^{-3}$ yr$^{-1}$;\footnote{This aligns well with the high occurrence of FRBs, $\sim 10^{5}$ FRBs/yr/Gpc$^{3}$ at a fiducial energy of $\sim 10^{39}$ ergs \citep{2023ApJ...944..105S}.}][]{2009A&A...499..653B} and possible connection with Galactic magnetar SGR 1935+2154 (see \S\ref{sec:intro}).
%and 1E 1547.0–5408
%\citep{2021ApJ...907....7I} which are reminiscent of FRBs\citep{2016MNRAS.457.3448I,2018ApJ...852...54K} which in 2020 emitted FRB-like radio bursts \citep{2020Natur.587...54C, 2020Natur.587...59B}.
Moreover, no significant statistical differences among the host galaxies of SNIb, SNIc and SNII have been noted in the literature \citep{2012MNRAS.424.1372A,2013MNRAS.436..774E,2014A&A...572A..38G}. 

To compare our local Universe FRB host sample with that of SNII hosts, we use the Palomar Transient Factory Core-collapse Supernova Host-galaxy catalog \citep[][]{2021ApJS..255...29S}.\footnote{The catalog can be accessed from here: \url{10.26093/cds/vizier.22550029}} We consider those SNII (1) discovered at z $<$ 0.1 and (2) for which independent and consistent spectroscopic redshift measurements for both the SNII and their associated host galaxies are available. To check for the host galaxy redshift, we use the Sloan Digital Sky Surveys (SDSS), Release 16 (DR16) catalog \citep{2020ApJS..249....3A}. This yields a sample of 83 SNII host galaxies, which forms the basis of our comparative analyses discussed below. 

\begin{figure}[h]
\centering     %%% not \center

\subfigure{\includegraphics[width=0.75\textwidth]{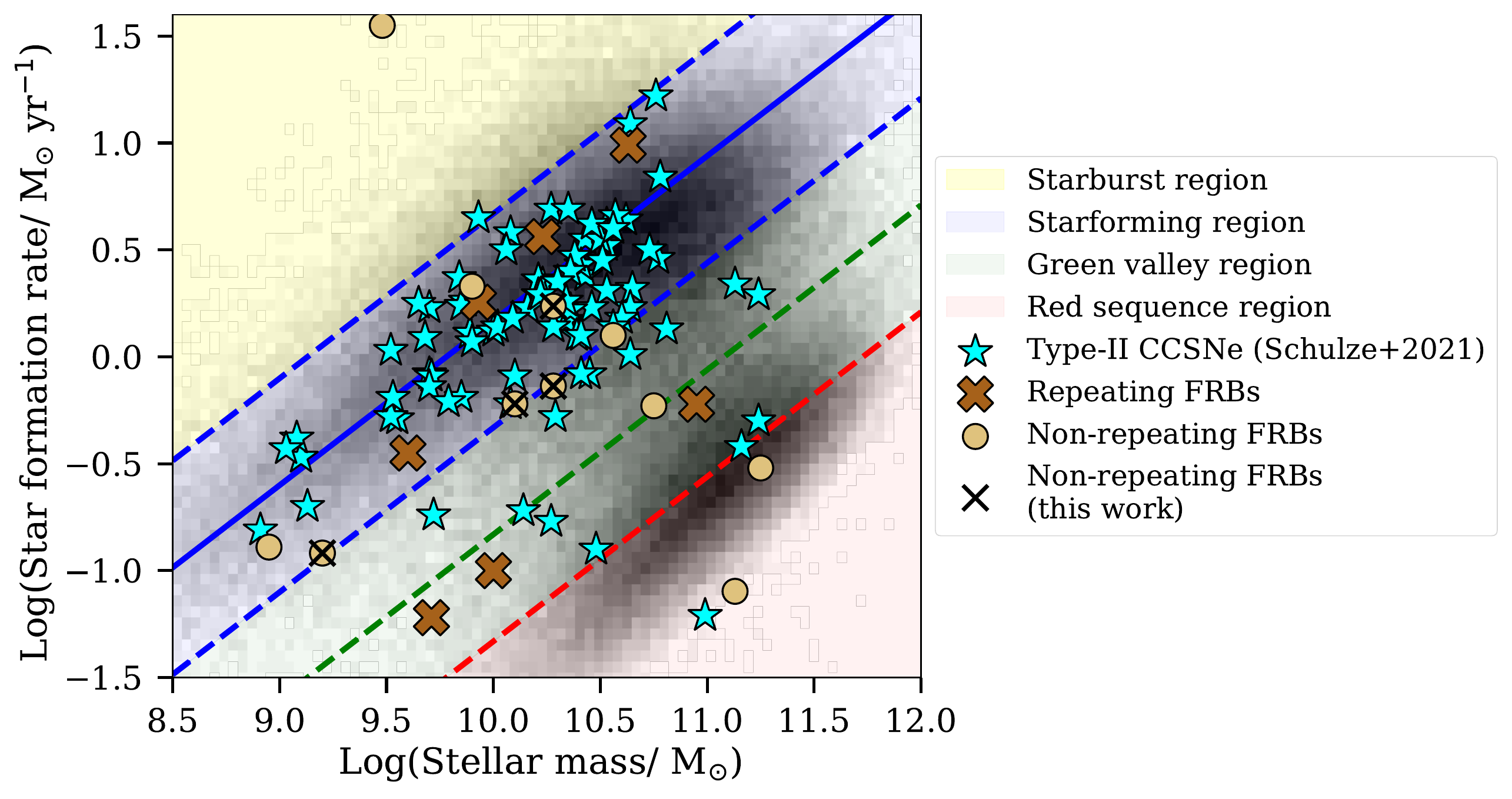}}\\

\subfigure{\includegraphics[width=0.40\textwidth]{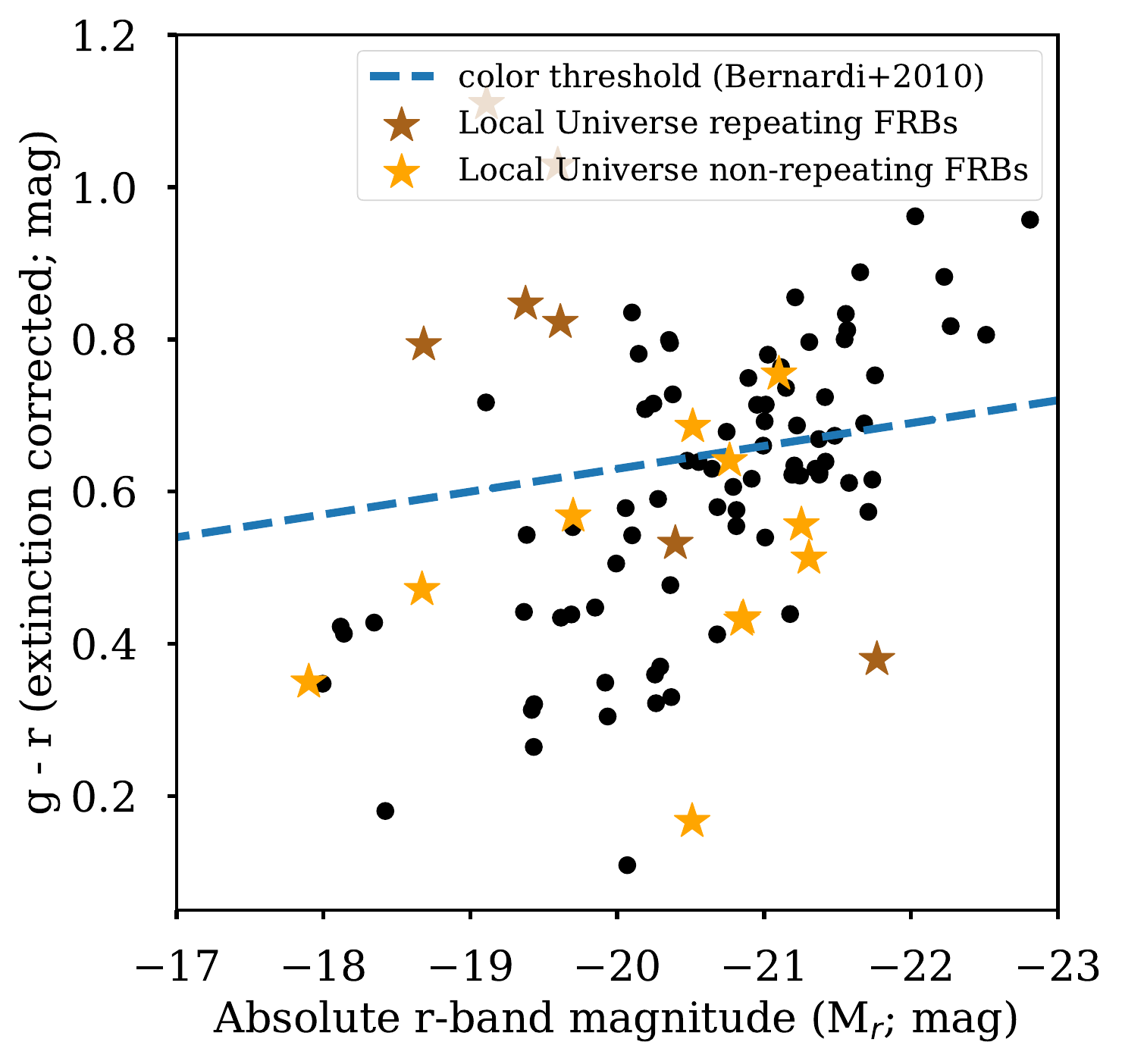}}\\

\caption{Comparison of notable physical properties of the host galaxies of 18 local Universe FRBs and 83 SNII. The sample of host galaxies of SNII (represented as cyan stars) are taken from the Palomar Transient Factory (PTF) Core-collapse Supernova Host-galaxy catalog \citep[][]{2021ApJS..255...29S}. Our local Universe FRB sample consists of 7 repeating FRBs (bold crosses) and 11 apparently non-repeating FRBs (circles). The 4 apparently non-repeating FRBs presented in this work can be differentiated from other apparently non-repeating FRBs via the crosses overlaid. Top plot: Star formation rate (SFR) versus total stellar mass (M$_{\ast}$) for local Universe FRBs and PTF SNII hosts. The background population of galaxies is taken from the Sloan Digital Sky Survey MPA-JHU DR7 catalog \citep[][]{2004MNRAS.351.1151B}. The solid blue line represents the star-forming main sequence \citep[SFMS;][]{2010ApJ...721..193P}. Following the framework developed by
\cite{2017ApJ...850...22L}, dashed lines are spaced 1$\sigma$ = 0.5 dex apart (the median scatter in SFMS), with the green valley falling 1-3$\sigma$ below the SFMS (the region between the lower blue dashed line and red dashed line). The galaxies below the red dashed line are found to be quiescent. These lines are used to define four regions that are shaded and labeled in the Figure. Bottom plot: Rest-frame color-magnitude diagram of the host galaxies of FRBs (stars) compared to those of Type II CCSNe (black circles). The rest-frame colors and absolute r-band magnitude of FRBs are taken from their discovery papers, or in cases where those values are missing, we estimate them using the host photometry magnitudes in the Pan-STARRS release 1 (PS1) survey catalog data \citep{2016arXiv161205560C}.}
\label{fig:ccsne_comparison}
\end{figure}

From the SFR versus stellar mass plot shown in Figure \ref{fig:ccsne_comparison}, it can be noted that most SNII host galaxies ($\approx$ 83\%) are located above the star-forming main-sequence (SFMS) region threshold (the dotted blue line that separates star-forming and green valley regions in Figure \ref{fig:ccsne_comparison}). This is expected as SNII mark the endpoints in the lives of short-lived (lifetime $\lesssim$ few $\times$ 10$^{7}$ yr) but massive stars (M $\gtrsim$ 8 M$_{\odot}$), making their detection more probable within galaxies undergoing active star formation. In comparison, we note the lower fraction of local Universe FRB host galaxies located in the star-forming region ($\approx$ 67\%). However, using the binomial statistics discussed in \S\ref{subsec:spiralhost} and \S\ref{subsec:preferred_channel}, the relatively lower presence of FRB hosts in the SFMS region, in contrast to SNII hosts, lacks statistical significance (p $>$ 0.07) and can be attributed to the modest sample size and unaccounted selection effects within our local Universe FRB host sample (see \S\ref{sec:hostdemographicsanalysis}). 
Similar insights emerge from the color-absolute magnitude plot shown in Figure \ref{fig:ccsne_comparison}. Notably, local Universe FRBs exhibit a discernible yet statistically modest surplus in the fraction of red-sequence galaxies (50\%) compared to SNII hosts (37\%) using the color-magnitude threshold defined by \cite{2010MNRAS.404.2087B} to select red-sequence galaxies (above the dotted blue line in Figure \ref{fig:ccsne_comparison}). Therefore, the similarity between the local Universe FRB and Type II CCSNe host galaxies, which is evident in their color, star formation rate, and stellar mass properties, suggests a common progenitor. This supports the assertion made in \S\ref{subsec:preferred_channel} that CCSNe likely serve as the predominant progenitor formation channel for FRBs, thereby arguing for the neutron star origins of the majority of local Universe FRBs.

However, it is still possible that the other channels discussed in \S\ref{subsec:preferred_channel} might also be contributing a small number of sources to the FRB population. For example, the source of the repeating FRB 20200120E is likely produced via a delayed-formation channel \citep{Kristan2021arXiv}. However, it is still unclear if FRB 20200120E is a true representative of the FRB population or just an outlier, as the majority of the bursts detected from the FRB 20200120E source exhibited lower energies \citep{2021ApJ...919L...6M, 2022NatAs...6..393N, 2022MNRAS.513.1858T, 2023MNRAS.520.2281N} compared to the bursts observed from SGR 1935+2154 on 28 April 2020 \citep{2020SGR,Bochenek2020}. Nonetheless, the detection of FRBs in red sequence spirals, such as FRB 20180814A \citep[tentative host;][]{2022arXiv221211941M} and FRB 20220509G \citep{2023ApJ...950..175S}, poses a challenge when attempting to account for them through sources formed via CCSNe, as CCSNe typically prefer star-forming late-type galaxies.
%would be challenging to explain using the sources formed via  CCSNe as they prefer star-forming late-type galaxies.
However, a small but non-negligible fraction of CCSNe is found to occur in quiescent spirals \citep{2008A&A...488..523H, 2011ApJ...730..110S,2022ApJ...927...10I}. Moreover, it is also possible that these two FRBs are associated with localized star-formation activity that is not captured by the global properties of their hosts, as seen in the case of FRB 20180916B \citep{2020Natur.577..190M, 2020arXiv201103257T}. Therefore, a larger FRB host sample is required to establish definitively whether the presence of quiescent spiral FRB hosts and GC sources is indeed inconsistent with the CCSNe progenitor scenario.%2010MNRAS.405...57S,2012ApJ...759..107K,2021ApJS..255...29,

\subsubsection{Comparison of the host properties of repeating and apparently non-repeating FRBs}
\label{subsec:comparison_r_nr_frbs}

\begin{figure*}[h]
\centering     %%% not \center

\subfigure{\label{fig7:a}\includegraphics[width=0.49\textwidth, height=7cm]{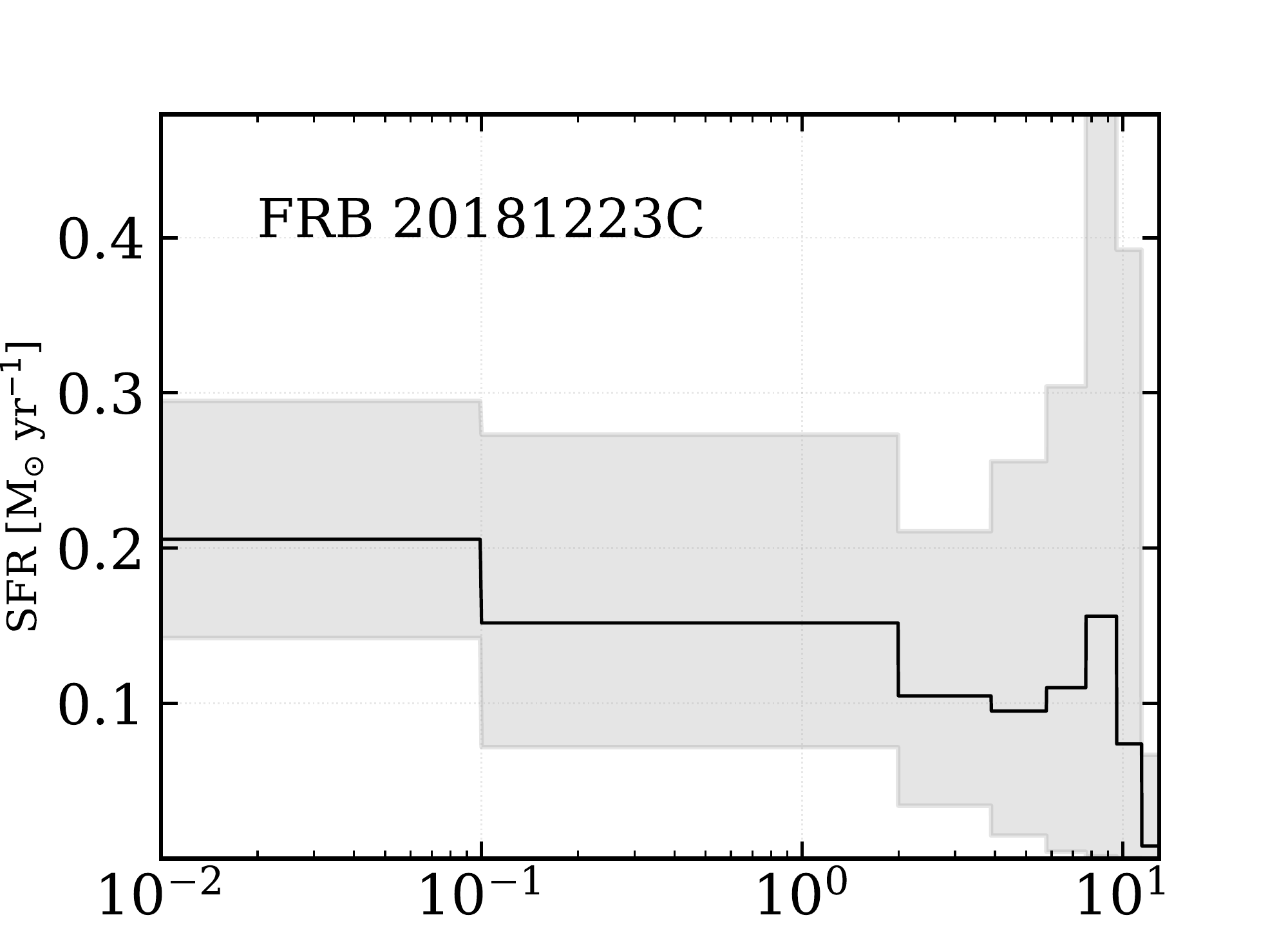}}
\subfigure{\label{fig7:b}\includegraphics[width=0.49\textwidth, height=7cm]{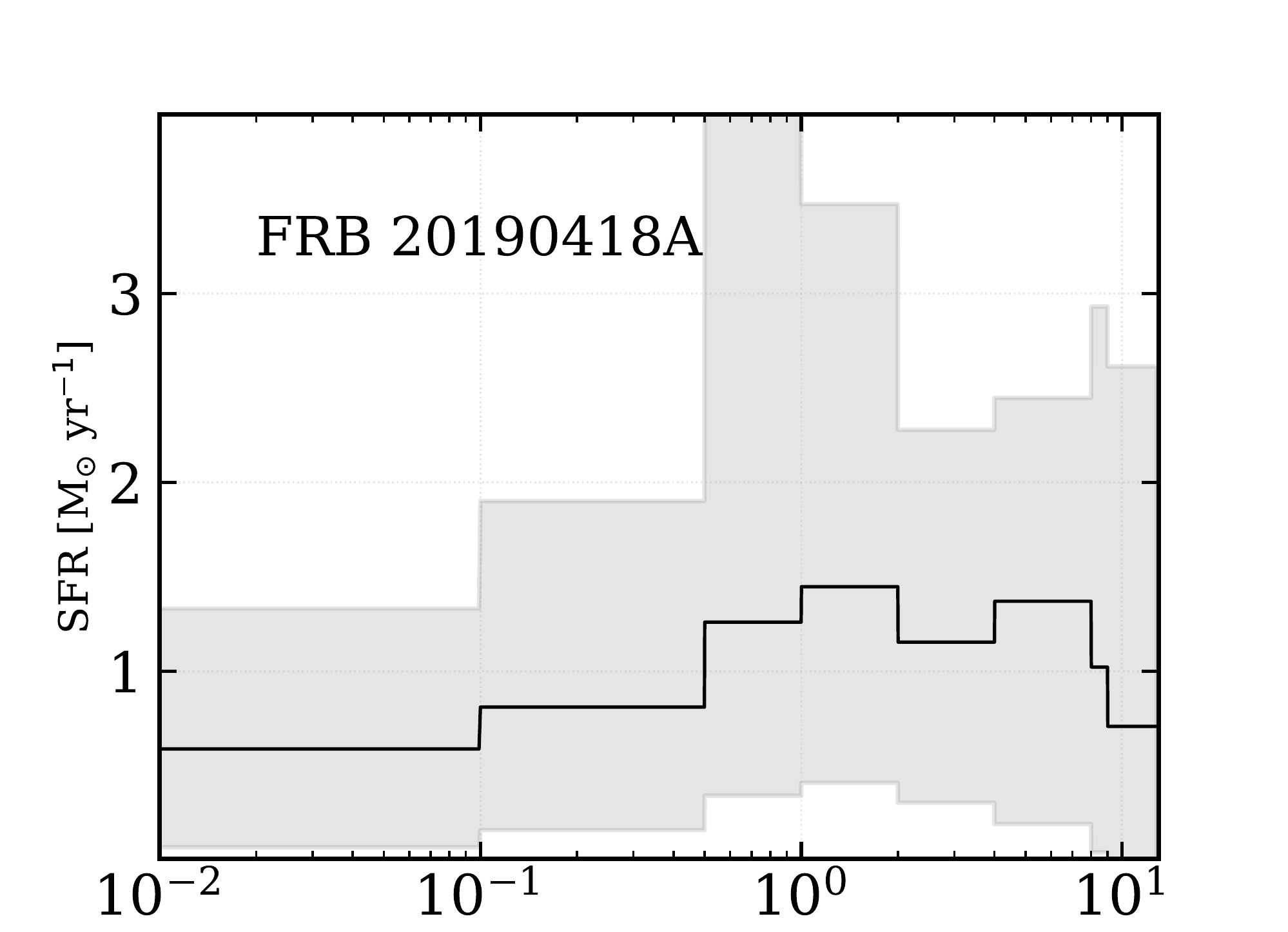}}\\

\subfigure{\label{fig7:c}\includegraphics[width=0.49\textwidth, height=7cm]{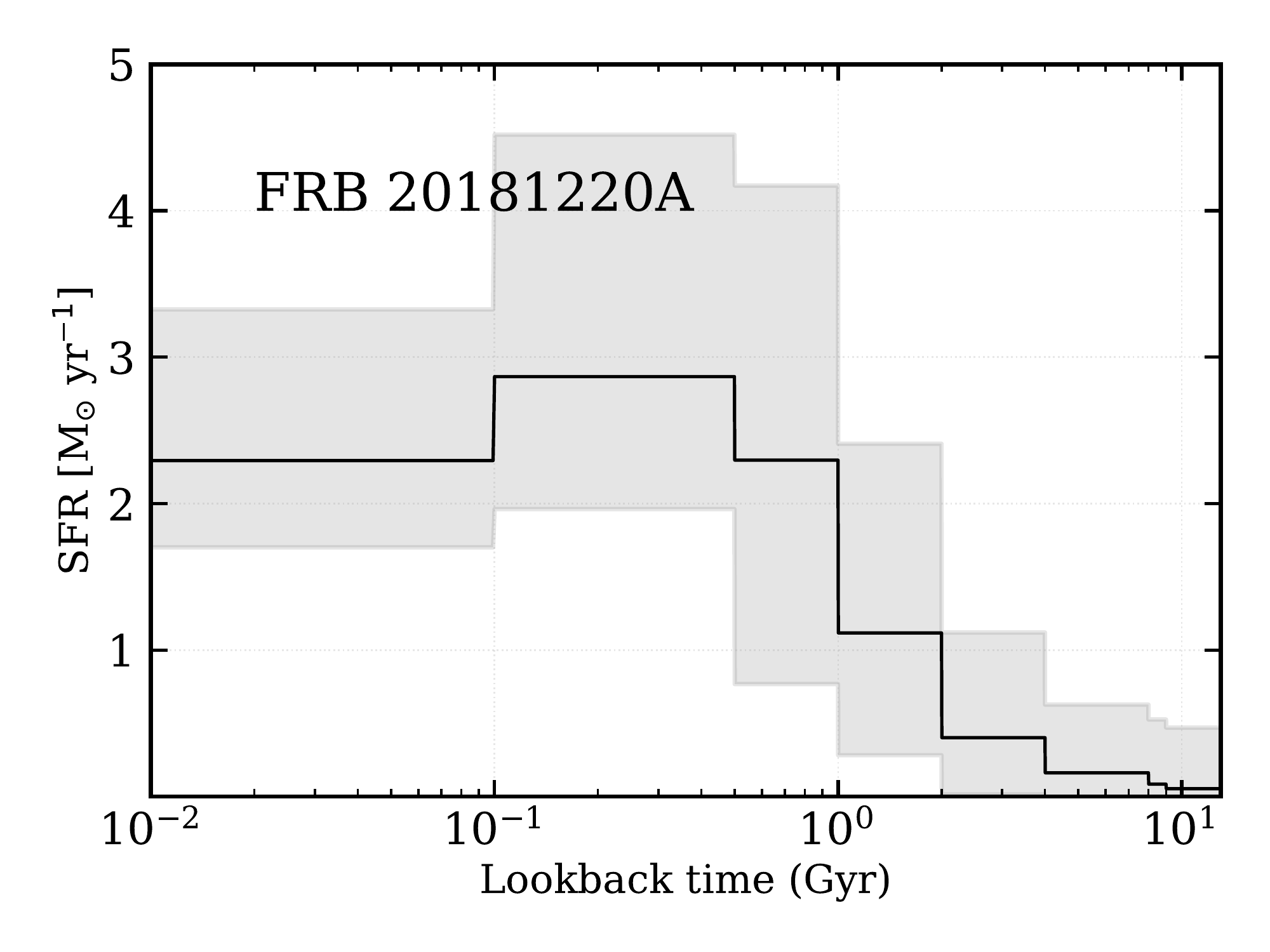}}
\subfigure{\label{fig7:d}\includegraphics[width=0.49\textwidth, height=7cm]{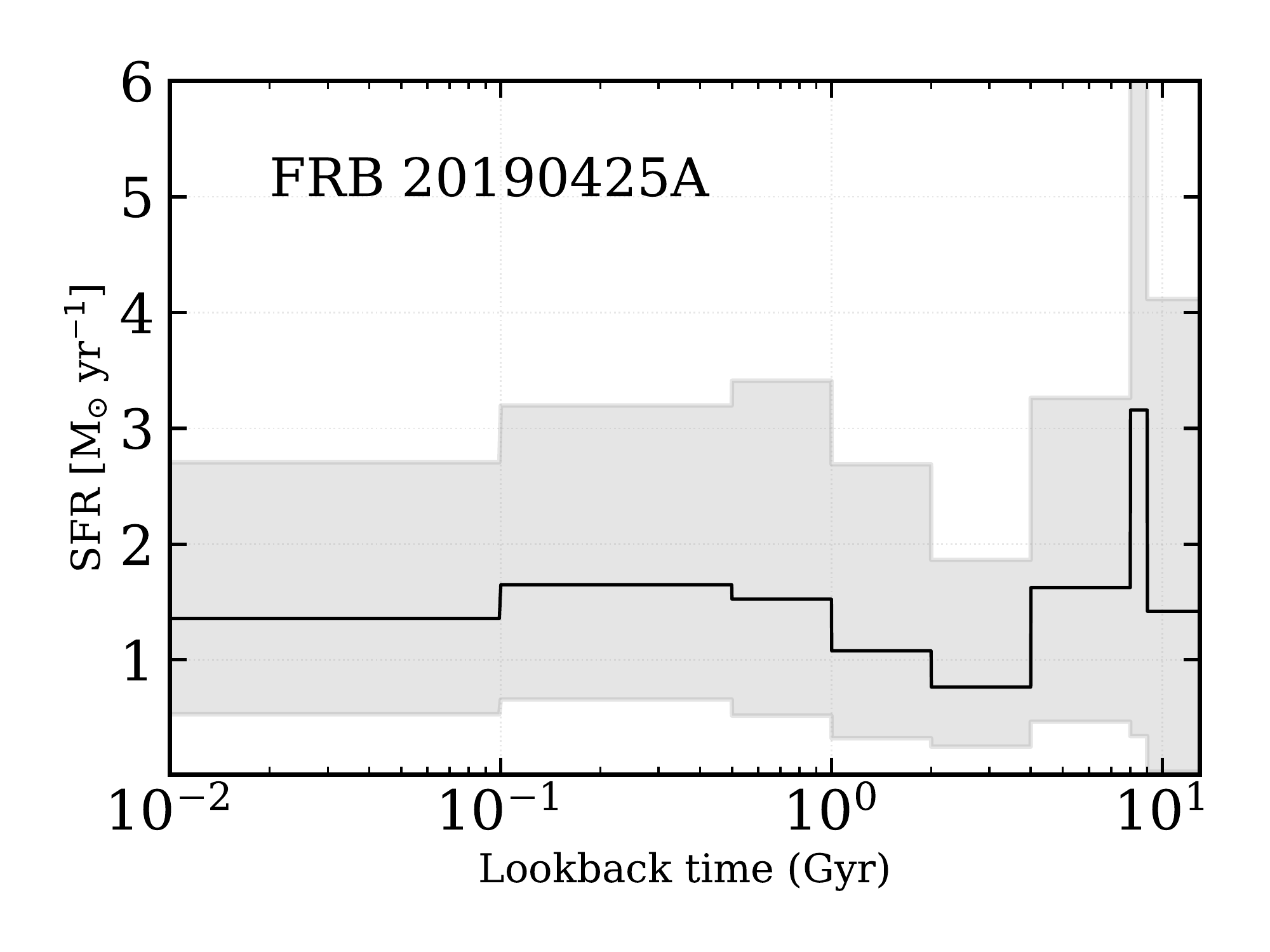}} 
\caption{Star formation histories (SFHs) of the four FRB host galaxies reported in this work. The x-axis is the look-back time (Gyr), such that the left-hand side is the present day, and the right-hand side is the age of the Universe at the redshift of the galaxy. These SFHs are estimated using {\tt Prospector}. For more details, see Appendix \ref{app:prospector}.}
\label{fig:SFH}
\end{figure*}

The vast majority of FRB sources have only been detected once,
%\citep{2022A&ARv..30....2P}, 
but a small sub-population ($\lesssim$ 3\%) is known to burst repeatedly \citep{firstchimefrbcatalog2021,2023ApJ...947...83C}. This naturally raises speculations that there are two distinct classes of FRBs \citep{2018ApJ...854L..12P, 2018NatAs...2..839C, 2019MNRAS.484.5500C}. 
Supporting evidence comes from the strong temporal and spectral differences (including in the DM distribution) of repeater and apparent non-repeater bursts \citep{2021ApJ...923....1P}.
%or dissimilar DM distribution of CHIME/FRB repeater and apparent non-repeater sources.  However, some observations blur the distinction.
%For example, based on burst morphology, one can make a case for apparently non-repeating and repeating FRBs to be two distinct classes.  but that does not necessarily require invoking distinct progenitor models \citep{2021ApJ...923....1P}. %However, recent observations have blurred the distinction between apparently non-repeating and repeating FRBs.
In terms of FRB energetics, we also find that repeating FRBs are capable of producing bursts of energy comparable to what is seen for apparently non-repeating localized FRBs \citep{2022ATel15817....1O, 2023arXiv230615505K}. Given that all other major observables of the two classes remain statistically indistinguishable for the two populations \citep{firstchimefrbcatalog2021,2021MNRAS.500.3275C,2022Univ....8..355Z}, it is not unreasonable to assume that they have a common origin \citep[for example, see][]{2023arXiv230617403J}. A similar inference can be made by comparing their host galaxies. For example, \cite{2022AJ....163...69B} and \cite{2023arXiv230205465G} found the global properties of repeating and apparently non-repeating FRB hosts to be statistically indistinguishable. 
However, it is important to note that the sample size of FRB hosts employed in their study may not be large enough to ensure the robustness of the inferred conclusions
%However, it is important to note that the sample of FRB hosts they employed is not large enough to assure robustness of inferred conclusions. 
For example, \cite{2023arXiv230205465G} note that the hosts of repeating FRBs generally extend to lower stellar masses and that the hosts of non-repeating FRBs arise in more optically luminous galaxies. 
But for our local Universe FRB sample, as can be inferred from Figure \ref{fig:ccsne_comparison}, the median masses of repeating and apparently non-repeating FRB hosts are comparable (log(M/M$_{\odot}$) = 10.1 and 10.2, respectively). However, we note that the median absolute r-band magnitude (M$_{\rm r}$) of apparently non-repeating FRB hosts is $\approx$ 1 AB mag larger than that of repeating FRBs, which was also noted by \cite{2023arXiv230205465G}. 
%regarding the difference in host luminosities. 
Moreover, in their sample of 23 FRBs consisting of six repeating FRBs and 17 apparently non-repeating FRBs, \cite{2023arXiv230205465G} found that four out of six repeating FRB hosts show peaks in their star-formation rates within a look-back time of 100 Myrs (often called time-delay), while for the apparently non-repeating FRBs, that ratio is only seven out of 17. Assuming that the FRB progenitor (zero-age main-sequence star) was born at the time of peak star formation, this might imply that repeating FRB sources are statistically younger than apparently non-repeating ones, as a more active central engine should correspond to a younger age. Using Prospector and a non-parametric SFH, we show in Figure \ref{fig:SFH} the SFHs of the four low-DM FRB host galaxies reported in this study over look-back time, which is estimated using the formalism discussed in Appendix \ref{app:prospector}.
For the four low-DM FRBs, we find that three FRBs show time-delays $\lesssim$ 100 Myrs, as can be inferred from Figure \ref{fig:SFH}, which is at odds with the aforementioned observation made by \cite{2023arXiv230205465G}. Thus, we clearly need a larger sample of both repeating and apparently non-repeating FRB hosts to truly see any statistical difference between the distributions of their delay times. 

%Thus, our preliminary conclusion from the host demographics is that local Universe FRB sources do not appear to track the cosmic stellar mass fraction in early- and late-type galaxies. https://arxiv.org/abs/2101.05144

\subsection{Are all local Universe FRBs repeating sources?}
\label{subsec:repition_rate}

Based on the high volumetric rate of local Universe CHIME bursts, \cite{2021ApJ...919L..24B} argued that the nearby CHIME FRBs are likely to be repeating sources. Therefore, we searched for repeat bursts from the four low-DM FRBs in CHIME data at epochs between 2019 September 30 and 2021 May 1, using the formalism employed by \cite{2023ApJ...947...83C} and found none. Briefly, we first calculate the effective Poisson burst rates of the four apparently non-repeating FRBs based on the effective exposure time within the baseband localization region (3$\sigma$) of the FRBs
%(within a full-width half maxima of CHIME/FRB formed beams at 600 MHz) 
and sensitivity \citep[using fluence threshold of 5 Jy ms, i.e., the average sensitivity of the CHIME/FRB system; ][]{2021ApJ...923....2J} and on the detection of one burst from each source. We then infer 68\% confidence intervals on each rate using the \cite{1991ApJ...374..344K} formalism. Note that we estimate the total on-sky exposure only for the upper transit (more sensitive) for the FRBs from the start of the experiment, 2018 August 28, through to 2021 May 1, from the CHIME/FRB exposure map produced and made public by \cite{2023ApJ...947...83C}. The exposure values of the four FRBs are stated in Table \ref{tab:FRB-params}. To remove the effect of cosmological distance-related selection effects that can bias against detecting repeat bursts from more distant FRBs, we compare the distance-independent burst rates of the four low-DM FRBs with those of well-localized repeating FRBs that have either been discovered or have had at least one burst detected by CHIME (in the case of FRB 20121102A).
We first convert the fluence sensitivity threshold for each FRB in this study into respective isotropic energies using the host redshifts reported in the discovery papers. For CHIME/FRB repeating sources such as FRBs 20200223B, 20190110C, 20180814A, 20180916B, 20181030A, 20190303A, and 20201124A, the burst rates at 5 Jy ms have been already reported by \cite{2023ApJ...947...83C}. For FRB 20121102A, we use the fluence sensitivity threshold of 7 Jy ms from \cite{2019ApJ...882L..18J} and z = 0.19273 from \cite{2017ApJ...834L...7T}. We then scale the burst rates to the fiducial FRB isotropic energy threshold of 10$^{39}$ erg s$^{-1}$. 

\begin{figure}[!h]
\centering
\includegraphics[width=.95\linewidth]{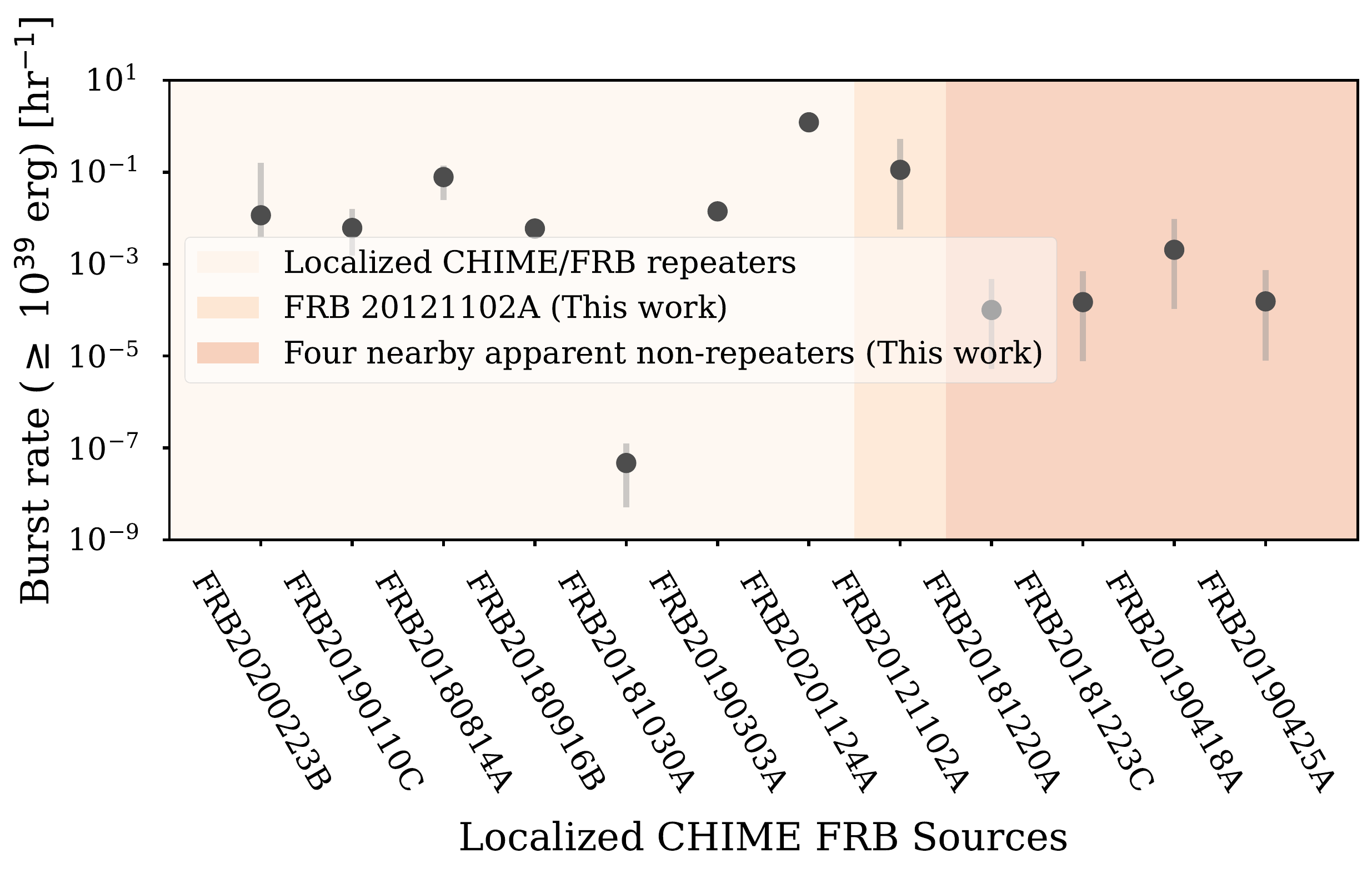}%\centering{}
\caption{Comparison of the effective Poisson burst rates of the four low-DM FRBs studied in this work with those of published CHIME/FRB repeating sources with known hosts.}
\label{fig:repetion_rate}
\end{figure}

The scaling involves the assumption of a power-law index of $-$1.5 for the cumulative burst energy distribution of each source, as assumed by \cite{2023ApJ...947...83C}. Figure \ref{fig:repetion_rate} shows the scaled burst rate of the localized repeating and the four apparently non-repeating sources, respectively. As shown in the plot, although the burst rates of the four low-DM FRBs are lower than those of localized CHIME repeating sources, they are still larger than the observed burst rates of FRB 20181030A. 
This suggests that there is a large variation in the burst rates of localized repeating sources, and thus it is still possible for the four low-DM FRBs to be repeating sources. Therefore, we encourage further follow-up of these sources using more sensitive radio telescopes. 
Finally, we note that the burst rate of FRB 20121102A in the CHIME band is the same as what is reported at 1.4 GHz \citep{2021MNRAS.500..448C}. This suggests that it is possible that FRB 20121102A has only been detected once in the CHIME band because of the comparatively low-sensitivity of CHIME \citep{2019ApJ...882L..18J}, rather than solely due to either an intrinsically lower burst rate at $\sim$ 600 MHz (i.e., the source's luminosity peaks at higher frequencies) or due to propagation effects. Further monitoring of the FRB 20121102A source would aid in determining the most probable reason among the aforementioned potential explanations.

\section{Summary \& Conclusions}
\label{sec:conclusion}

We present the host galaxies of four apparently non-repeating FRBs, FRBs 20181223C, 20190418A, 20191220A, and 20190425A, that were first reported in CHIME/FRB Catalog-1. These FRBs are selected based on a planned hypothesis testing framework where only FRBs with low DM-excess ($<$ 100 pc cm$^{-3}$), high Galactic latitude ($\lvert$b$\rvert> 10\degrees$), and saved baseband data were considered. The baseband data facilitated the estimation of much more precise localization regions than the ones reported by \cite{firstchimefrbcatalog2021} using the CHIME/FRB baseband localization pipeline \citep{2021ApJ...910..147M}. From our host follow-up studies, we identify only one host galaxy candidate in the deepest available archival images for each FRB. We then robustly associate the identified host candidates to their respective FRBs with chance association probabilities $<$ 10\%, after correcting for the look-elsewhere effect. 

We also infer major host properties and star formation histories of the host galaxies using {\tt Prospector} with a non-parametric star formation history model. Additionally, we search for possible multi-messenger and multi-wavelength counterparts, including gravitational wave events and compact persistent radio sources, within the 2$\sigma$ baseband localization region of the FRBs and find none. Based on this null result, we derived a 3$\sigma$ upper limit on the luminosity of compact radio sources at 3 GHz, effectively eliminating the possibility of a compact radio source similar to those observed in the cases of FRB 20121102A and FRB 20190520B.
We note that the host of FRB 20190425A matches to what \cite{2023NatAs...7..579M} proposed based on the suggested association between GW 190425 and FRB 20190425A. However, the association of GW 190425 with UGC 10667 has already been shown to be unlikely by \cite{2023arXiv230600948B}, which renders the proposed GW-FRB association unlikely. 

We also conduct an FRB host demographic analysis, considering only local Universe FRB hosts (z$<$0.1) reported until July 31, 2023. Our choice of the maximum redshift limit is guided by the DM-excess criterion within our planned hypothesis testing framework. This results in a sample of 18 local Universe FRB hosts, including the four reported in this study. We find all 18 FRB hosts to be spiral or late-type galaxies. The predominance of late-type galaxies suggests that the sources of FRBs are significantly younger than those of sGRBs and SNIa.
More importantly, among all major FRB source formation channels proposed in the literature, core-collapse supernovae appear to be the dominant one. Moreover, our local Universe FRB host sample remains broadly consistent with the hosts of core-collapse supernovae. However, it is still possible that other formation channels contribute a limited number of sources to the overall FRB population. 
%However, the exact fraction of their contribution remains uncertain. 
Furthermore, we compare the host properties of repeating and apparently non-repeating FRBs in our local Universe FRB sample. Our analysis does not reveal any significant evidence of discernible differences in their stellar populations.
%This lends support to the notion that both types share a common origin. 

Finally, considering their close proximity to Earth, we search the CHIME/FRB database for repeat bursts, between 2018 July 1 and 2021 May 1, from the four apparently non-repeating FRBs and find none. Nonetheless, their burst rates fall within the range observed for localized CHIME/FRB repeating sources. Therefore, we encourage continued monitoring of these FRBs using more sensitive radio telescopes.

\facility{CHIME, Gemini North, GTC, Magellan Baade Telescope}

\software{ Prospector \citep{Leja2017,prospect2019}, FSPS \citep{Conroy2009}, MARZ \citep{Hinton}, dynesty \citep{2020MNRAS.493.3132S}, IRAF \citep{tody1986iraf,tody1993iraf}, Fermi GBM data tools \citep{GbmDataTools}, Astropy \citep{astropy:2013,astropy:2018}, APLpy \citep{2012ascl.soft08017R}, SAOImage DS9 \citep{2003ASPC..295..489J}, NumPy \citep{harris2020array}, Matplotlib \citep{Hunter:2007}}

\acknowledgements
We thank the Dominion Radio Astrophysical Observatory, operated by the National Research Council Canada, for gracious hospitality and expertise. We acknowledge that CHIME is located on the traditional, ancestral, and unceded territory of the Syilx/Okanagan people. CHIME is funded by a grant from the Canada Foundation for Innovation (CFI) 2012 Leading Edge Fund (Project 31170) and by contributions from the provinces of British Columbia, Qu\'ebec and Ontario. The CHIME/FRB Project is funded by a grant from the CFI 2015 Innovation Fund (Project 33213) and by contributions from the provinces of British Columbia and Qu\'ebec, and by the Dunlap Institute for Astronomy and Astrophysics at the University of Toronto. Additional support was provided by the Canadian Institute for Advanced Research (CIFAR), McGill University and the McGill Space Institute via the Trottier Family Foundation, and the University of British Columbia. The Dunlap Institute is funded through an endowment established by the David Dunlap family and the University of Toronto. Research at Perimeter Institute is supported by the Government of Canada through Industry Canada and by the Province of Ontario through the Ministry of Research \& Innovation.  FRB research at UBC is supported by an NSERC Discovery Grant and by the Canadian Institute for Advanced Research.  The CHIME/FRB baseband system is funded in part by a Canada Foundation for Innovation John R. Evans Leaders Fund award to I.H.S. The National Radio Astronomy Observatory is a facility of the National Science Foundation (NSF) operated under a cooperative agreement by Associated Universities, Inc.  
This work is based on data obtained as part of the Canada-France Imaging Survey, a CFHT large program of the National Research Council of Canada and the French Centre National de la Recherche Scientifique. Based on observations obtained with MegaPrime/MegaCam, a joint project of CFHT and CEA Saclay, at the Canada-France-Hawaii Telescope (CFHT) which is operated by the National Research Council (NRC) of Canada, the Institut National des Science de l'Univers (INSU) of the Centre National de la Recherche Scientifique (CNRS) of France, and the University of Hawaii.
This work uses observations made with the Gran Telescopio Canarias (GTC), installed at the Spanish Observatorio del Roque de los Muchachos of the Instituto de Astrofísica de Canarias, in the island of La Palma.
%cite NASA/IPAC database
This research has made use of the NASA/IPAC Extragalactic Database (NED), which is operated by the Jet Propulsion Laboratory, California Institute of Technology, under contract with the National Aeronautics and Space Administration. 
%SIMBAD and %VizieR
This research has made use of the SIMBAD database and VizieR catalog access tool operated at CDS, Strasbourg, France.
%This research made use of Photutils, an Astropy package for detection and photometry of astronomical sources (Bradley et al. 2023). This research made use of PetroFit (Geda et al. 2022), a package based on Photutils, for calculating Petrosian properties and fitting galaxy light profiles.

M.B. is a McWilliams fellow and an International Astronomical Union Gruber fellow. M.B. also receives support from the McWilliams seed grant. V.\,M.\,K. holds the Lorne Trottier Chair in Astrophysics \& Cosmology, a Distinguished James McGill Professorship, and receives support from an NSERC Discovery grant (RGPIN 228738-13), from an R. Howard Webster Foundation Fellowship from CIFAR, and from the FRQNT CRAQ. 
K.S. is supported by the NSF Graduate Research Fellowship Program.
B.\,C.\,A. is supported by an FRQNT Doctoral Research Award.
A.M.C. is funded by an NSERC Doctoral Postgraduate Scholarship. 
F.A.D is supported by the UBC Four Year Fellowship.
The Dunlap Institute is funded through an endowment established by the David Dunlap family and the University of Toronto. B.M.G. acknowledges the support of the Natural Sciences and Engineering Research Council of Canada (NSERC) through grant RGPIN-2022-03163, and of the Canada Research Chairs program. 
K.W.M. holds the Adam J. Burgasser Chair in Astrophysics and is supported by NSF grants (2008031, 2018490).
C. L. is supported by NASA through the NASA Hubble Fellowship grant HST-HF2-51536.001-A awarded by the Space Telescope Science Institute, which is operated by the Association of Universities for Research in Astronomy, Inc., under NASA contract NAS5-26555.
A.P. is funded by the NSERC Canada Graduate Scholarshops -- Doctoral program.
A.B.P. is a Banting Fellow, a McGill Space Institute~(MSI) Fellow, and a Fonds de Recherche du Quebec -- Nature et Technologies~(FRQNT) postdoctoral fellow.
Z.P. is a Dunlap Fellow.

\bibliographystyle{aasjournal}
\bibliography{ref.bib}

\newpage
\appendix

%--------------- Appendix A ---------------------

\section{Stellar Population Synthesis Using {\tt Prospector}}
\label{app:prospector}

To determine the properties of the stellar population in the FRB host galaxies, we employ a python-based Bayesian inference code, {\tt Prospector} \citep{Leja2017,prospect2019}. It generates model spectral energy distributions (SEDs) using stellar population synthesis models defined within the framework of the Flexible Stellar Populations Synthesis (FSPS) stellar populations code \citep{Conroy2009}. By fitting the archival photometry data with {\tt Prospector} using a nested sampling fitting routine called {\tt dynesty} \citep{2020MNRAS.493.3132S}, we obtain posterior distributions for the stellar population properties of interest, which include the age of the galaxy at the time of observation (with the maximum allowed value being the age of the universe at the redshift of the FRB host), total mass formed, stellar metallicity (log(Z/Z$_{\odot}$)), and V-band optical depth.

\begin{table}[h]

\caption{ Free parameters and their associated priors for the {\tt Prospector} `continuity$\_$flex$\_$sfh' model.}
\label{tab:sfhmodel}
\begin{center}
\hspace{-1.in}
%\resizebox{0.95\textwidth}{!}{ 
\begin{tabular}{ p{3cm}p{6.5cm}p{5.3cm}}\toprule
\textbf{Parameter} & \textbf{Description} & \textbf{Prior}\\\midrule 
log(M/M$_{\odot}$)                                                        & total stellar mass formed                                                      & uniform: min=8, max = 12.0                                                                        \\
log(Z$_{*}$/Z$_{\odot}$)                                                  & stellar metallicity                                                            & clipped normal: min = $-$0.5, max= 1.0, mean and $\sigma$ following \cite{2019ApJ...877..140L} mass-metallicity prior \\
%log(Z$_{*}$/Z$_{\odot}$)$_{gas}$                                          & gas-phase metallicity                                                          & uniform: min = -2.0, max=0.5                                                                      \\
$\hat{\tau}_{\lambda, 2}$ & diffuse dust optical depth                                                     & uniform: min = 0.0 mag, max = 2.5 mag                                                             \\
n                                                                         & slope of \cite{2013ApJ...775L..16K} dust attenuation curve                                & uniform: min = $-$1.0, max = 4.0                                                                  \\
f$_{\rm AGN}$                                                             & fraction of total AGN luminosity relative to the bolometric stellar luminosity & log uniform: min = 10$^{-5}$, max=3                                                               \\
$\tau_{\rm AGN}$                                                          & optical depth of the AGN dust torus                                            & log uniform; min = 5, max=150                                                                     \\
log(SFR$_{\rm ratio, young}$)                                             & ratio of SFR in youngest bin to last flex bin                                  & Student t (mean=0, scale=0.3, df=2)                                                                                       \\
log(SFR$_{ratio, old}$)                                                   & 3-vector array; ratio of SFR in old bins to first flex bin                     & Student t (mean=0, scale=0.3, df=2)                                                                                      \\
log(SFR$_{ratio}$)                                                        & 4-vector array; ratio of SFR in flex bins                                      & Student t (mean=0, scale=0.3, df=2)                                                                                                                \\
 \hline
\end{tabular}
\end{center}
\end{table}

In all our fits using {\tt Prospector}, we adopt the \cite{2003PASP..115..763C} IMF and utilize a non-parametric continuity$\_$flex$\_$sfh
with 9 star formation history (SFH) bins \citep{2019ApJ...876....3L}. In this model, the edges of the time bins are adjusted so that for a given set of SFRs, the same amount of mass forms in each bin. The SFR from both old and young populations in each time bin are determined using the ``continuity” prior, which places a Student-t prior on the log of the ratio of the SFR in adjacent bins (“log SFR$_{ratio}$”). This prior encourages smooth SFHs, where the SFR does not jump significantly between each time bin. However, sharp burst or quenching events are still allowed. Note that the edges of the first and last time bins are fixed, which are $\approx$ 10 Gyr (to cover the first 1.5 Gyr of the galaxy’s history) and 10 Myr (to capture the most recent SFR) lookback time, respectively, in our runs. However, the edges of the other 7 time bins are allowed to vary. These bins are used to capture the SFR at certain epochs in the SFH such that each bin forms an equal stellar mass. Note that this allows for more flexibility in the duration, start time, and end time of the recent burst: the SFR can change at an arbitrary time, as opposed to only changing at the edges of fixed time bins. In our case, we use default student-t prior values of its three parameters, which are described in Table \ref{tab:sfhmodel}.
%However, after repeated run, we fixed even remaining 7 bins to optimal edge values we get in our trail runs (all in lookback time): 100 Myr,500 Myr, 1 Gyr, 2 Gyr, 4 Gyr, 8 Gyr, and 9Gyr. Note that our fitting results do not depend on this choice; 
    
We also fix the model redshifts to the spectroscopic redshifts of the FRB host galaxies and add nebular emission \citep{2017ApJ...840...44B} to the SED using the default fixed parameters in {\tt Prospector}. To ensure realistic masses corresponding to a given stellar metallicity, we constrain the total mass formed and stellar metallicities using the \cite{2005MNRAS.362...41G} mass-metallicity relation.
For measuring dust attenuation, we employ the \cite{2013ApJ...775L..16K} model, which includes a sampled parameter determining the offset from the \cite{2000ApJ...533..682C} attenuation curve. Furthermore, following \cite{2020MNRAS.494..529W}, we fix the birth-cloud optical depth to the same value as the diffuse optical depth. This implies that young stars are attenuated twice as much as old stars. We also set the dust emission parameters such that the warm dust fraction is fixed to 0.01, the minimum radiation field is fixed to 1.0, and the Polycyclic Aromatic Hydrocarbon (PAH) mass fraction is fixed to 2\%, as proposed by \cite{2019ApJ...876....3L}. furthermore, because all four FRB host galaxies have available 2MASS and WISE data, we incorporate the \cite{2007ApJ...657..810D} infrared dust emission model, which consists of a three-components.

Given that all four FRB host galaxies have WISE magnitudes available, we also included the AGN dust torus emission described by
two parameters, f$_{\rm AGN}$ and $\tau_{\rm AGN}$ \citep{2008ApJ...685..147N,2018ApJ...854...62L}, to account for a possible AGN contribution to the mid-IR part of the SED. It is worth noting that our fitting process does not indicate evidence of a significant AGN contribution in our FRB hosts' SED (f$_{\rm AGN} < 1$\%). This is also evident from the WISE color-color diagnostics discussed in \S\ref{subsec:spiralhost}.

Finally, to calculate the stellar mass, present-day SFR, and mass-weighted age for each host, we follow the methods outlined in 
\cite{2021ApJ...919L..24B}.

\begin{figure*}
\centering     %%% not \center

\subfigure{\label{fig9:b}\includegraphics[width=0.95\textwidth]{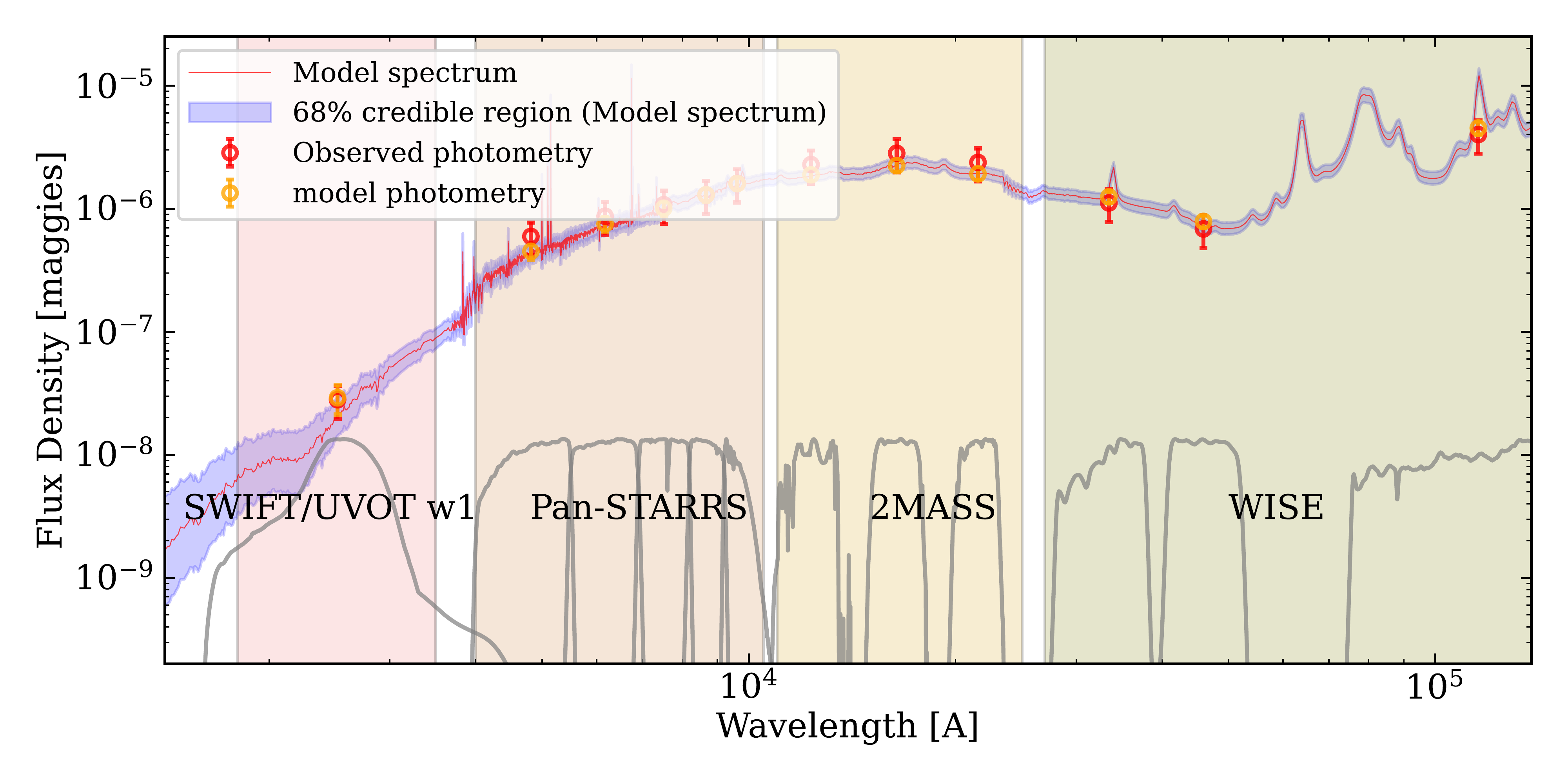}}\\

\subfigure{\label{fig9:c}\includegraphics[width=0.95\textwidth]{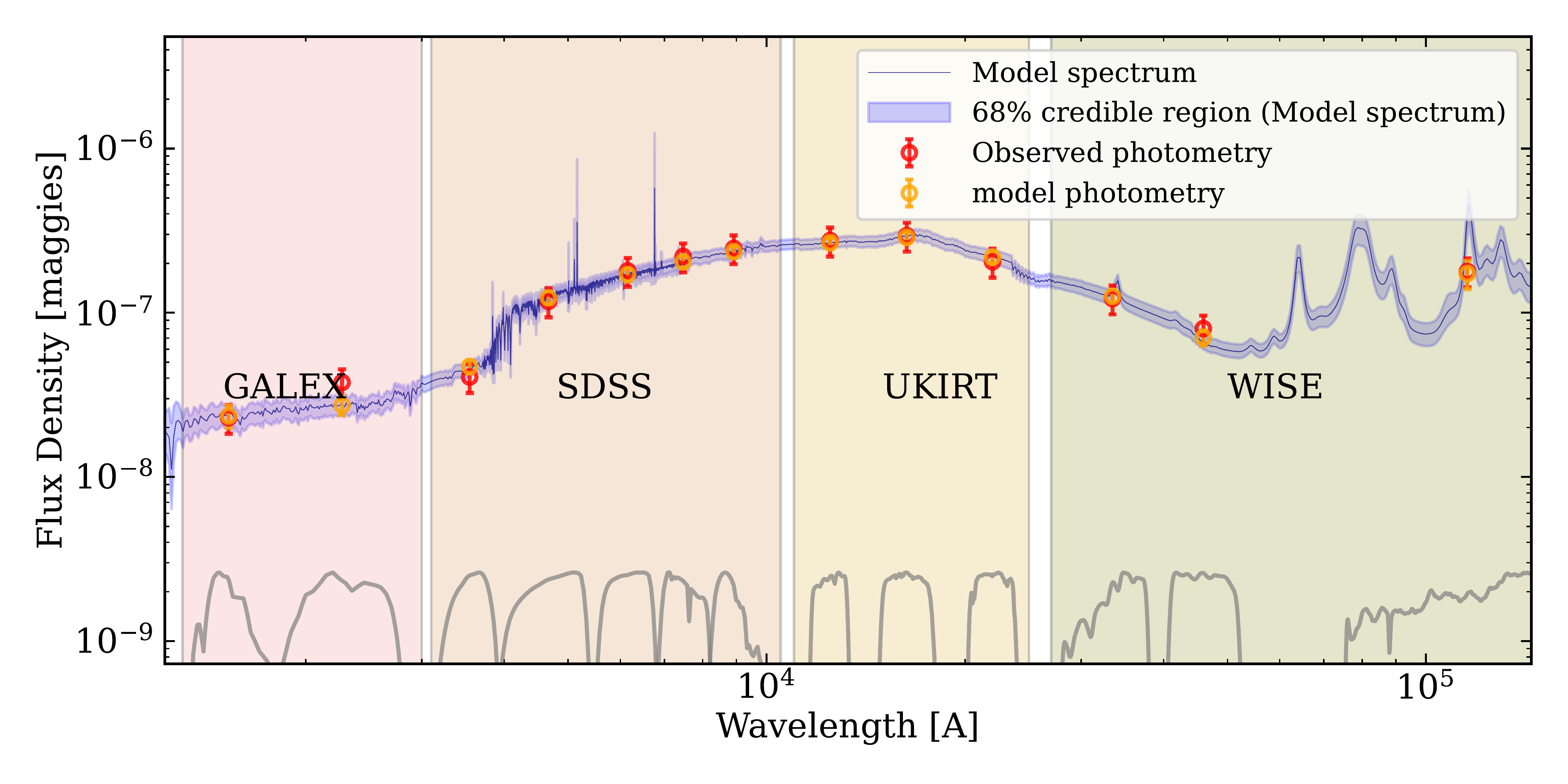}}\\

\caption{The best-fitted SED model of the host galaxies of FRB 20181220A (top) and FRB 20181223C (bottom). The flux density of the host galaxies in different wavelength bands are plotted along with the best-fit {\tt Prospector} model spectrum. To assess the quality of the {\tt Prospector} models, the modelled and actual photometry data are also shown. For more information, see Appendix \ref{app:prospector}.}
\label{fig:SEDS_1}
\end{figure*}

\begin{figure*}
\centering     %%% not \center

\subfigure{\label{fig10:a}\includegraphics[width=0.95\textwidth]{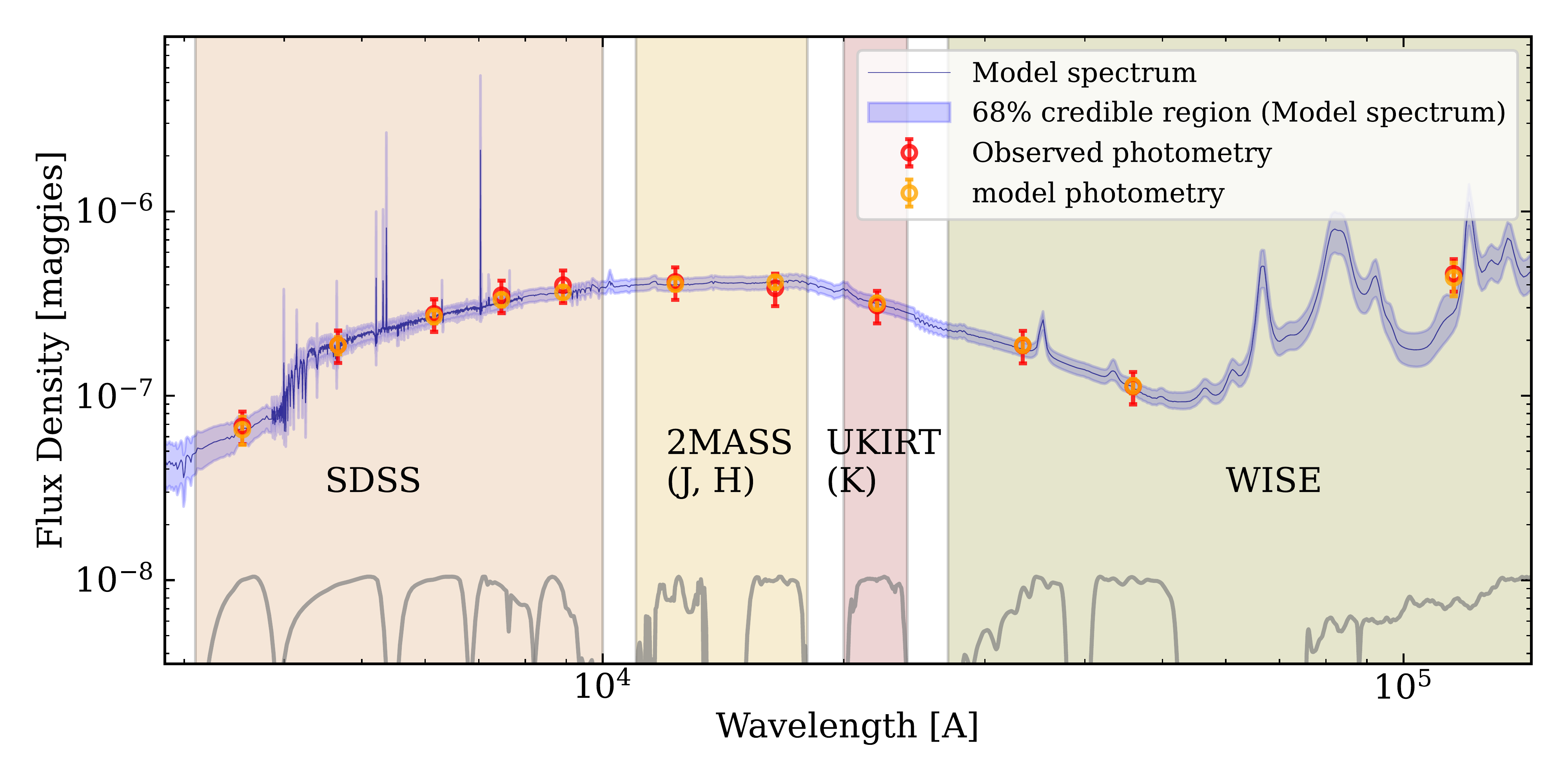}} \\

\subfigure{\label{fig10:d}\includegraphics[width=0.95\textwidth]{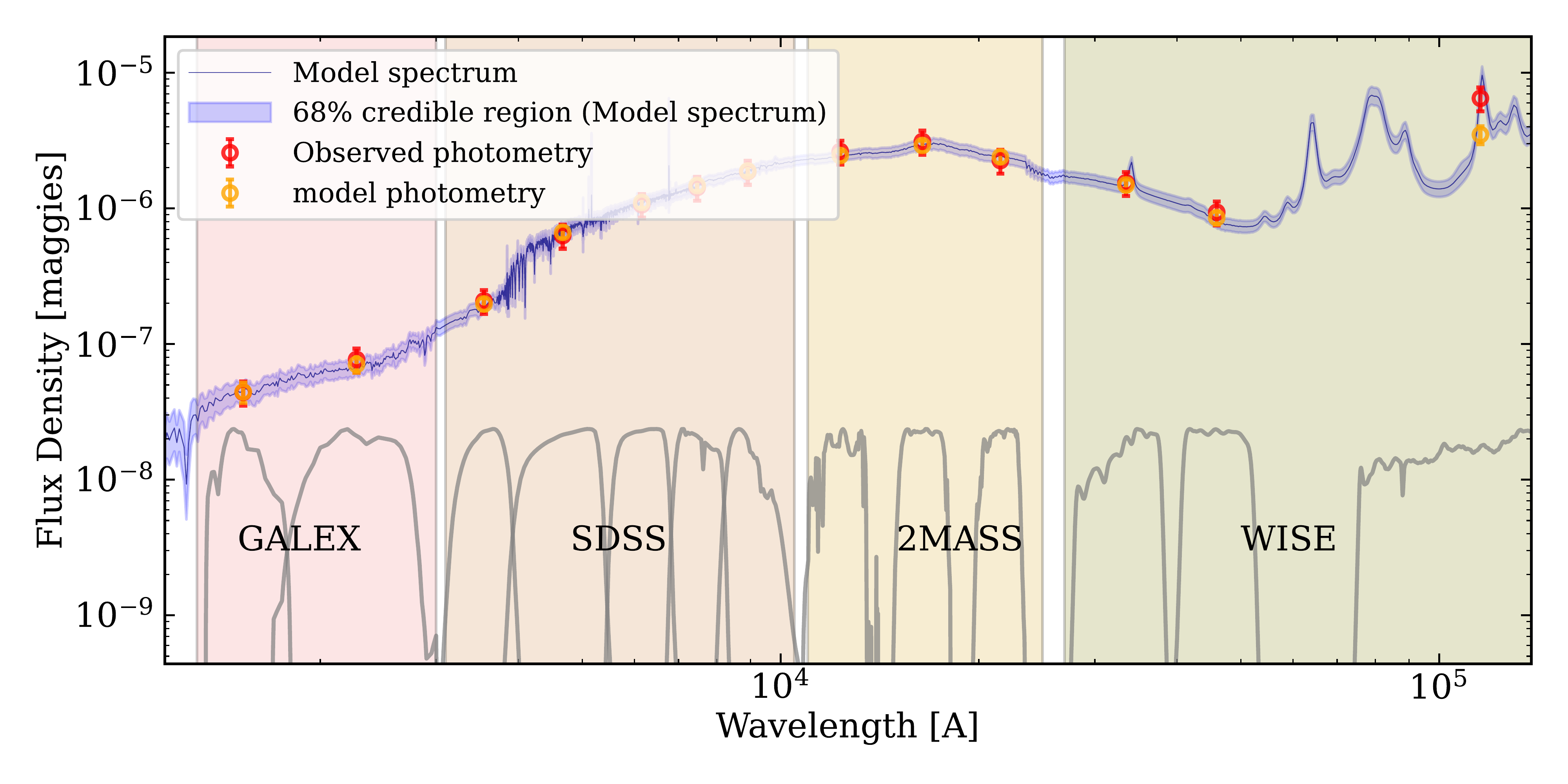}} \\
\caption{The best fitted SED model of the host galaxies of FRB 20190418A (top) and FRB 20190425A (bottom). Rest is same as Figure \ref{fig:SEDS_1}.}
\label{fig:SEDS_2}
\end{figure*}

\begin{table}[h]
\begin{center}
\hspace{-1.in}
%\resizebox{0.95\textwidth}{!}{ 
\caption{12 broadband filters used to model the SED of FRB 20181220A.}
\label{sed:maggies_FRB20181220A}
\begin{tabular}{@{} lccc @{}}%{ccccc} 
\toprule
\textbf{Instrument}$^c$ & \textbf{Filter} & \textbf{Effective Wavelength} & \textbf{Flux density}$^{a,b}$\\
& & \AA & maggies \\ \midrule %\textbf{z$_{\mathrm{phot}}$}& \textbf{$\sigma_{\mathrm{phot-err}}$}&
%Source &  Ra & Dec & gmag & rmag & zmag & photoz & photoz_err\\
SWIFT/UVOT$^{a}$ & w1 & 2600 & 2.80 $\times 10^{-8}$ \\
Pan-STARRS DR2$^{d}$ & g & 3546 & 5.95 $\times 10^{-7}$  \\
 & r & 4670 & 8.67 $\times 10^{-7}$ \\
  & i & 6156 & 1.08 $\times 10^{-6}$ \\
 & z & 7472 & 1.29 $\times 10^{-6}$ \\
 & y & 8917 & 1.60 $\times 10^{-6}$  \\
2MASS & J & 12319 & 2.28 $\times 10^{-6}$   \\
 & H & 16420 & 2.82 $\times 10^{-6}$   \\
 & Ks & 21567 & 2.38 $\times 10^{-6}$   \\
WISE & W1 & 33461 & 1.11 $\times 10^{-6}$   \\
 & W2 & 45952 & 6.84 $\times 10^{-7}$  \\
 & W3 & 115526 & 4.00 $\times 10^{-6}$   \\\bottomrule 
 \hline
%\footnotetext{Thus far}
\end{tabular}
%}
\end{center}

$^{a}$  Note that 1 maggie is defined as the flux density in Janskys divided by 3631. Fluxes at $ \lambda < 100000$ \AA~are corrected for Galactic extinction according to the prescription of \cite{schlafly2011ApJ}.\\
$^b$ All broadband fluxes are assigned a 20\% fractional uncertainty.\\
$^c$ For all instruments' flux densities, the aperture used for estimating respective magnitudes is larger than twice the effective size of the galaxies.

$^{d}$ The SWIFT/UVOT w1 flux density is extinction and inclination angle corrected taken from the Swift/UVOT Serendipitous Source Catalog \citep{2014Ap&SS.354...97Y}.
\end{table}

\begin{table}[h]
\begin{center}
\hspace{-1.in}
%\resizebox{0.95\textwidth}{!}{ 
\caption{13 broadband filters used to model the SED of the FRB 20181223C host.}
\label{sed:maggies_FRB20181223C}
\begin{tabular}{@{} *4l @{}}%{ccccc} 
\toprule
\textbf{Instrument}$^c$ & \textbf{Filter} & \textbf{Effective Wavelength} & \textbf{Flux density}$^{a,b}$\\
& & \AA & maggies \\ \midrule %\textbf{z$_{\mathrm{phot}}$}& \textbf{$\sigma_{\mathrm{phot-err}}$}&
%Source &  Ra & Dec & gmag & rmag & zmag & photoz & photoz_err\\

GALEX$^{a}$ & FUV & 1549 & 2.29 $\times 10^{-8}$ \\
 & NUV & 2304 & 3.77 $\times 10^{-8}$  \\
SDSS$^{b}$ & u & 3546 & 4.44 $\times 10^{-8}$  \\
 & g & 4670 & 1.30 $\times 10^{-7}$ \\
  & r & 6156 & 2.01 $\times 10^{-7}$ \\
 & i & 7472 & 2.48 $\times 10^{-7}$ \\
 & z & 8917 & 2.26 $\times 10^{-7}$  \\
UKIRT$^{c}$ & J & 12480 & 2.70 $\times 10^{-7}$   \\
 & H & 16310 & 2.92 $\times 10^{-7}$   \\
 & K & 22010 & 2.04 $\times 10^{-7}$   \\
WISE$^{b}$ & W1 & 33461 & 1.44 $\times 10^{-7}$   \\
 & W2 & 45952 & 8.44 $\times 10^{-8}$  \\
 & W3 & 115526 & 2.69 $\times 10^{-7}$   \\\bottomrule 
 \hline
%\footnotetext{Thus far}
\end{tabular}
%}
\end{center}

$^{a}$ From the SPRING catalogue \citep{2023A&A...671A.118C}.\\
$^b$ SDSS and WISE filters' flux densities are obtained from the aperture-matched photometry catalog of nearby galaxies by \cite{2015ApJS..219....8C}.\\
$^{c}$ From United Kingdom Infrared Telescope (UKIRT) Infrared Deep Sky Survey (UKIDSS) Data Release 9 \citep{2013yCat.2319....0L} where kron magnitudes are estimated using the same aperture size employed by \cite{2015ApJS..219....8C}.\\
\end{table}

\begin{table}[ht]
\begin{center}
\hspace{-1.in}
%\resizebox{0.95\textwidth}{!}{ 
\caption{11 broadband filters used to model the SED of FRB 20190418A.}
\label{sed:maggies_FRB20190418A}
\begin{tabular}{@{} lccc @{}}%{ccccc} 
\toprule
\textbf{Instrument} & \textbf{Filter} & \textbf{Effective Wavelength} & \textbf{Flux density}\\
& & \AA & maggies \\ \midrule %\textbf{z$_{\mathrm{phot}}$}& \textbf{$\sigma_{\mathrm{phot-err}}$}&
%Source &  Ra & Dec & gmag & rmag & zmag & photoz & photoz_err\\

SDSS & u & 3546 & 6.83 $\times 10^{-8}$  \\
 & g & 4670 & 1.88 $\times 10^{-7}$ \\
  & r & 6156 & 2.78 $\times 10^{-7}$ \\
 & i & 7472 & 3.51 $\times 10^{-7}$ \\
 & z & 8917 & 3.98 $\times 10^{-7}$  \\
2MASS & J & 12319 & 4.14 $\times 10^{-7}$   \\
 & H & 16420 & 3.83 $\times 10^{-7}$   \\
UKIRT & WFCAM K & 22010 & 3.03 $\times 10^{-7}$   \\
WISE & W1 & 33461 & 1.87 $\times 10^{-7}$   \\
 & W2 & 45952 & 1.12 $\times 10^{-7}$  \\
 & W3 & 115526 & 4.59 $\times 10^{-7}$   \\\bottomrule 
 \hline
%\footnotetext{Thus far}
\end{tabular}
%}
\end{center}
\end{table}

\begin{table}[h]
\begin{center}
\hspace{-1.in}
%\resizebox{0.95\textwidth}{!}{ 
\caption{14 broadband filters used to model the SED of FRB 20190425A.}
\label{sed:maggies_FRB20190425A}
\begin{tabular}{@{} lccc @{}}%{ccccc} 
\toprule
\textbf{Instrument} & \textbf{Filter} & \textbf{Effective Wavelength} & \textbf{Flux density}$^{a,b}$\\
& & \AA & maggies \\ \midrule %\textbf{z$_{\mathrm{phot}}$}& \textbf{$\sigma_{\mathrm{phot-err}}$}&
%Source &  Ra & Dec & gmag & rmag & zmag & photoz & photoz_err\\

GALEX$^{a}$ & FUV & 1549 & 4.41 $\times 10^{-8}$ \\
 & NUV & 2304 & 7.70 $\times 10^{-8}$  \\
SDSS$^b$ & u & 3546 & 2.08 $\times 10^{-7}$  \\
 & g & 4670 & 6.31 $\times 10^{-7}$ \\
  & r & 6156 & 1.07 $\times 10^{-6}$ \\
 & i & 7472 & 1.43 $\times 10^{-6}$ \\
 & z & 8917 & 1.87 $\times 10^{-6}$  \\
2MASS$^{c}$ & J & 12319 & 2.46 $\times 10^{-6}$   \\
 & H & 16420 & 2.71 $\times 10^{-6}$   \\
 & Ks & 21567 & 1.92 $\times 10^{-6}$   \\
WISE$^{b}$ & W1 & 33461 & 1.55 $\times 10^{-6}$   \\
 & W2 & 45952 & 9.37 $\times 10^{-7}$  \\
 & W3 & 115526 & 6.49 $\times 10^{-6}$   \\
 & W4 & 220783 & 4.65 $\times 10^{-6}$\\\bottomrule 
 \hline
%\footnotetext{Thus far}
\end{tabular}
%}
\end{center}
$^a$ GALEX calibrated and extinction corrected fluxes are taken from \cite{2017ApJS..230...24B}.\\
$^b$ SDSS and WISE filters' flux densities are obtained from the aperture-matched photometry catalog of nearby galaxies by \cite{2015ApJS..219....8C}.\\
$^{c}$ For 2MASS fliter magnitudes, we used Kron magnitudes from \cite{2006AJ....131.1163S} with the same aperture size used by \cite{2015ApJS..219....8C}, thus ensuring accurate and comparable measurements.
\end{table}

Tables \ref{sed:maggies_FRB20181220A}, \ref{sed:maggies_FRB20181223C}, \ref{sed:maggies_FRB20190418A}, and \ref{sed:maggies_FRB20190425A} list the different broadband filter fluxes used in our SED fitting for the host galaxies of FRBs 20181220A, 20181223C, 20190418A, and 20190425A, respectively. Note that all flux densities are estimated after correcting for Milky Way extinction. All derived host galaxy properties from this analysis are provided in Table \ref{tab: A-galaxy-properties}. The quoted uncertainties in all cases are 1$\sigma$ values.

% ----------------- Appendix B -------------------

\section{Spectroscopic observations of FRB 20181223C}
\label{sec:mos_frb20181223c}

\label{sec:GTC_obs}
For our GTC observation, we only consider sources with r-band magnitude $<$ 22 AB as at the maximum estimate redshift of the FRB 20181223C host = 0.089 (see Table \ref{tab:FRB-params}), an FRB 121102-like host would have an apparent r-band magnitude $\approx$ 21.1 AB mag. Thus, we obtained the spectroscopic redshift of five host galaxy candidates in the FRB 20181223C 2-$\sigma$ localization region (Sources 1, 3, 5, 6 and 9 in Fig. \ref{fig:FOV}) using observations performed with the Optical System for Imaging and low-Intermediate-Resolution Integrated Spectroscopy (OSIRIS) instrument at the GTC on 2021 February 18.\footnote{Program GTCMULTIPLE3B-20BMEX.}
To optimize the observations, we separated the targets in two observing blocks (OBs) to use both the long-slit and the multi-object spectroscopy (MOS) capabilities of OSIRIS. 

For the MOS OB, we designed a mask with the OSIRIS Mask Designer Tool \citep{mask1,mask2}, using four fiducial stars and catalog coordinates of the host galaxy candidates. The mask contained rectangular slitlets with lengths between 10\arcsec\ and 20\arcsec. 

The observations were performed during dark/gray time under clear conditions, with seeing of $\sim$1\arcsec\ and airmass between 1.05 and 1.3. 
Three 1200-s and two 1000-s exposures were taken in the respective MOS and long-slit configurations.  
We used the R500B grism covering the spectral range 3600$-$7200 {\AA}. The slit width was 1.5\arcsec\ in both observing modes. The spectral resolution was $\sim$21 \AA. 

The spectra from both OBs were reduced using the GTCMOS pipeline \citep{gtcmos} which uses IRAF routines \citep{tody1986iraf,tody1993iraf}. To calibrate the flux we used the spectrophotometric standard G191-B2B \citep{std2,standard,massey1988} observed during the same night as the target galaxies. For the wavelength calibration we used arc-lamp spectra of Ne, Hg and Ar. 
The rms errors of the resulting solutions were $<$2 Å. 

We extracted the calibrated spectra from the resulting 2D product, identified lines for each of the five observed galaxies, and estimated their redshifts. The results were verified by comparing the extracted spectra with the galaxy templates from the Manual and Automatic Redshifting
Software \citep[MARZ;][]{Marz}. 

For Source 2, we took spectroscopic observation using the IMACS wide-field imaging spectrograph on the 6.5-m Magellan Baade Telescope at Las Campanas Observatory (LCO). The observations were taken on 2022 February 21, with an observing air-mass from 1.8$-$2.0 and seeing at approximately 0\arcsec.8. We took five 300-second exposures using a 0.9\arcsec long slit-mask and a 150 lines/mm Grism. Using {\tt IRAF}, the frames were corrected using bias and flats frames.
%which are takenon the same night. 
Arc lamp exposures from He, Ne, and Ar were used in order to obtain a wavelength calibration spectrum, which was then used to reshape the corrected science frames. Finally, the spectroscopic redshift of Source 2  was estimated to be $z_{\rm spec}$ = 0.23932 $\pm$ 0.00002 using H$\alpha$, H$\beta$, [OIII], and [SII] emission lines.

The corresponding redshifts are presented in Table \ref{tab:r2galaxies}. 

Finally, in Table \ref{tab:r2galaxies}, we have listed the photometric redshifts of the remaining three galaxies (Sources 4, 7, and 8) from the photometric redshift catalog of DESI extragalactic sources \citep{Zou2019ApJS}. Note that all three galaxies have a 3$\sigma$ lower limit on the redshift $> 0.089$. Hence, they are very unlikely to be associated with the FRB. 

\begin{deluxetable}{lllll}[h]
\tablecaption{
 Galaxies from the DESI Legacy catalog in the 2$\sigma$ localization region of FRB 20181223C. Redshifts have been measured using the GTC telescope. Galaxy SDSS J120340.98+273251.4, a.k.a.\ Source 1, is the most probable host in the field given its low redshift. The spectroscopic redshift of Source 2 is estimated using Magellan telescope.
 \label{tab:r2galaxies}
}
\tablehead{
 \colhead{\textbf{\#}} & \colhead{\textbf{RA (J2000)}} & \colhead{\textbf{Dec (J2000)}} & \colhead{\textbf{r mag}}\tablenotemark{a} & \colhead{\textbf{redshift}}
}
\startdata
1 & 12$^h$03$^m$40.98$^s$ & 27\degr39\arcmin51.4\arcsec & 16.88 & 0.03024(1)  \\
2$^{c}$ & 12$^h$03$^m$40.04$^s$ & 27\degr32\arcmin44.5\arcsec & 19.90 & 0.23932(2) \\
3 & 12$^h$03$^m$44.53$^s$ & 27\degr32\arcmin55.9\arcsec & 20.20 & 0.274(1) \\
4 & 12$^h$03$^m$46.14$^s$ & 27\degr33\arcmin33.9\arcsec & 22.9 & 0.86(7)$^{b}$  \\
5 & 12$^h$03$^m$46.72$^s$ & 27\degr32\arcmin46.9\arcsec & 20.79 & 0.371(1) \\
6 & 12$^h$03$^m$42.00$^s$ & 27\degr33\arcmin19.8\arcsec & 21.39 & 0.514(1) \\
7 & 12$^h$03$^m$41.82$^s$ & 27\degr33\arcmin47.7\arcsec & 22.2 & 0.54(4)$^{b}$\\ %0.237(1) \\
8 & 12$^h$03$^m$41.97$^s$ & 27\degr33\arcmin37.7\arcsec & 22.3 & 0.45(9)$^{b}$\\ %0.455(1)\\
9 & 12$^h$03$^m$43.36$^s$ & 27\degr32\arcmin12.2\arcsec & 19.06 & 0.347(1)\\
\enddata
\tablenotetext{a}{SDSS model magnitude in r-band (AB scale).}
\tablenotetext{b}{Photometric redshift from the DESI DR9 catalog.}
\tablenotetext{c}{the spectroscopic redshift of Source 2 is estimated using 1D spectra obtained using 6.5-m Magellan Baade Telescope at Las Campanas Observatory (LCO).}
\end{deluxetable}

%--------------- Appendix C ---------------------

\section{Local Universe FRB Host sample}
\label{app:hostsample}

\begin{figure}[h]
\centering
\includegraphics[width=.32\textwidth,height=3.5cm]{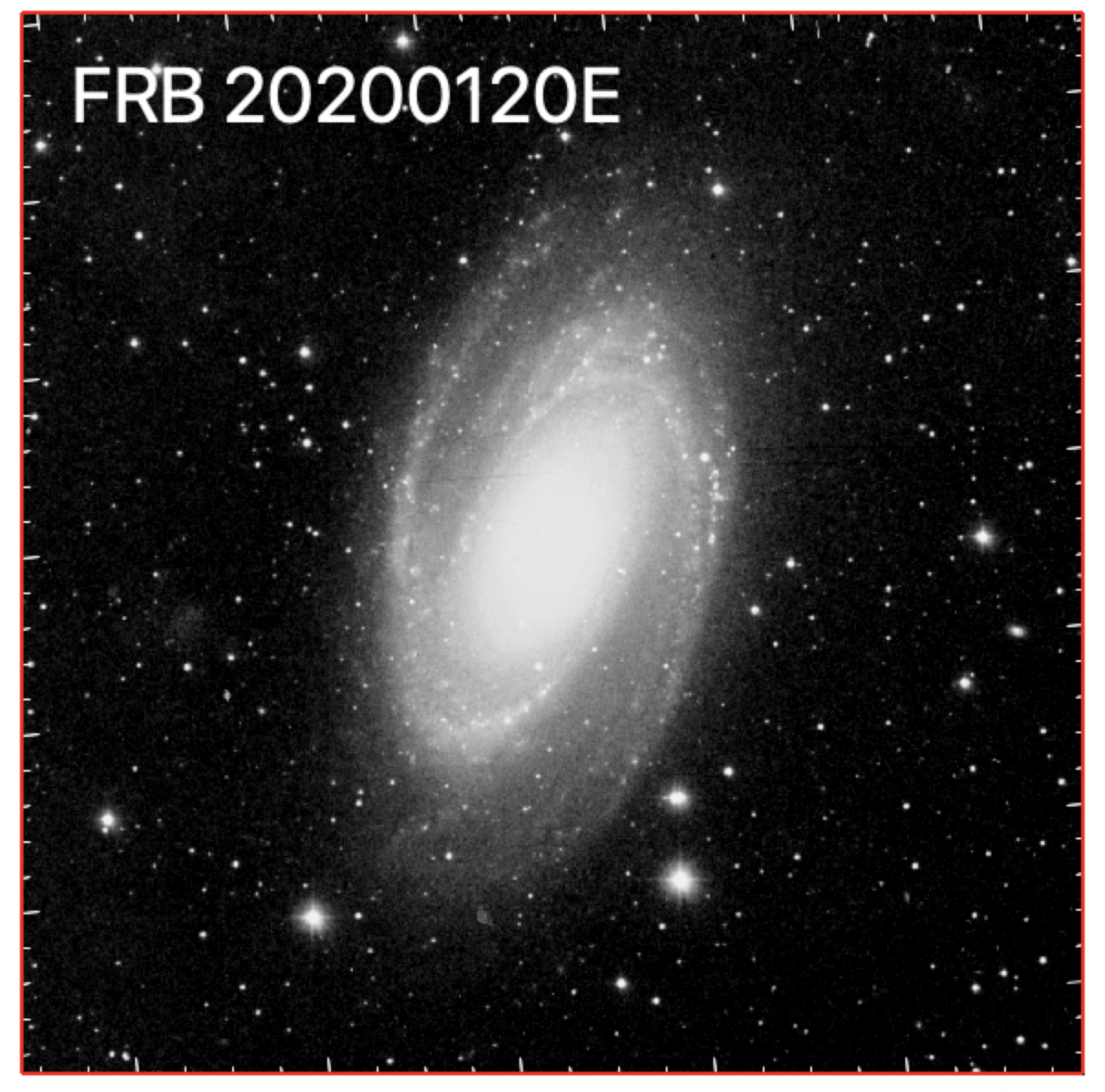}\hfill%
\includegraphics[width=.32\textwidth,height=3.5cm]{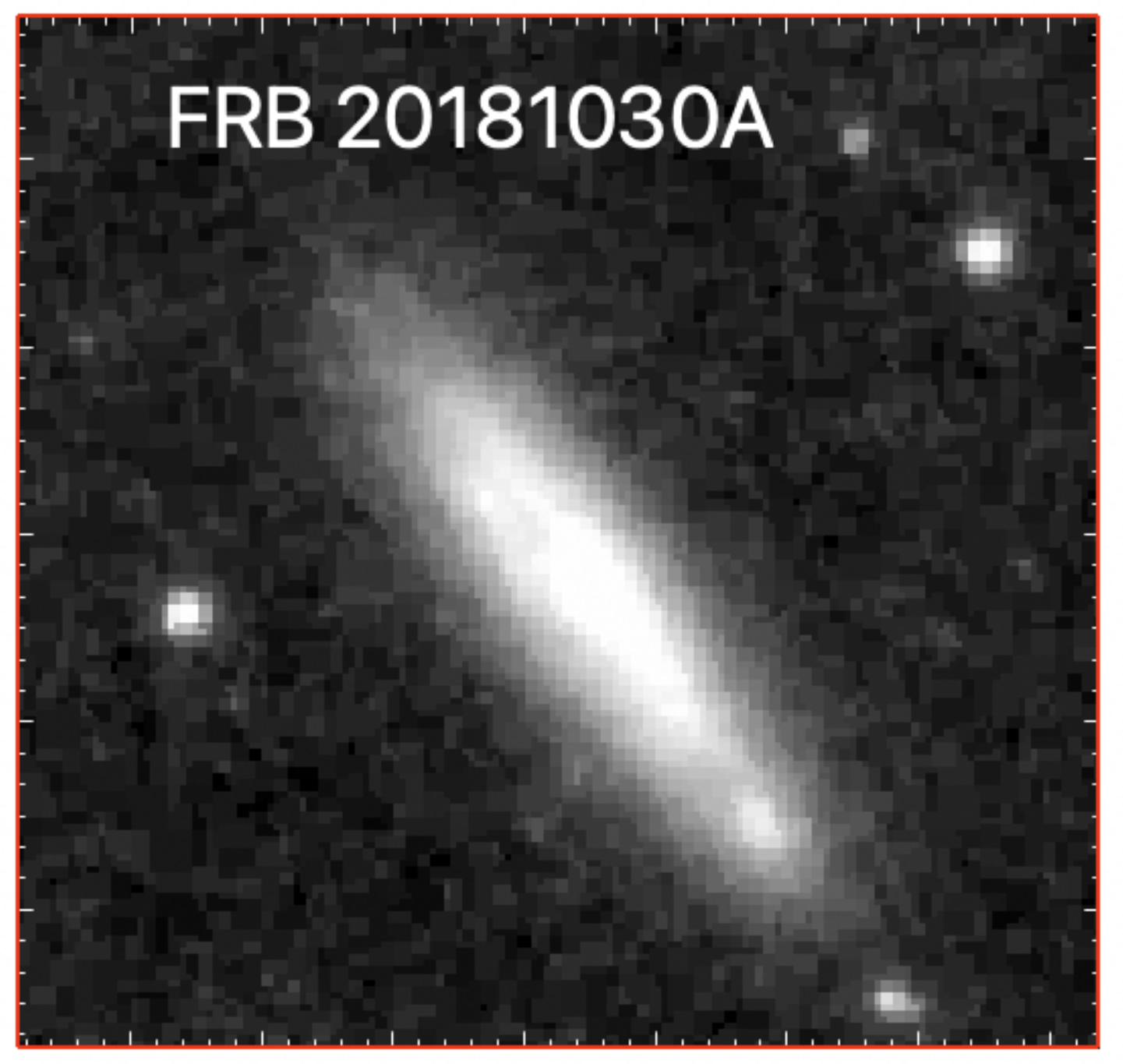}\hfill%
\includegraphics[width=.32\textwidth,height=3.5cm]{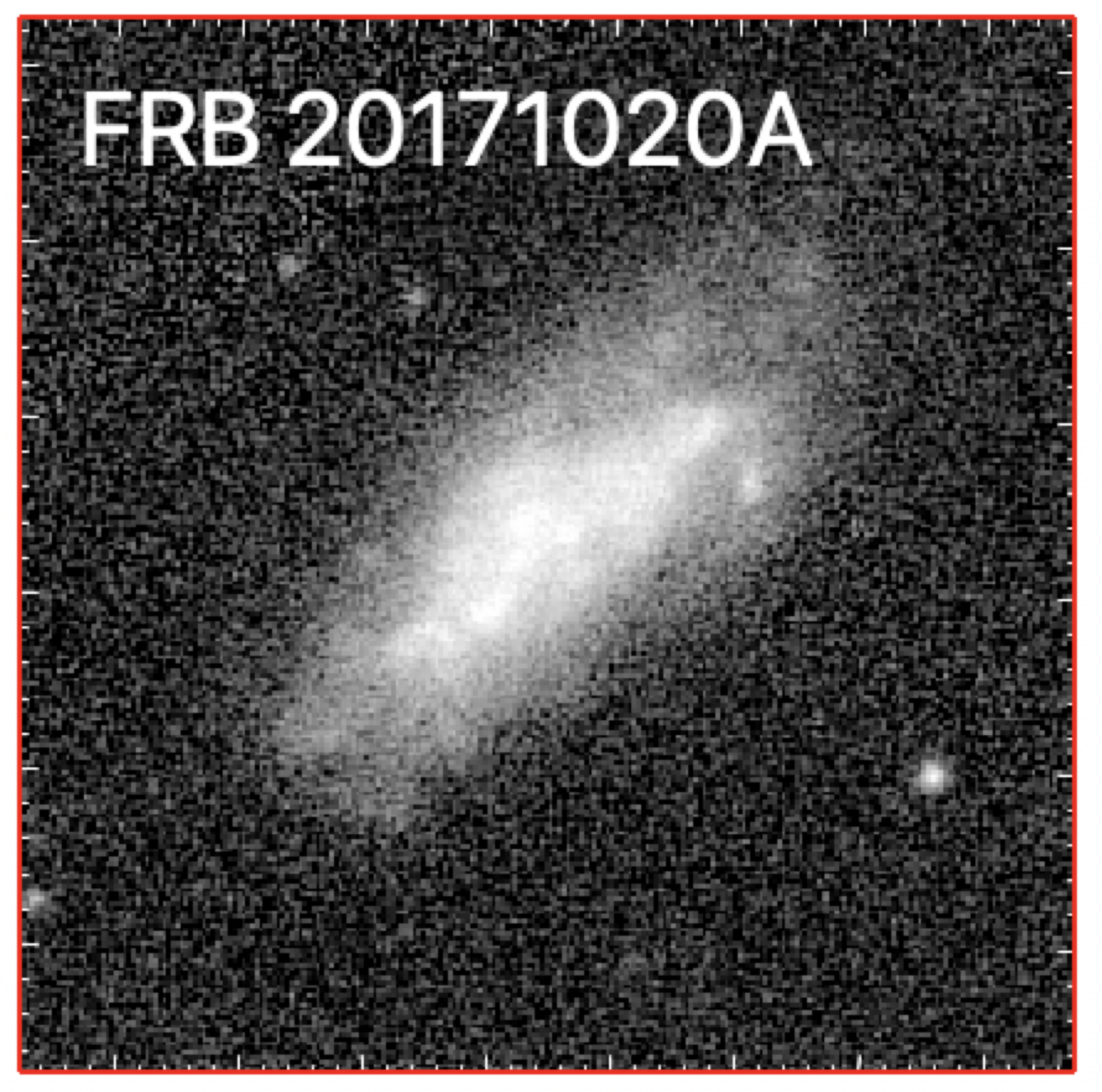}\\
\includegraphics[width=.32\textwidth,height=3.5cm]{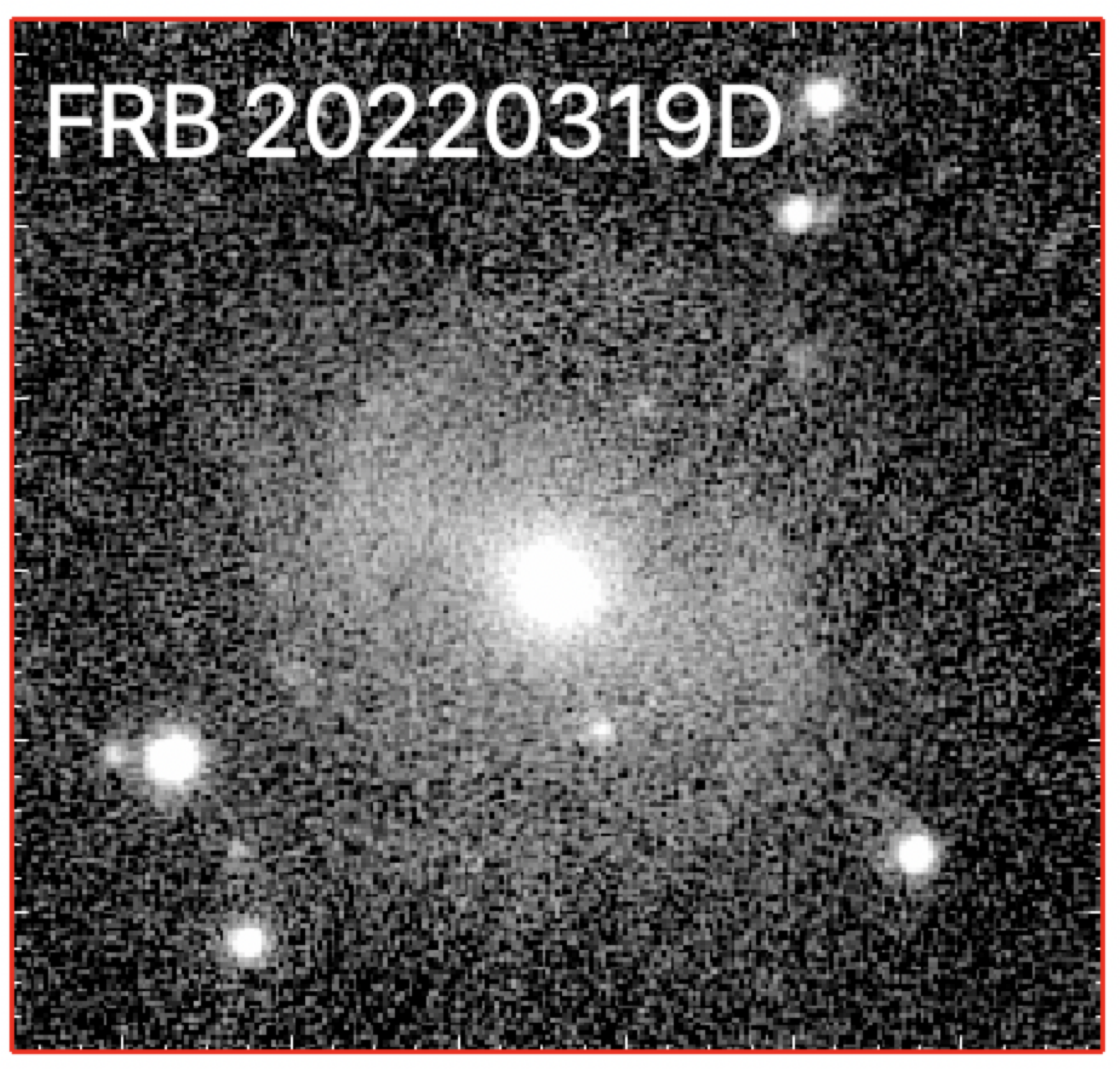}\hfill%
\includegraphics[width=.32\textwidth,height=3.5cm]{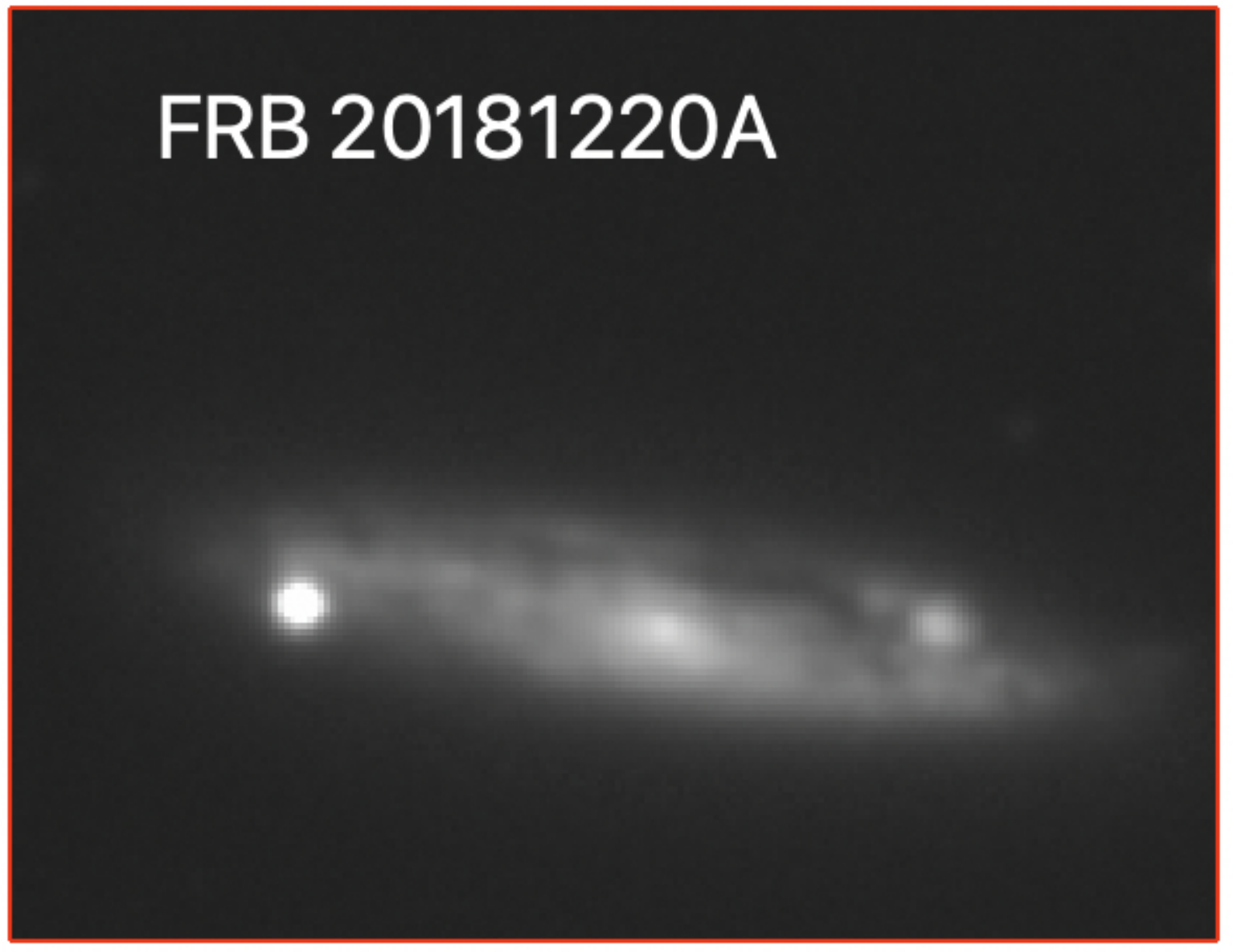}\hfill%
\includegraphics[width=.32\textwidth,height=3.5cm]{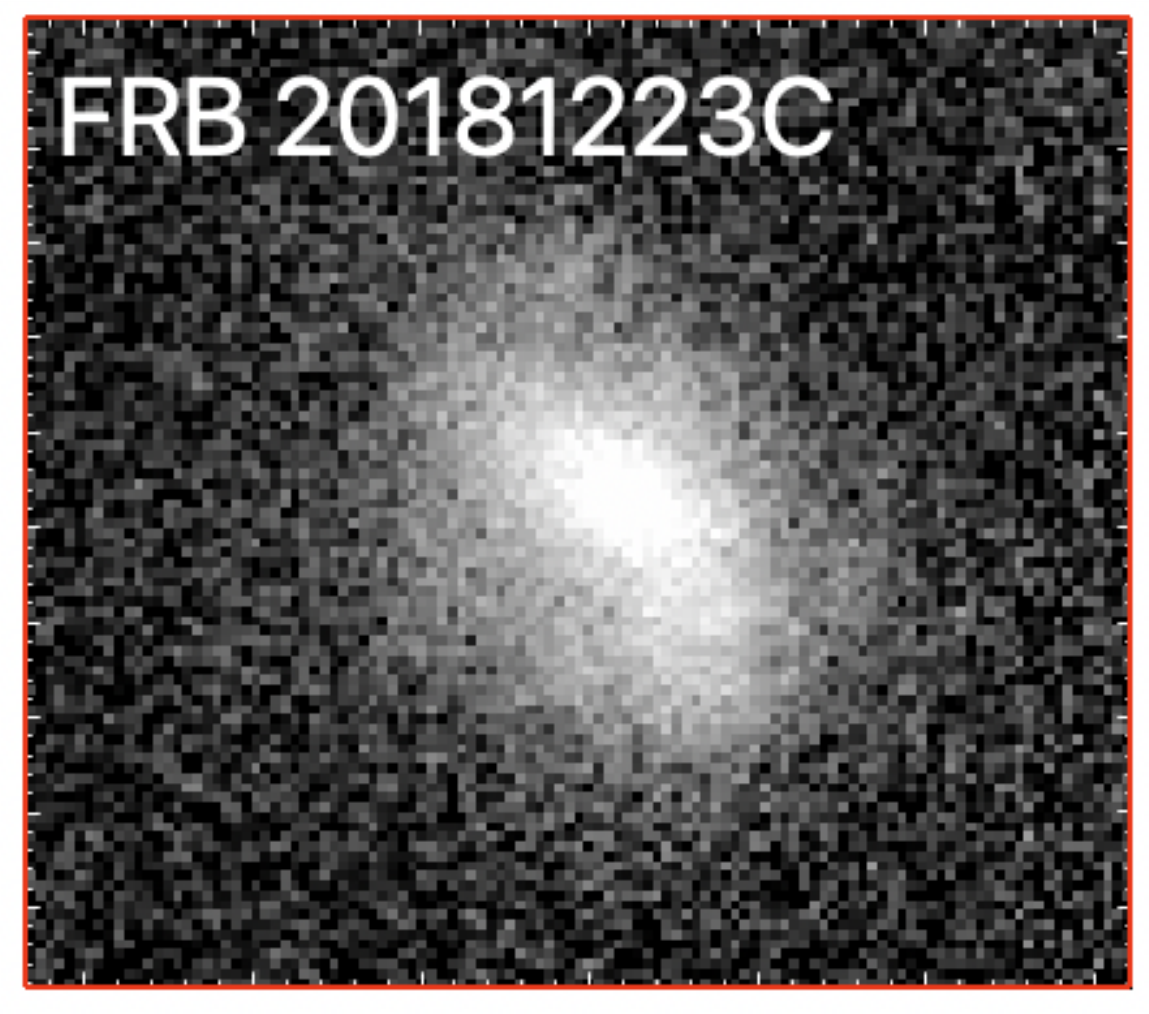}\\
\includegraphics[width=.32\textwidth,height=3.5cm]{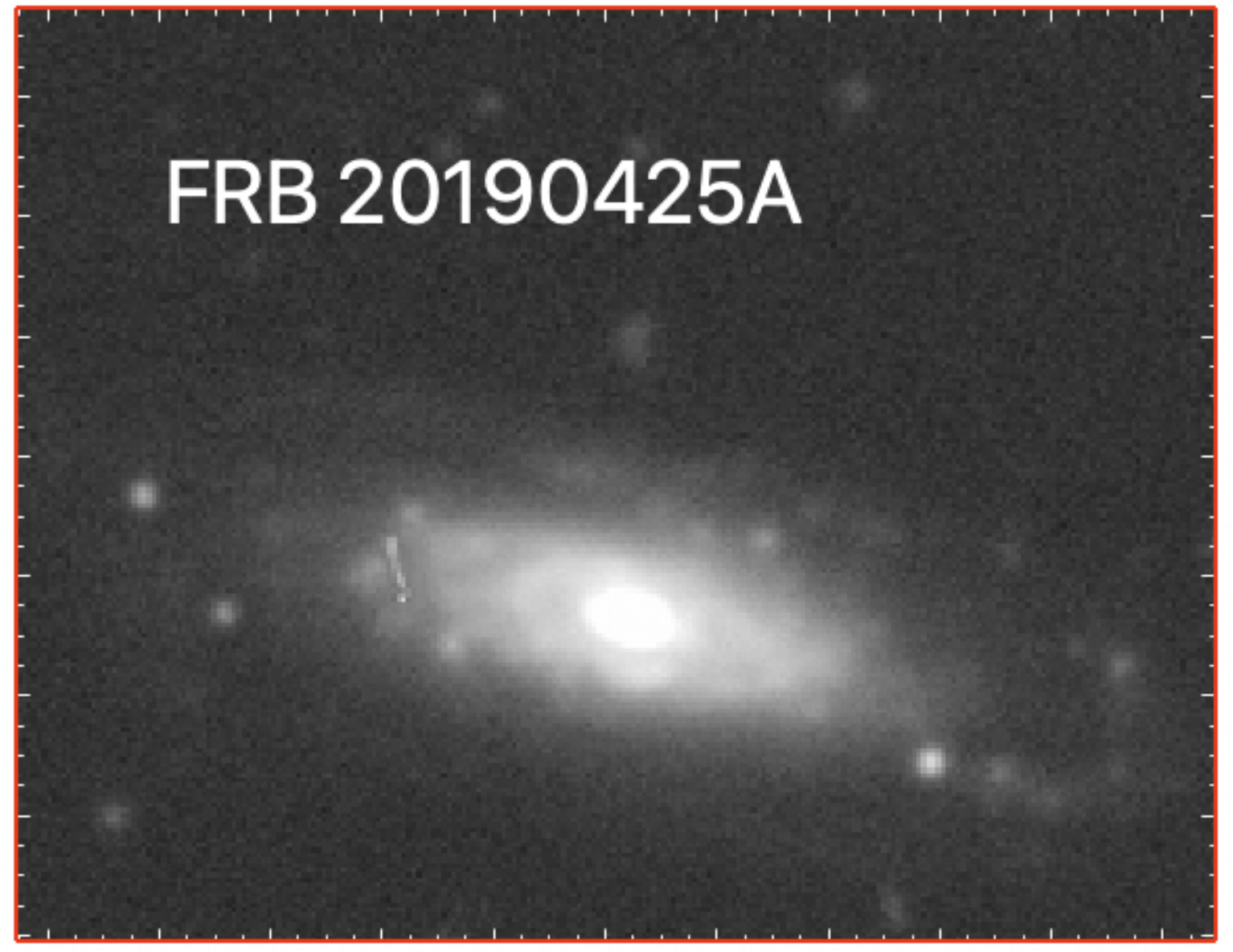}\hfill%
\includegraphics[width=.32\textwidth,height=3.5cm]{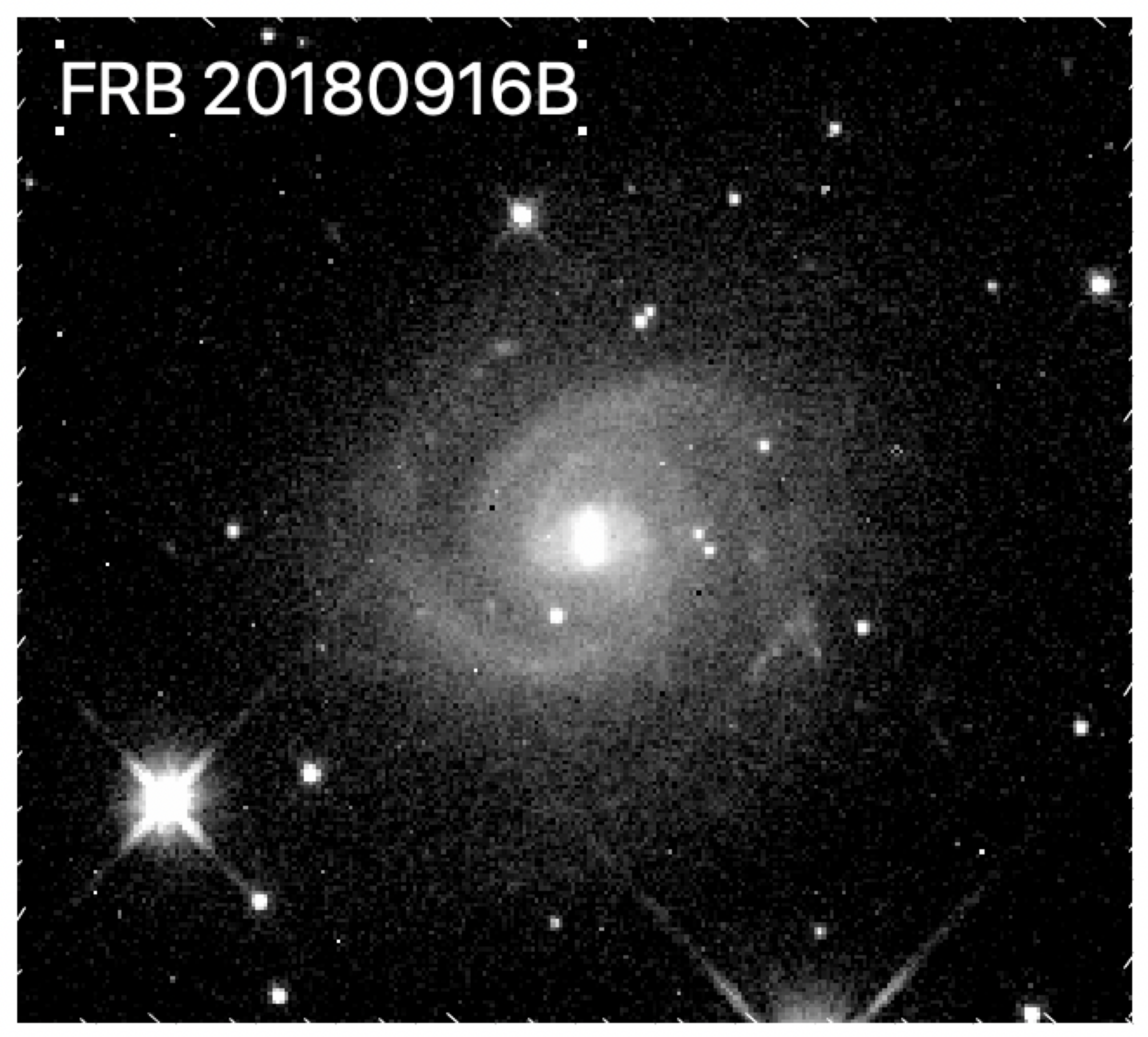}\hfill%
\includegraphics[width=.32\textwidth,height=3.5cm]{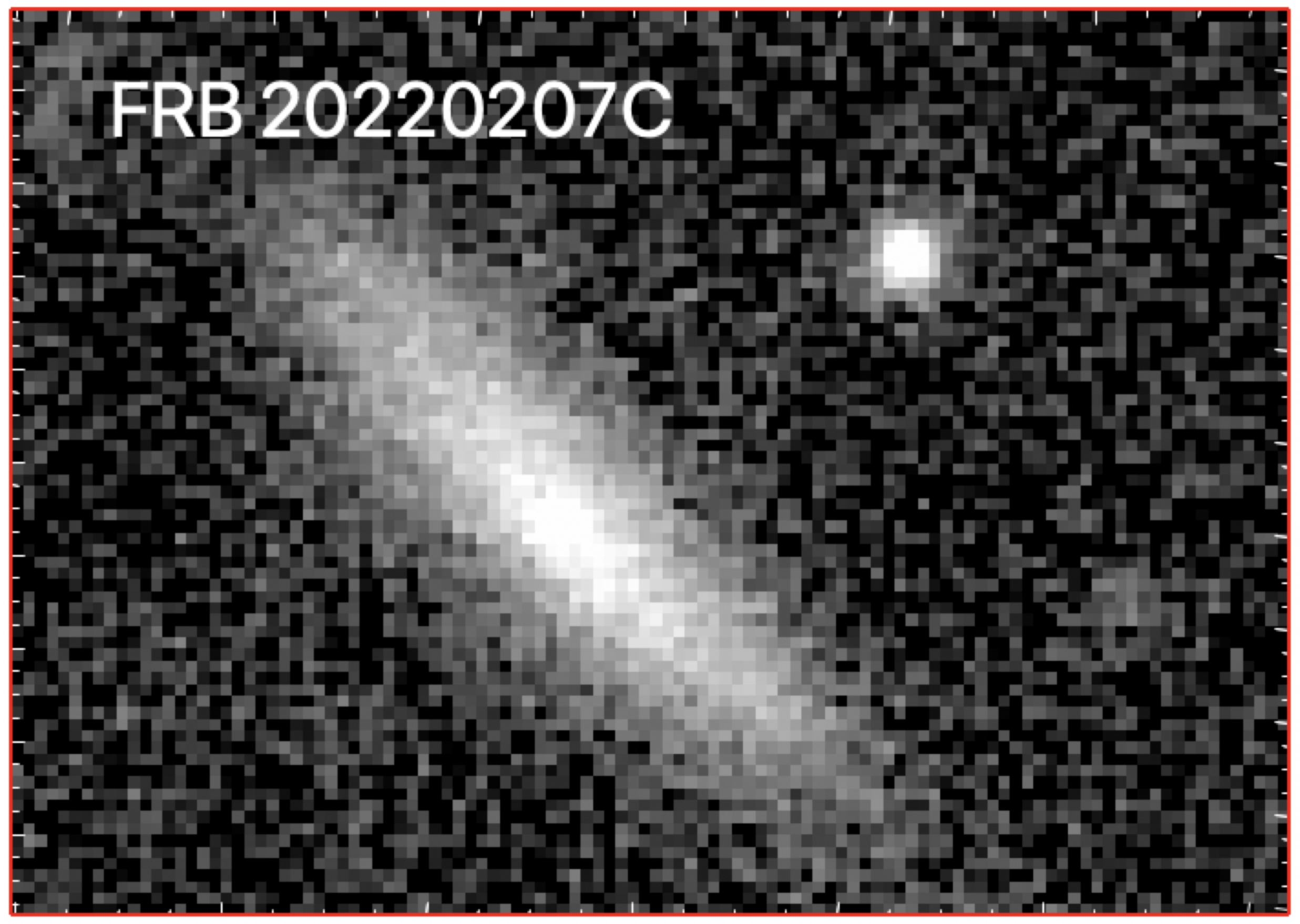}\\
\includegraphics[width=.32\textwidth,height=3.5cm]{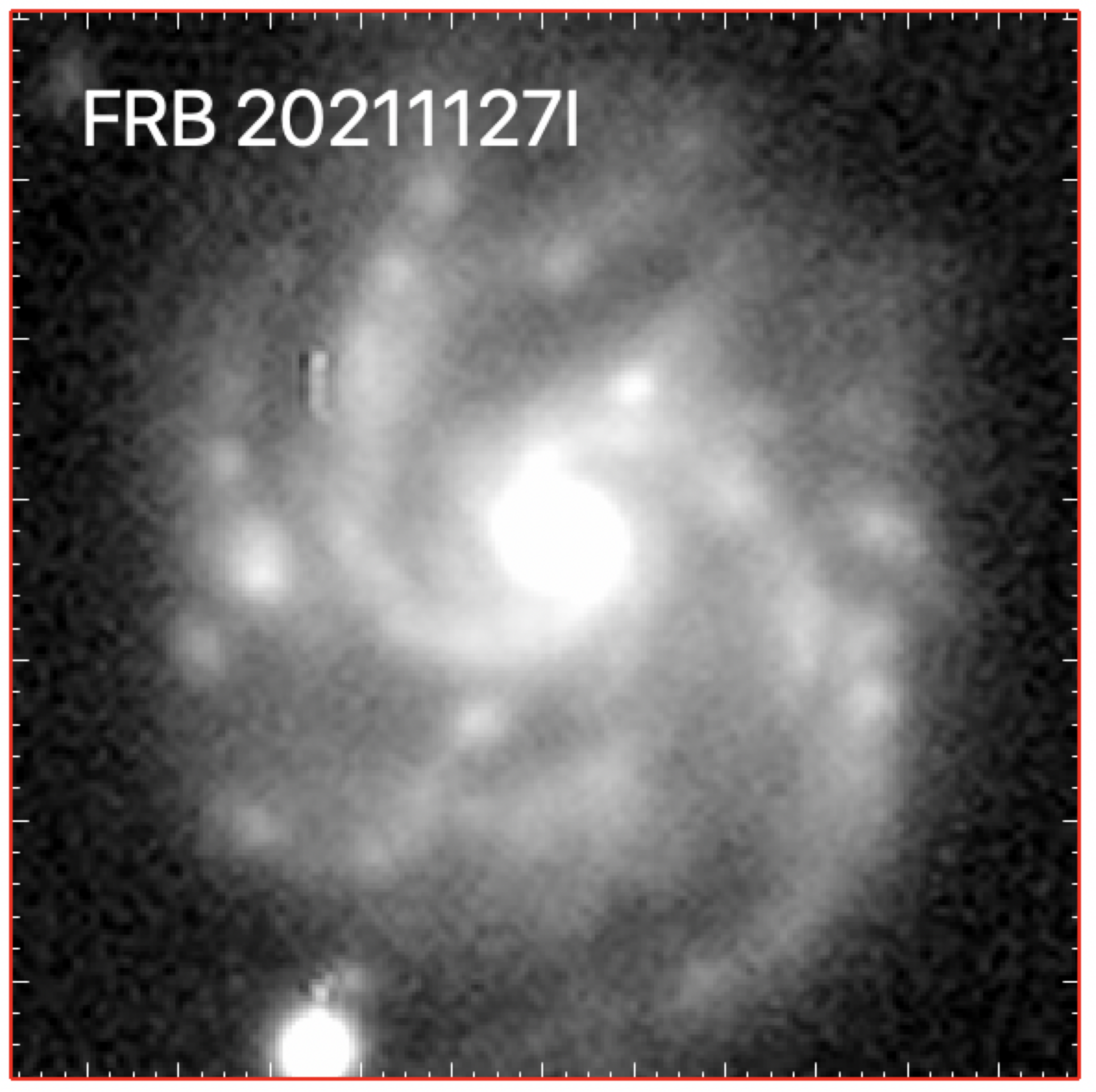}\hfill%
\includegraphics[width=.32\textwidth,height=3.5cm]{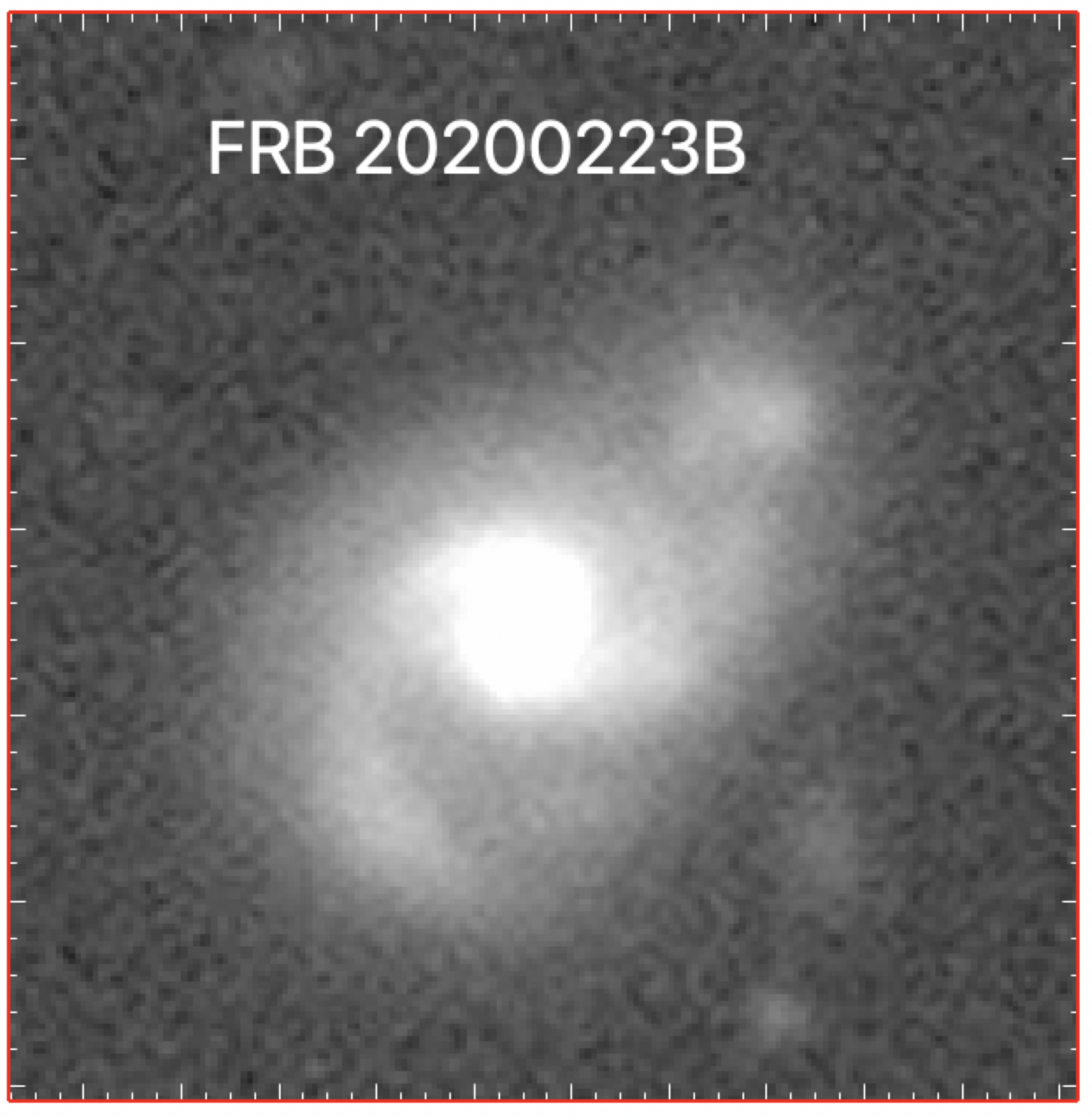}\hfill%
\includegraphics[width=.32\textwidth,height=3.5cm]{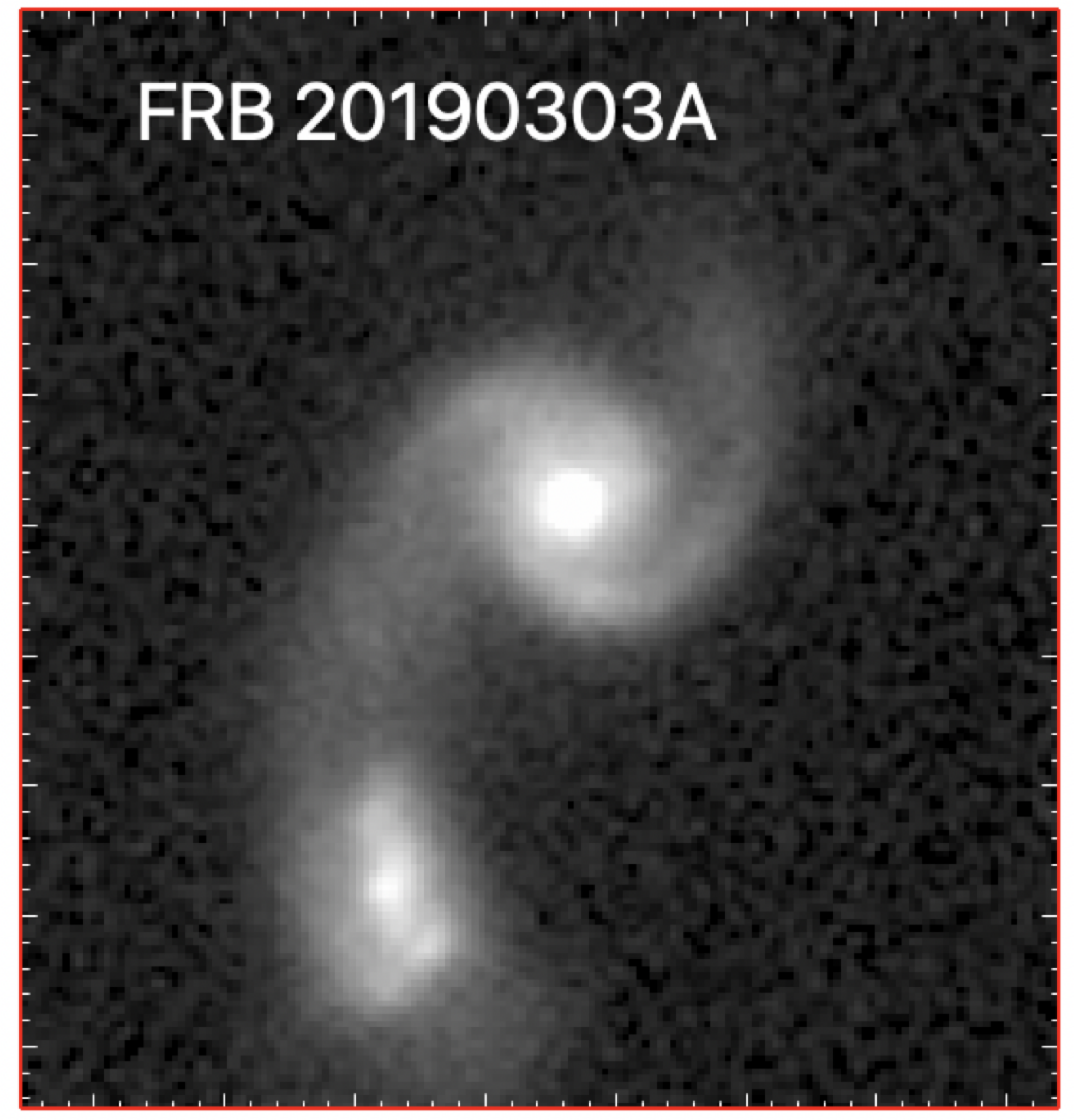}
\includegraphics[width=.32\textwidth,height=3.5cm]{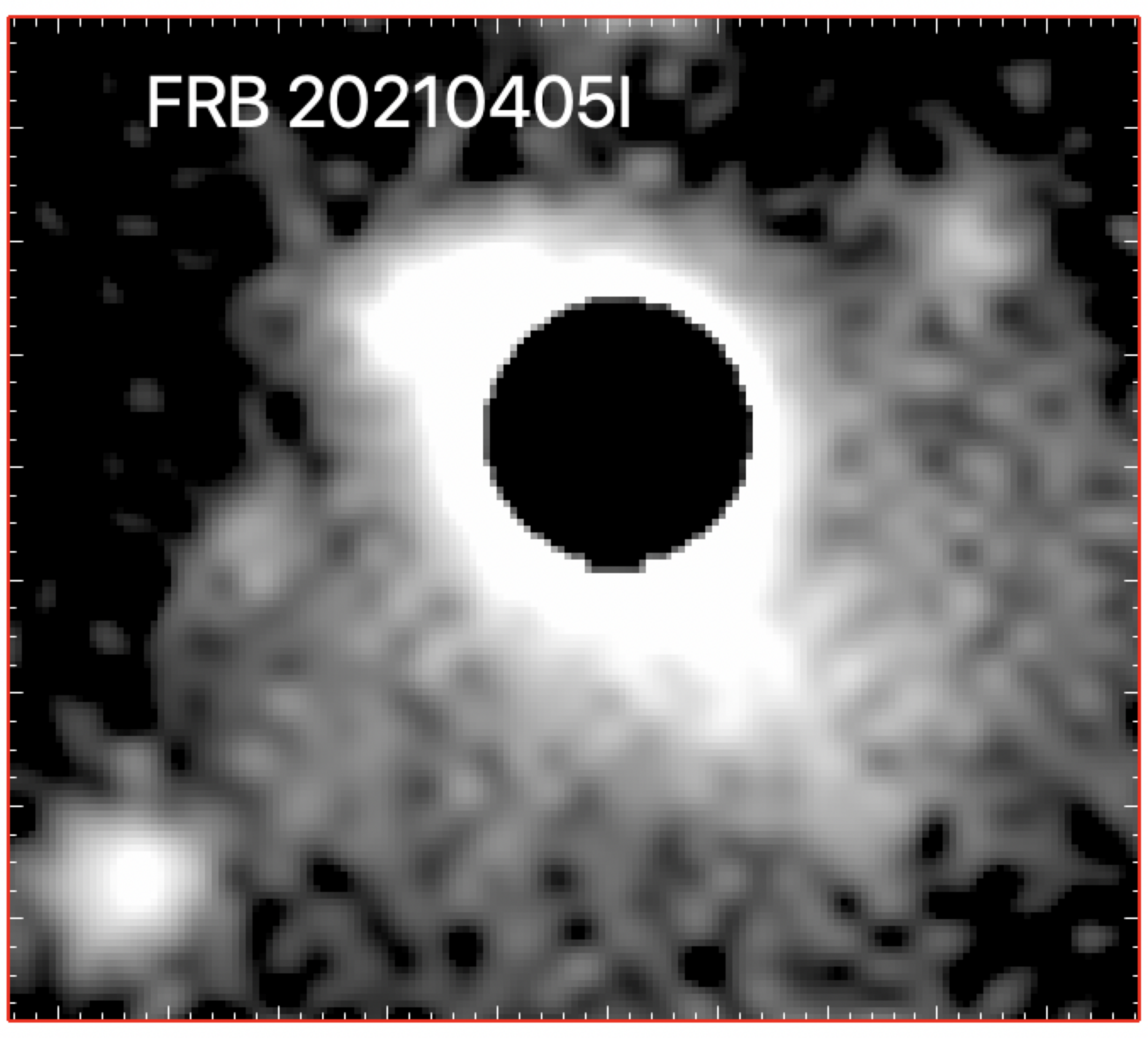}\hfill%
\includegraphics[width=.32\textwidth,height=3.5cm]{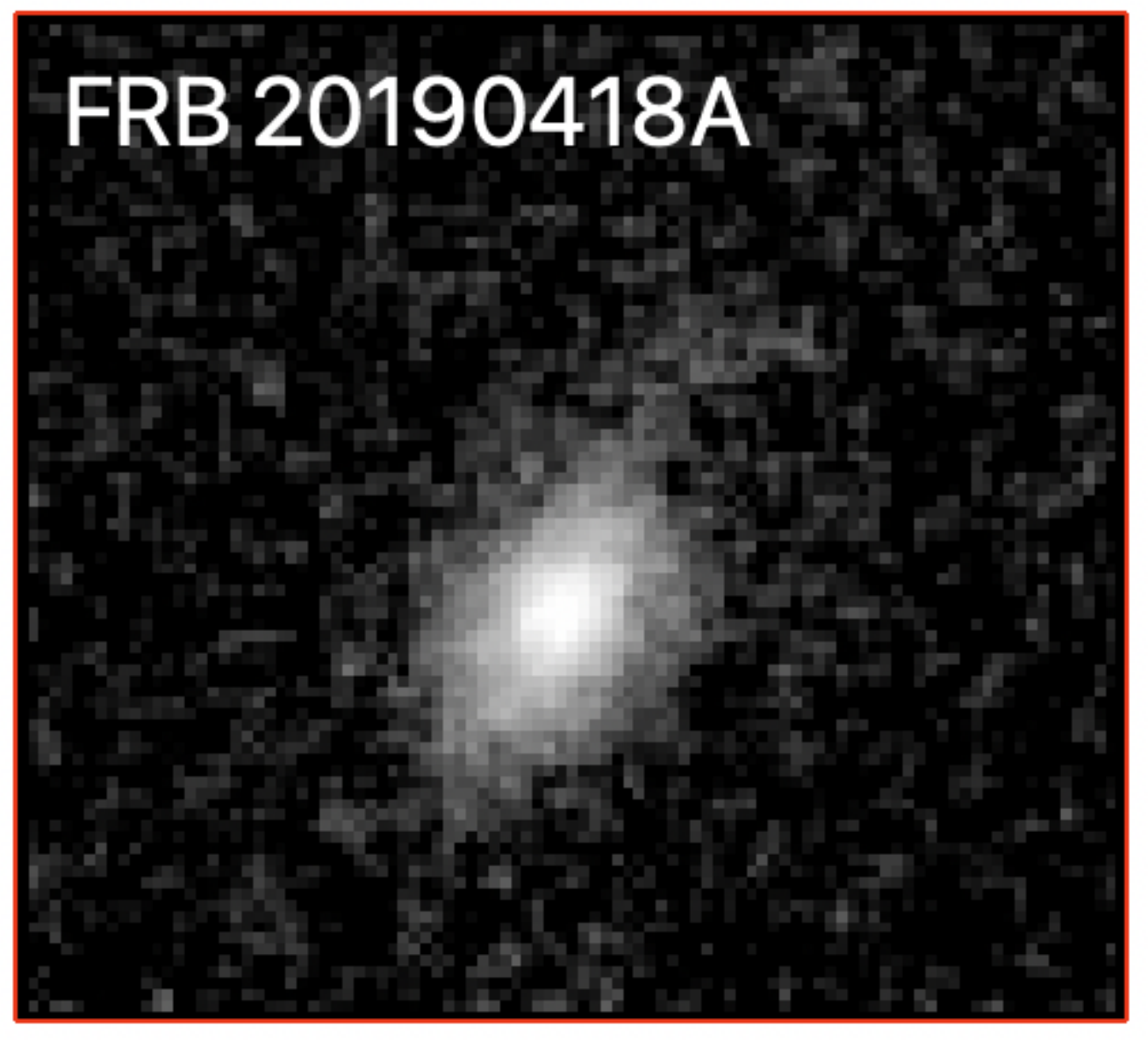}\hfill%
\includegraphics[width=.32\textwidth,height=3.5cm]{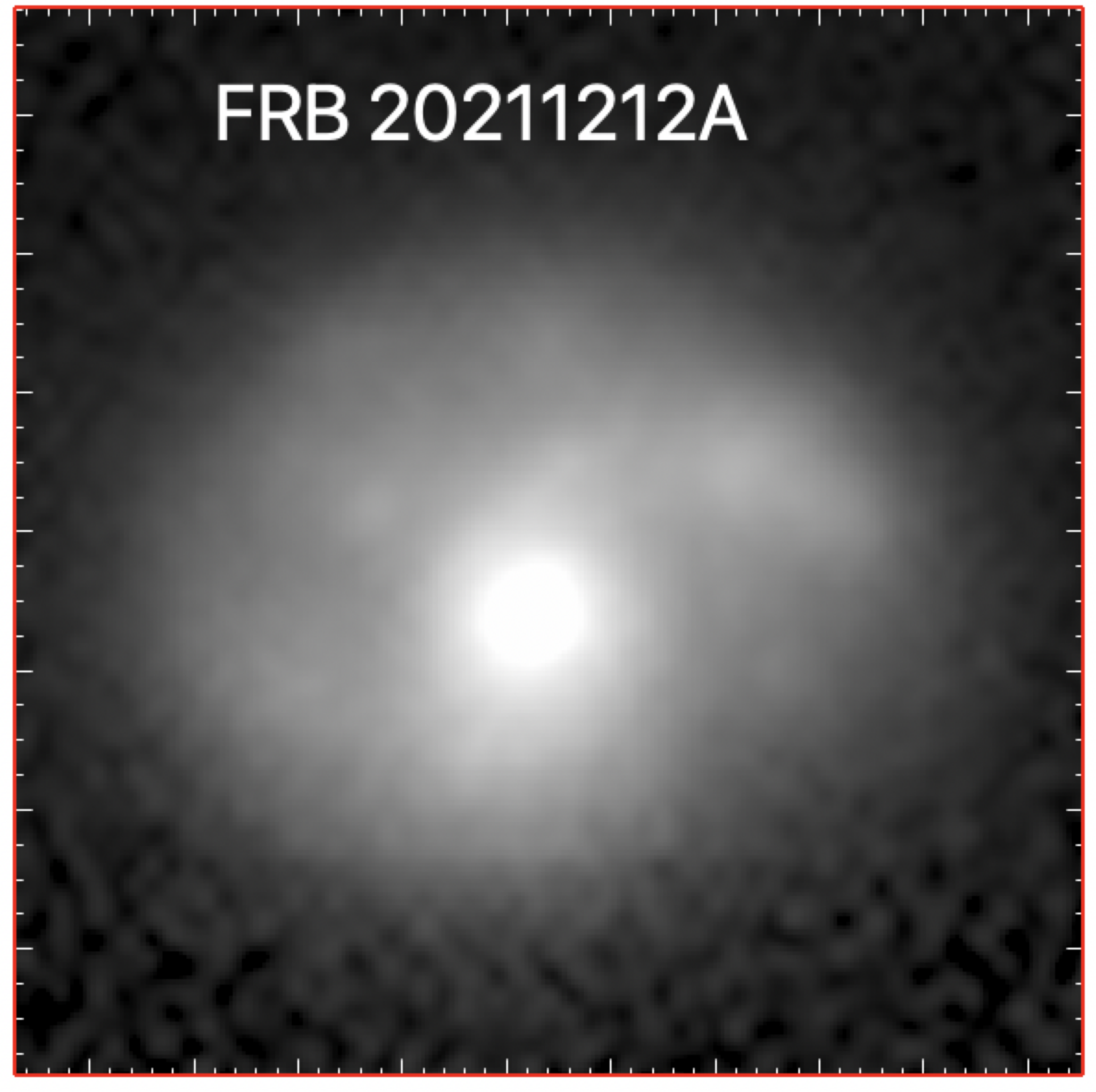}\\
\includegraphics[width=.32\textwidth,height=3.5cm]{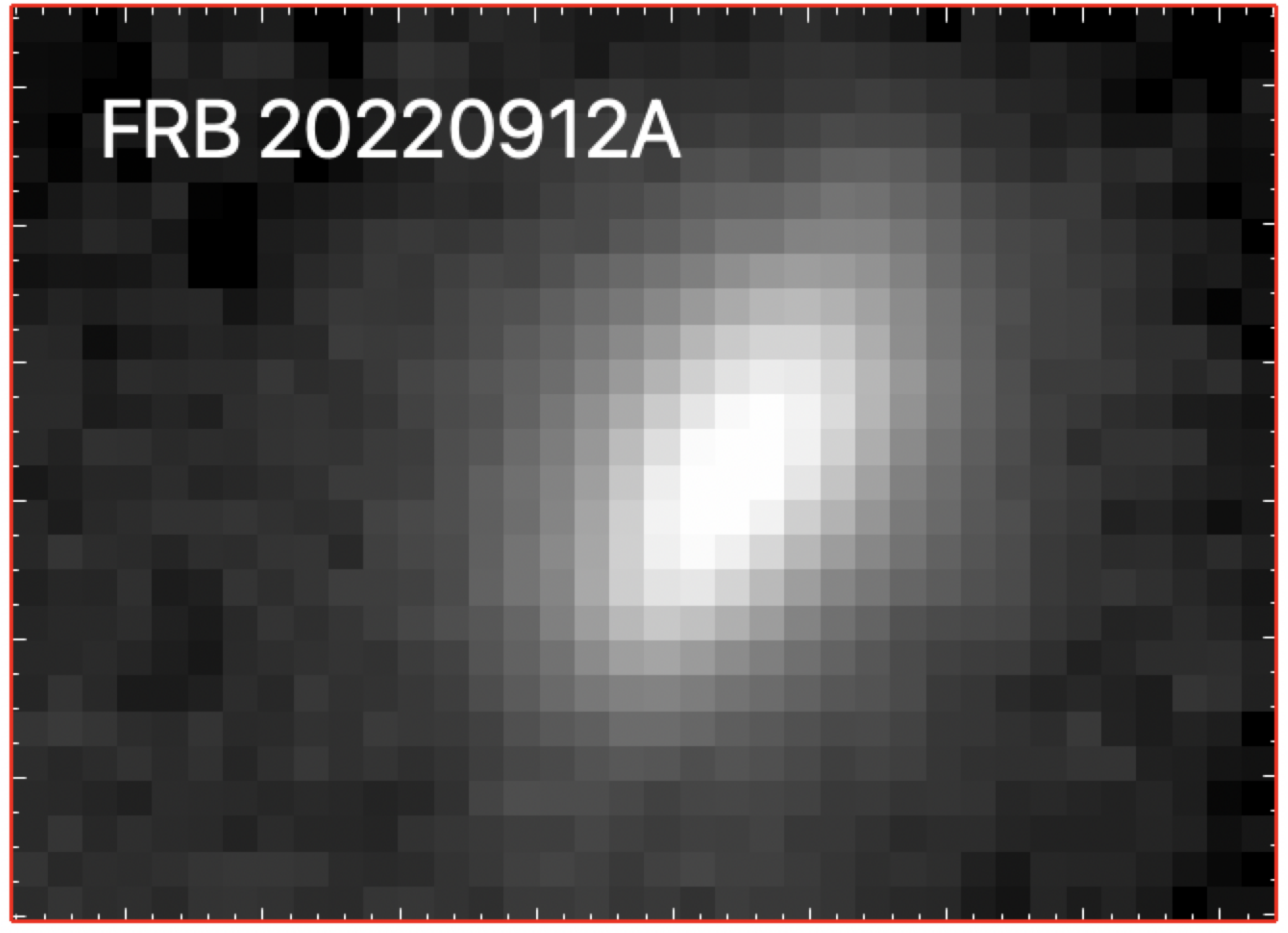}\hfill%
\includegraphics[width=.32\textwidth,height=3.5cm]{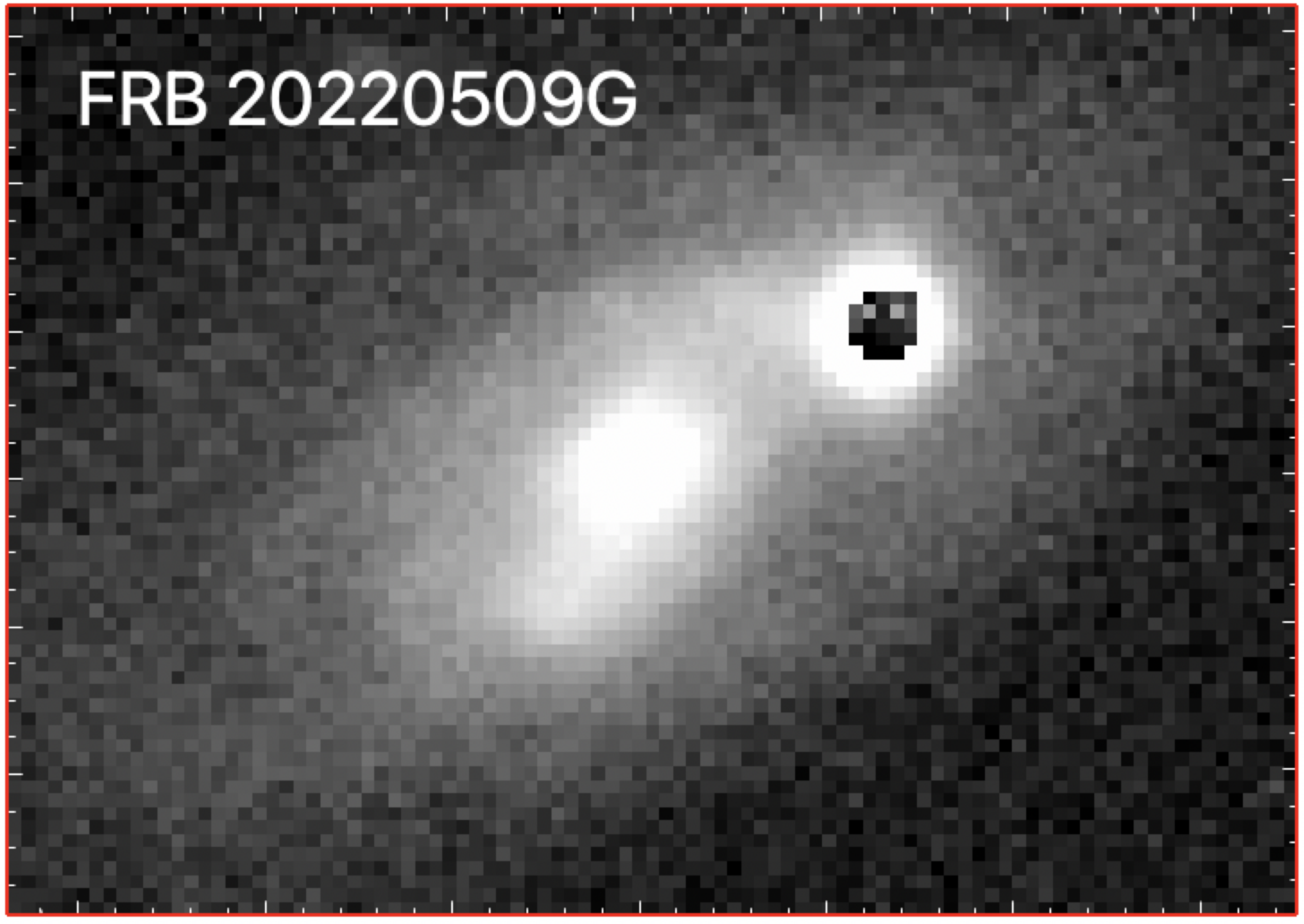}\hfill%
\includegraphics[width=.32\textwidth,height=3.5cm]{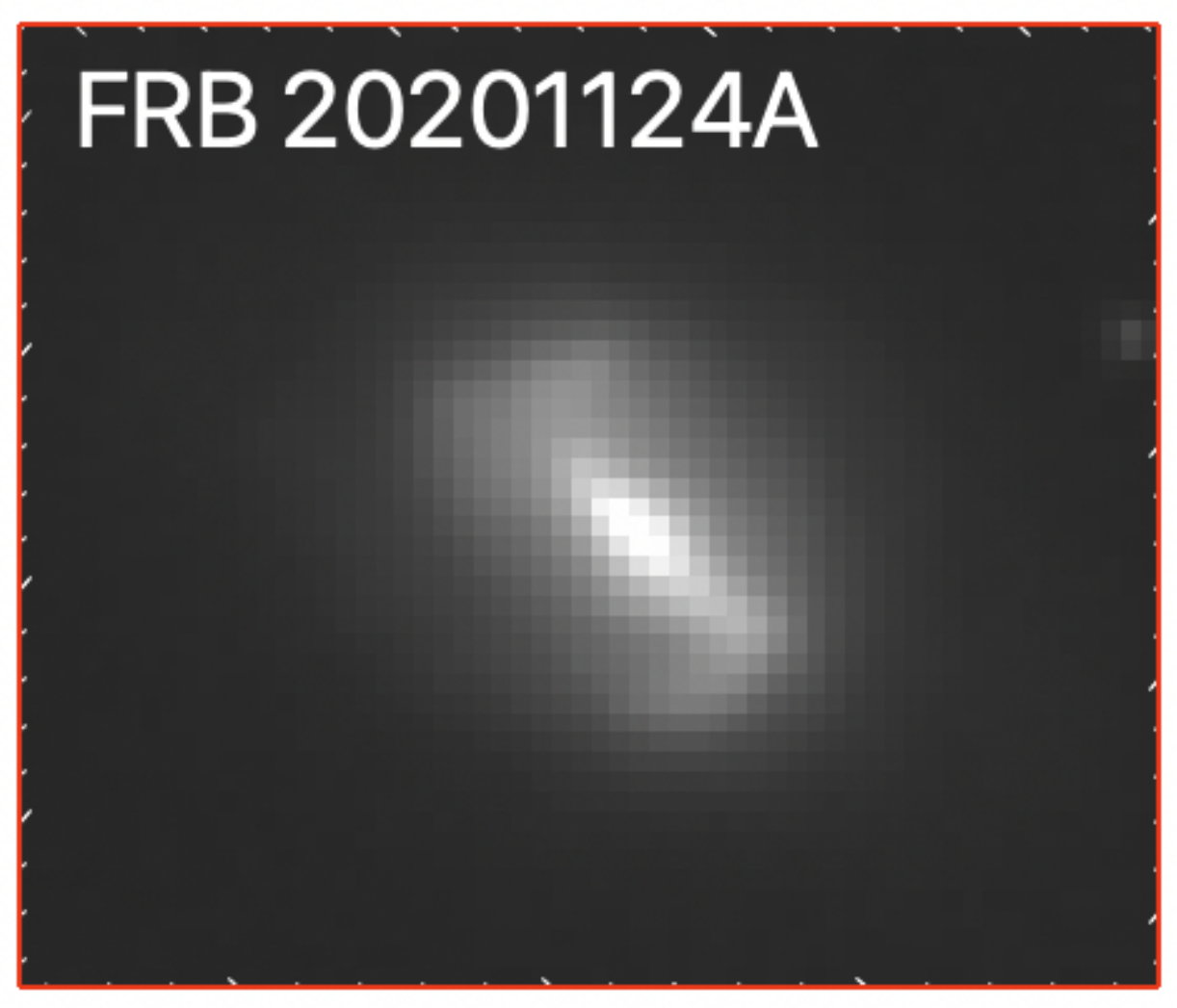}
\caption{Host galaxies of 18 local Universe FRBs used in this study, with black circles masking stars overlapping FRB 20210405I and FRB 20220509G hosts. For additional information, see Appendix \ref{app:hostsample}.}
\label{fig:host_images}
\end{figure}

In this section, we delve into major observational evidence pertaining to the morphology and structural characteristics of the galaxies in our sample of robustly associated local Universe FRB hosts as discussed in \S\ref{sec:hostdemographicsanalysis}, which lends support to their classification as spiral galaxies. As detailed in Section \ref{sec:hostdemographicsanalysis}, the presence of discernible spiral arms serves as the most compelling evidence for identifying a host galaxy as a spiral. However, in situations where the resolution or sensitivity of archival optical images is insufficient to reveal the distinctive spiral arm features, alternative structural attributes, such as the presence of a well-defined disk coupled with ongoing star formation, prove invaluable in distinguishing late-type spirals from early-type disk galaxies, notably lenticular galaxies \citep{2009ARA&A..47..159B,2015A&A...577A..97V}.

Lenticular galaxies have a structure that appears intermediate between elliptical galaxies (characterized by a dominant bulge component) and spiral galaxies (distinguished by their disk morphology but lacks spiral arms) \citep{1979JRASC..73..198V}. Their structural resemblance to elliptical galaxies is further accentuated by the fact that they have used up or lost most of their interstellar matter and therefore have very little ongoing star formation \citep{2009ApJ...702.1502V}. This unique characteristic aligns them with elliptical galaxies.
In instances where a clear disk is noted with a weak or non-existent bulge and the host shows ongoing star formation, we consider the galaxy as a late-type spiral instead of a lenticular galaxy, as suggested by \cite{1998ARA&A..36..189K} and \cite{2015MNRAS.451.2933B}.

We now discuss below major morphological features of each FRB host galaxy in our sample:

\vspace{12pt}

\noindent\textbf{FRB 20200120E}: FRB 20200120E is the closest extragalactic repeating FRB detected by the CHIME/FRB project \citep{bhardwaj2021}, which is localized to a globular cluster of M81 \citep{Kristan2021arXiv}, a nearby grand design spiral galaxy at 3.6 Mpc \citep{2019MNRAS.488..590B} with with P$_{\rm cc}$ $<$ 0.1\% \citep{Kristan2021arXiv}. Figure \ref{fig:host_images} shows the Digital Sky Survey r-band image of M81.

\noindent\textbf{FRB 20181030A}: FRB 20181030A is the second closest extragalactic repeating FRB first reported by \cite{abb+19c}. Subsequently, using CHIME/FRB baseband localization of the FRB, \cite{2021ApJ...919L..24B} reported the host galaxy of the FRB to be NGC 3252, a Sd-type spiral galaxy at 20 Mpc \citep{2018ApJ...864..123M,2019MNRAS.487.2061H}, with P$_{\rm cc} <$ 1\%. The Pan-STARRS r-band image of NGC 3252 is presented in Figure \ref{fig:host_images}.

\noindent\textbf{FRB 20171020A}: FRB 20171020A is a low-DM apparently non-repeating FRB (DM = 114 pc cm$^{-3}$) discovered by ASKAP. \cite{2018ApJ...867L..10M} identified the likely host of the FRB to be ESO 601-36, a Sc-type spiral galaxy \citep{1982euse.book.....L} at a redshift of z = 0.00867 \citep{2004MNRAS.350.1195M} with a P(O\textbar x) $\approx$ 98\% \citep{2023arXiv230517960L}. The Pan-STARRS r-band image of ESO 601-36 is shown in Figure \ref{fig:host_images}.

\noindent\textbf{FRB 20220319D}: FRB 20220319D is an apparently non-repeating FRB discovered and localized by the DSA-110 to a barred spiral galaxy IRAS 02044+7048 \citep{1995MNRAS.277..125H}, at 50 Mpc \citep{2023arXiv230101000R} with a chance association probability $\leq$ 10$^{-4}$. In Figure \ref{fig:host_images}, we showed the SDSS r-band image of the host galaxy.

\noindent\textbf{FRB 20181220A}: Discussed in \S\ref{sec:frb181220a}. The Pan-STARRS r-band image of the host galaxy, 2MFGC 17440, is displayed in Figure \ref{fig:host_images}. 

\noindent\textbf{FRB 20181223C}: Discussed in \S\ref{sec:frb181223c}. The DESI r-band image of the host galaxy, SDSS J120340.98+273251.4, is shown in Figure \ref{fig:host_images}.

\noindent\textbf{FRB 20190425A}: Discussed in \S\ref{sec:frb190425a}. The DESI r-band image of UGC 10667, the host galaxy of FRB 20190425A, is shown in Figure \ref{fig:host_images}.

\noindent\textbf{FRB 20180916B}: FRB 20180916B is a repeating FRB discovered by \cite{abb+19c}, which was later localized to the nearby massive nearly face-on spiral galaxy SDSS J015800.28+654253.0 \citep[see Figure \ref{fig:host_images}; from][]{2021ApJ...908L..12T} at redshift z = 0.0337 \citep{2020Natur.577..190M} with P$_{\rm cc}$ $\leq$ 0.1\%. 

\noindent\textbf{FRB 20220207C}: FRB 20220207C is an apparently non-repeating FRB localized by the DSA-110 to a nearby galaxy, PSO J310.1977+72.8826 \citep{2023arXiv230703344L}. From the Pan-STARRS r-band optical image of the host shown in Figure \ref{fig:host_images}, it is a nearly edge-on disk-dominated galaxy with the index of the best-fitted S\'ersic profile $\approx$ 0.5. Using the formalism discussed by \cite{2006ApJ...639L...1P}, the galaxy is likely to have the Hubble T-type morphological classification $>$ 2, suggesting the host to be a late-type spiral galaxy. 
%Moreover, we find that the galaxy is fitted well using a Sersic profile of index $\approx$ 0.6 and concentration index = 2. 
Finally, \cite{2023arXiv230703344L} noted that the host has a high star-formation rate (2.1 M$_{\odot}$/yr) which along with its disk-dominated morphology supports it to be a late-type spiral galaxy. %Therefore, the galaxy is likely a late-type spiral galaxy.

\noindent\textbf{FRB 20211127I}: FRB 20211127I is an apparently non-repeating FRB localized by ASKAP \citep{2023arXiv230205465G} to a nearby grand design spiral galaxy, 6dFGS J131914.0-185017,  at z = 0.04695 with a P(O\textbar x) $>$ 99\% \citep{2023ApJ...949...25G}. The DESI r-band image of the host is shown in Figure \ref{fig:host_images}.

%\noindent\textbf{FRB 20201123A}: FRB 20201123A is an apparently non-repeating FRB which is discovered and localized with the MeerKAT telescope to a galaxy at z = , EQ J173438.3-504550.4   In the Dark Energy Camera Plane Survey (DECaPS) (Saydjari et al. 2022) r-band image presented in Rajwade et al. 2023, the host clearly is a diskgalaxy with warped disk with ongoing star-formation. Therefore, we argue that the host is likely a late-type spiral galaxy.

\noindent\textbf{FRB 20200223B}: 
FRB 20200223B is a repeating FRB reported by \cite{2023ApJ...947...83C}. Using the CHIME/FRB baseband localization, \cite{2023arXiv230402638I} identified 
LEDA 1847306 or SDSS J003304.68+284952.6 (the SDSS r-band image is shown in Figure \ref{fig:host_images}), a Scd-type spiral galaxy \citep{2016ApJS..223...20K}, as its most likely host at z = 0.060235 with a P(O\textbar x) $>$ 90\%. 
%It is classified as a spiral galaxy due to the clear presence of spiral arms .

\noindent\textbf{FRB 20190303A}: FRB 20190303A is a repeating FRB detected by the CHIME/FRB project \citep{fonseca2020nine}. Using the sub-arcminute localization precision, \cite{2022arXiv221211941M} associated the FRB to a pair of merging spiral galaxies (the SDSS r-band image of the merger system is shown in Figure \ref{fig:host_images}), SDSS J135159.17+480729.0 and J135159.87+480714.2 \citep{1962MCG...C01....0V}, located at a redshift of z = 0.064 \citep{2012ApJS..203...21A} with a P(O\textbar x) $>$ 99\%.

\noindent\textbf{FRB 20210405I}: FRB 20210405I is an apparently non-repeating FRB detected using MeerKAT \citep{2023arXiv230209787D} and was also observed commensally with the ThunderKAT survey \citep{2016mks..confE..13F} which facilitated its sub-arcsecond localization and subsequent association to a nearby galaxy named 2MASS J1701249-4932475 at a redshift of z = 0.066 with P$_{\rm cc}$ $< 1$\%. \cite{2023arXiv230209787D} identify the morphology of the galaxy to be a spiral in the DECaPS2 DR2 r-band image \citep{2023ApJS..264...28S} as shown in Figure \ref{fig:host_images}.
%where most of the galaxy's disk is obscured by high Galactic extinction (A$_{v}$ = 2.41). 
Moreover, the high inferred star-formation rate of the host as inferred by \cite{2023arXiv230209787D} is consistent with it being a star-forming spiral galaxy. 

%\noindent\textbf{FRB 20180814A}: extended disk structure which fitted well with a sersic profile of index n = 1.3. 

\noindent\textbf{FRB 20190418A}: Discussed in \S\ref{sec:frb190418a}. The UKIDSS-DR9 GCS K-band image of the host is shown in Figure \ref{fig:host_images}.

\noindent\textbf{FRB 20211212A}: FRB 20211212A is an apparently non-repeating FRB discovered by ASKAP \citep{2023arXiv230205465G}. Using the sub-arcsecond localization of the FRB, \cite{2023arXiv230205465G} localized it to a star-forming spiral galaxy, SDSS J102924.22+012139.2, at z = 0.0707 \citep{2016ApJS..223...20K}, as can be inferred from the r-band DESI image of the host shown in Figure \ref{fig:host_images}.

\noindent\textbf{FRB 20220912A}: FRB 20220912A is a highly active repeating FRB that was discovered by the CHIME/FRB project \citep{2022ATel15679....1M}. The FRB was localized to a disk-dominated galaxy with a moderate SFR ($\approx$ 0.1 M$_{\odot}$/yr) at z = 0.077 by \cite{2023ApJ...949L...3R}. This is supported by the fact that the best-fitted S\'ersic profile has an index $= 1$ suggesting it is a disk galaxy with a low degree of central concentration \citep[bulge-to-total
luminosity ratio $\leq$ 0.2;][]{2016ApJ...824..112D,2015MNRAS.450..763M}. This suggests that the FRB host likely has a Hubble sequence classification $> 0$ \citep{2011A&A...532A..74B}, which makes it likely a late-type spiral galaxy. The GTC r-band image of the host galaxy is shown in Figure \ref{fig:host_images}.

\noindent\textbf{FRB 20220509G}: Discussed in Appendix \ref{sec:FRB20220509G}.
The CFIS DR3 r-band image of the FRB 20220509G host is shown in Figure \ref{fig:host_images}.

\noindent\textbf{FRB 20201124A}: FRB 20201124A is a repeating FRB first detected by the CHIME/FRB project \citep{2022ApJ...927...59L}. It was localized to a galaxy named SDSS J050803.48+260338.0 at z = 0.098 by ASKAP \citep{2021ApJ...919L..23F}, the Very Large Array \citep{2022MNRAS.513..982R}, and the European Very Long Baseline Interferometry Network \citep{2022ApJ...927L...3N} with P$_{\rm cc}$ $\leq$ 0.1\%. \cite{2022Natur.609..685X}, using the 10-m Keck telescopes, identified SDSS J050803.48+260338.0 as a barred spiral galaxy (shown in Figure \ref{fig:host_images}).

%--------------- Appendix C ---------------------

\section{Nature of the host of FRB 20220509G}
\label{sec:FRB20220509G}

\cite{2023ApJ...950..175S} and \cite{2023ApJ...949L..26C} claimed that the FRB 20220509G host is a quiescent, elliptical (hence, early-type) galaxy. This would make it the only FRB  localized to an elliptical galaxy, and an outlier among all the local Universe FRBs considered in \S\ref{subsec:spiralhost}. However, the FRB host is classified as a spiral based on the WISE color-color classification discussed in  \S\ref{subsec:spiralhost}. Moreover, there is an indication of a bar-like extended structure in the host galaxy Pan-STARRS image presented by \cite{2023ApJ...950..175S}. However, the tentative detection of a ``bar-like" structure is not strong enough to claim that it is a spiral galaxy. Therefore, we searched for a deeper image in different optical archival surveys, and found a deeper r-band image of the FRB FOV in
the Canada-France Imaging Survey (CFIS) legacy survey data release 3 \citep{2017ApJ...848..128I,2019ApJ...887..148F} with a median point source depth of 5$\sigma$ $=$ 25 AB magnitude.\footnote{The FRB FOV image is accessed via the Canadian Astronomy Data Centre: \url{https://www.cadc-ccda.hia-iha.nrc-cnrc.gc.ca/en/search/}}

The FRB 20220509G host galaxy detected in the CFIS DR3 r-band image is shown in Figure \ref{fig:frb20220509g_cfis}, where we clearly see the host's extended spiral arm and bar-like morphological features. Hence, we classify the FRB host as a spiral, and hence, a late-type galaxy, in \S\ref{subsec:spiralhost}.

%-------------- Appendix D ----------------------

\section{Pantheon sample of Type Ia SUPERNOVAE}
\label{sec:typeIa_frb}

\begin{figure}[h]
\centering
\includegraphics[width=.6\linewidth]{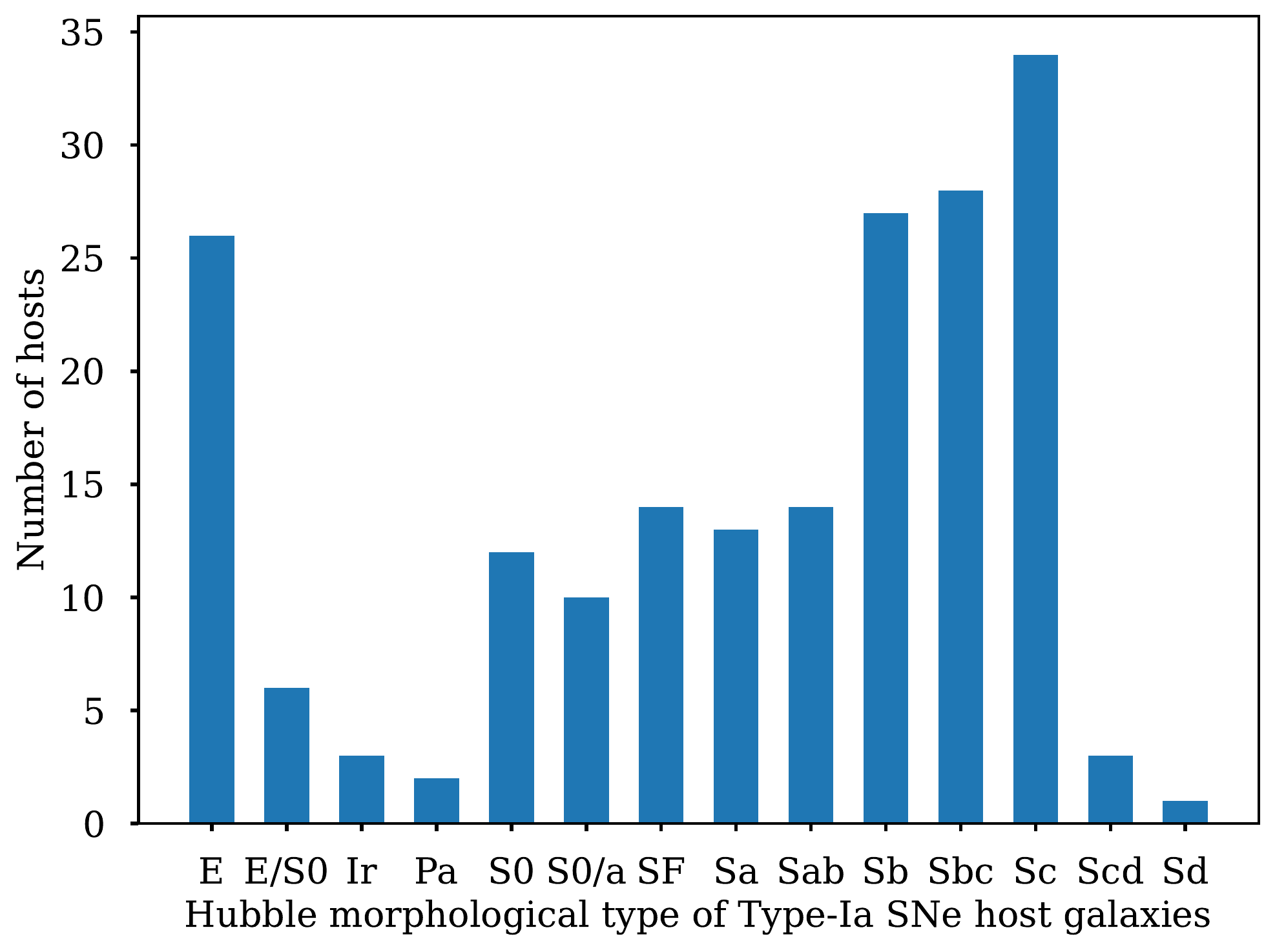}%\centering{}
\caption{Distribution of the host galaxies of the Pantheon SN Ia sub-sample according to their morphological classes as discussed in Appendix \ref{sec:typeIa_frb}. For more detail on the classification scheme, see \cite{2020MNRAS.499.5121P}.}
\label{fig:ia_sne_fraction}
\end{figure}

The Pantheon cosmological supernova sample \citep{2018ApJ...859..101S} is the largest spectroscopic sample of Type Ia Supernova (SN Ia) available, comprising a total of 1047 SNIa with redshifts spanning from 0.01 $<$ z $<$ 2.3. For a subset of 330 SNIa from the Pantheon sample, \cite{2020MNRAS.499.5121P} reported their host morphology. They classified visually the 330 galaxies into the following five morphological classes: spheroid (E/S0), spheroid+disk (S0/Sa), disk (Sb/Sbc/Sc), disk+irregular (Sc/Scd), irregular (Scd/Ir). Furthermore, \cite{2020MNRAS.499.5121P} categorize galaxies with morphological classes `E', `E/S0', `Pa', and `S0' as early-type galaxies, and classes `S0/a',`Sa', `Sab', `Sb', `Sbc', `Sc', `Scd', `Sd',`SF', and `Ir' as late-type galaxies. For our analysis, we consider a subset of those SNIa host galaxies with z $\leq$ 0.1, which gives us a sample of 193 SNIa host galaxies, consisting of 46 early-type and 147 late-type galaxies. Figure \ref{fig:ia_sne_fraction} shows the distribution of 193 SN Ia host galaxies in each of the aforementioned morphological categories. Based on this, we estimate the fraction of `late-type' galaxies in the sample to be 0.76. This fraction serves as the null hypothesis against which we conduct a comparison with the local Universe FRB host sample, as elaborated in \S\ref{subsec:preferred_channel}.

%\section{Comparing FRB and Core-collapse Supernova Local Universe Hosts}
%\label{sec:CCSNe_frb}

\end{document}